\documentclass[a4paper,11pt]{article}
\usepackage{aaskaiid}
\usepackage{orcidlink}

\def\swift{{\em Swift}}
\def\fermi{{\em Fermi}}

\newcommand{\skalow}{SKA-Low}
\newcommand{\skamid}{SKA-Mid}

\title{Rapid Response Triggering for Radio Transients with the SKA Observatory}
\ShortTitle{Rapid response radio follow-up}

\author[1,2]{G. E. Anderson\orcidlink{0000-0001-6544-8007}}
\ShortName{Anderson et al.} 
\author[3,4]{S. I. Chastain\orcidlink{0000-0003-3507-335X}}
\author[5,6]{A. P. Curtin\orcidlink{0000-0002-8376-1563}}
\author[1]{K. Gourdji\orcidlink{0000-0002-0152-1129}}
\author[7]{N. Hurley-Walker\orcidlink{0000-0002-5119-4808}}
\author[7]{C. W. James\orcidlink{0000-0002-6437-6176}}
\author[8,1]{R. F. Mandow\orcidlink{0000-0001-5131-522X}}
\author[9,10]{B.~Marcote\orcidlink{0000-0001-9814-2354}}
\author[11]{D. Oberoi\orcidlink{0000-0002-4768-9058}}
\author[12]{K. M. Rajwade\orcidlink{0000-0002-8043-6909}}
\author[2,1]{K. Rose \orcidlink{0000-0002-7329-3209}}
\author[13,10]{A. Rowlinson\orcidlink{0000-0002-1195-7022}}
\author[14]{F. Sch\"ussler\orcidlink{0000-0003-1500-6571}}

\affiliation[1]{Australia Telescope National Facility, CSIRO, Space and Astronomy, PO Box 1130, Bentley, WA  6151, Australia}
\affiliation[2]{Sydney Institute for Astronomy, School of Physics, The University of Sydney, NSW 2006, Australia}
\emailAdd{gemma.anderson.astro@gmail.com}
\affiliation[3]{$^1$Department of Physics and Astronomy, University of New Mexico, 210 Yale Blvd NE, Albuquerque, NM, 87106, USA}
\affiliation[4]{Department of Physics \& Astronomy, Texas Tech University, PO Box 41051, Lubbock, TX, 79409, USA}
\affiliation[5]{Department of Physics, McGill University, 3600 rue University, Montr\'eal, QC H3A 2T8, Canada}
\affiliation[6]{Trottier Space Institute, McGill University, 3550 rue University, Montr\'eal, QC H3A 2A7, Canada}
\affiliation[7]{International Centre for Radio Astronomy Research, Curtin University, Kent St, Bentley WA 6102, Australia}
\affiliation[8]{Department of Mathematics and Physical Sciences, Macquarie University, NSW 2109, Australia}
\affiliation[9]{Joint Institute for VLBI ERIC, Oude Hoogeveensedijk 4, 7991~PD Dwingeloo, The Netherlands.}
\affiliation[10]{ASTRON, Netherlands Institute for Radio Astronomy, Oude Hoogeveensedijk 4, 7991~PD Dwingeloo, The Netherlands.}
\affiliation[11]{National Centre for Radio Astrophysics, Tata Institute of Fundamental Research, S. P. Pune University Campus, Pune, India}
\affiliation[12]{Astrophysics, University of Oxford, Denys Wilkinson Building, Keble Road, Oxford OX1 3RH, UK}
\affiliation[13]{Anton Pannekoek Institute for Astronomy, University of Amsterdam, Science Park 904, P.O. Box 94249, 1090GE Amsterdam, The Netherlands}
\affiliation[14]{IRFU, CEA, Université Paris-Saclay, F-91191 Gif-sur-Yvette, France}

\abstract{Rapid-response triggering is when a telescope is able to automatically respond to an external or internal astronomical transient alert, causing it to rapidly repoint at that position in the sky to catch its earliest radio emission. 
Both SKA-Low and SKA-Mid will have the ability to perform rapid-response triggering observations on externally detected transients as well as those detected within the data streams. 
We first give a brief overview of those radio instruments with active rapid-response observing modes.
We then describe the different science cases motivating the need for this observing capability on SKAO and how the additional sensitivity afforded by the SKAO will enable us to answer fundamental questions relating to particle acceleration, transient central engines, coherent emission models and outflow physics in astrophysical systems spanning the range from the Sun to the high redshift Universe. Several suggestions relating to existing technologies and necessary SKAO system requirements are described. Through this chapter, we aim to ensure this is an existing, common and useful capability for the SKA Observatory.}

\begin{document}
\newcommand{\actaa}{Acta Astron.} 
\newcommand{\araa}{Annu. Rev. Astron. Astrophys.} 
\newcommand{\aar}{Astron. Astrophys. Rev.} 
\newcommand{\ab}{Astrobiol.} 
\newcommand{\aj}{Astron. J.} 
\newcommand{\apj}{Astrophys. J.} 
\newcommand{\apjl}{Astrophys. J. Lett.} 
\newcommand{\apjs}{Astrophys. J. Suppl. Ser.} 
\newcommand{\ao}{Appl. Opt.} 
\newcommand{\apss}{Astrophys. Space Sci.} 
\newcommand{\aap}{Astron. Astrophys.} 
\newcommand{\aapr}{Astron. Astrophys. Rev.} 
\newcommand{\aaps}{Astron. Astrophys. Suppl.} 
\newcommand{\baas}{Bull. Am. Astron. Soc.} 
\newcommand{\caa}{Chinese Astron. Astrophys.} 
\newcommand{\cjaa}{Chinese J. Astron. Astrophys.} 
\newcommand{\cqg}{Class. Quantum Gravity} 
\newcommand{\gal}{Galaxies} 
\newcommand{\gca}{Geochim. Cosmochim. Acta} 
\newcommand{\icarus}{Icarus} 
\newcommand{\jcap}{J. Cosmol. Astropart. Phys.} 
\newcommand{\jgr}{J. Geophys. Res.} 
\newcommand{\jgrp}{J. Geophys. Res.: Planets} 
\newcommand{\jqsrt}{J. Quant. Spectrosc. Radiat. Transf.} 
\newcommand{\memsai}{Mem. Soc. Astron. Italiana} 
\newcommand{\mnras}{Mon. Not. R. Astron. Soc.} 
\newcommand{\nat}{Nature} 
\newcommand{\nastro}{Nat. Astron.} 
\newcommand{\ncomms}{Nat. Commun.} 
\newcommand{\nphys}{Nat. Phys.} 
\newcommand{\na}{New Astron.} 
\newcommand{\nar}{New Astron. Rev.} 
\newcommand{\physrep}{Phys. Rep.} 
\newcommand{\pra}{Phys. Rev. A} 
\newcommand{\prb}{Phys. Rev. B} 
\newcommand{\prc}{Phys. Rev. C} 
\newcommand{\prd}{Phys. Rev. D} 
\newcommand{\pre}{Phys. Rev. E} 
\newcommand{\prl}{Phys. Rev. Lett.} 
\newcommand{\psj}{Planet. Sci. J.} 
\newcommand{\planss}{Planet. Space Sci.} 
\newcommand{\pnas}{Proc. Natl Acad. Sci. USA} 
\newcommand{\procspie}{Proc. SPIE} 
\newcommand{\pasa}{Publ. Astron. Soc. Aust.} 
\newcommand{\pasj}{Publ. Astron. Soc. Jpn} 
\newcommand{\pasp}{Publ. Astron. Soc. Pac.} 
\newcommand{\rmxaa}{Rev. Mexicana Astron. Astrofis.} 
\newcommand{\sci}{Science} 
\newcommand{\sciadv}{Sci. Adv.} 
\newcommand{\solphys}{Sol. Phys.} 
\newcommand{\sovast}{Soviet Ast.} 
\newcommand{\ssr}{Space Sci. Rev.} 
\newcommand{\uni}{Universe} 

\maketitle

\section{Introduction}

Rapid-response triggering is an operational mode that enable telescopes to automatically and rapidly repoint and begin observing an astronomical event upon receiving a alert.
Such an alert could be broadcast by dedicated instruments that search for astronomical transient signals or be internally generated when a transient is detected within a telescope's own datastream, providing an internal trigger for a different observing mode.
Some of the first automated transient alerts were of gamma-ray bursts (GRBs) detected by the Burst And Transient Source Experiment \citep[{\it BATSE};][]{fishman94} transmitted via BACODINE, the Real-Time BATSE Gamma-Ray Burst Coordinates Distribution Network system \citep{barthelmy95}.
This set the scene for GRBs to be the main target for rapid follow-up experiments, which were crucial for localising multi-wavelength counterparts \citep[e.g.][]{vanparadijs97}, enabling the confirmation of their extragalactic origin and likely progenitors. 
As more high-energy missions were launched and multi-messenger facilities built, transient alert networks have expanded to ensure rapid and global dissemination. BACODINE evolved into NASA's General Coordinates Network (GCN),\footnote{https://gcn.nasa.gov/} which transmits transient alerts over a Kafka\footnote{https://kafka.apache.org/} client from a plethora of missions, including (but not limited to) the \textit{Neils Gehrels Swift Observatory} \citep[hereafter \swift;][]{gehrels04,barthelmy05}, the \textit{Fermi Gamma-ray Space Telescope} \citep[hereafter \fermi;][]{atwood09,meegan09}, \textit{Einstein Probe} \citep{yuan22ep}, the \textit{Space-based multi-band astronomical Variable Objects Monitor} \citep[{\it SVOM}:][]{gonzalez18}, LIGO/Virgo/KAGRA \citep[LVK;]{abbott20}, and the IceCube Neutrino Observatory \citep{aartsen17}. 

Meanwhile, all-sky optical transient surveys including, Sloan Digital Sky Survey-II Supernova Survey \citep{sako08}, Pan-STARRS \citep{kaiser10}, the (intermediate) Palomar Transient Factory \citep[PTF;][]{rau09,law09,cao16,masci17}, Zwicky Transient Facility \citep[ZTF;][]{bellm19}, SkyMapper \citep{scalzo17}, the All-Sky Automated Survey for Supernovae \citep{kochanek17}, and the Gaia Science Alerts project \citep{hodgkin21} have revolutionized optical transient discovery and alert dissemination. These techniques have paved the way for transient alert Brokers, which are now managing the overwhelming number of transients detected nightly by the Rubin Observatory\footnote{https://rubinobservatory.org/} Legacy Survey of Space and Time \citep[LSST;][see Section~\ref{sec:alerts} for further details on the Brokers]{2019ApJ...873..111I}.\footnote{https://rubinobservatory.org/for-scientists/data-products/alerts-and-brokers} 
All of these missions provide amazing and diverse opportunities for studying the very earliest physics of astronomical transients via rapid-response follow-up with SKAO. 

While optical and X-ray rapid-response follow-up of transients is common \citep[e.g. \swift{} performs automatic follow-up of GRBs using the onboard X-ray Telescope (XRT) and Ultra-violet Optical Telescope (UVOT),][]{burrows05,roming04}, few radio facilities have participated in similar observing campaigns, likely due to their paucity and historical lack of sensitivity. 
The first rapid-response experiment was conducted with the Cambridge Low Frequency Synthesis Telescope (CLFST) in the 1990s, which triggered on {\it BATSE}-detected GRBs at 151\,MHz to search for associated prompt radio emission \citep{green95,dessenne96}.
However, it was the discovery of fast radio bursts \citep[FRBs;][]{lorimer07} that re-inspired rapid-response radio experiments after those conducted with CLFST \citep[e.g.][]{bannister12,palaniswamy14,kaplan15} as many models suggest these mysterious transients could be associated with GRBs \citep[][]{rowlinson19anderson}.
In 2012, higher frequency ($15$\,GHz) rapid-response triggering began with the Arcminute Microkelvin Imager Large Array \citep[AMI-LA;][]{zwart08} based in Cambridge in the UK. This program used AMI-LA to trigger automatic follow-up of \swift{} GRBs \citep{staley13}.  
The AMI-LA Rapid-Response Mode \citep[ALARRM;][]{staley13,anderson18} program could be on target within minutes, only constrained by its slew speed, resulting in the detection of GRB reverse shock evolution \citep{anderson14}, the first unbiased radio catalogue of GRBs from minutes to weeks post-burst \citep{anderson18}, the detection of a radio flare that was simultaneous with a \swift-detected high-energy (X-ray and $\gamma$-ray) superflare from a local M dwarf star \citep{fender15}, and radio flares associated with X-ray binary outbursts \citep{bright20,fender23}.
These historical programs have inspired SKAO precursors to be equipped with rapid-response capabilities. 
As a result, this mode of operation also forms part of the SKAO design baseline.\footnote{https://www.dropbox.com/scl/fi/4aiatav4gqxh3hwom7901/SKA-TEL-SKO-0001075-02\_DesignBaselineDescription.pdf?rlkey=8qme7b9spoo21s0hlda28li0t\&e=1\&st=vjy1oa2k\&dl=0}
In this chapter, we describe the radio telescopes that are currently operating a rapid-response observing mode and the lessons we have learned for the SKAO. We also provide science cases for rapid-response triggering with the SKAO for different source and multi-messenger event types, recommending SKAO rapid-response observing strategies and synergies. Finally, we describe the current triggering technology and landscape. 

\section{Current radio telescopes with rapid-response capabilities}


For the purposes of this chapter, we classify a radio telescope (operating within the \skalow\ and \skamid\ frequency range) as having a rapid-response capability if its response to an alert causes all or part of the instrument to repoint (change its sky position) and/or activates a different or additional observing mode. 
We note that while other instruments and programs have searched for prompt emission associated with GRBs through all-sky monitoring and serendipitous searches, they are not rapid-response observations so we have not detailed them here. However, please see the following references for details on some of these programs \citep{obenberger14,obenberger15,kuiack21,shulevski22,curtin23,curtin24}.

\subsection{MWA}

The Murchison Widefield Array \citep[MWA;][]{tingay13,wayth18,tingay26pp} began rapid-response triggering in 2015 \citep{kaplan15}. 
The MWA's wide field of view makes it useful for targeting transients with poor positional localisations \citep[e.g.][]{abbott16,kaplan16}. 
The theoretical hardware minimum for MWA to repoint with the new MWAX correlator \citep{morrison23} once a transient trigger has been received is 8-16 seconds, making it the fastest repointing rapid-response telescope in the world.\footnote{https://mwatelescope.atlassian.net/wiki/spaces/MP/pages/552435733/MWA+Rapid-Response+Triggering}. 
The MWA back-end trigger web services\footnote{https://mwatelescope.atlassian.net/wiki/spaces/MP/pages/24972656/Triggering+web+services} are described in \citet{hancock19voevent}, which accept parameters such as the transient's position and the setup of the observation.
These include the observing frequencies and whether you want the full array, sub-arrays or apodising (using a subset of dipoles within each tile). If the target is in the sky, the MWA scheduler will be automatically updated and begin observations.
It will also select a suitable calibrator source to observe following the triggered observation. 
There is also a Sun suppression algorithm that places the Sun within a primary beam null while optimising for the transient's position within the primary beam for daytime observing. 
The triggers can either be sent directly to the telescope or via the transient alert parsing tool TRACE-T\footnote{https://github.com/ADACS-Australia/TraceT2} (see future details in Section~\ref{sec:trigg_service}).
Rapid-response observations can be conducted in the standard imaging mode, which has a minimum intregration time of 250\,ms, and in a voltage capture mode \citep{tremblay15} with a temporal resolution of 781.25\,ns \citep{morrison2023mwax}. It also has a voltage capture ring buffer capable of storing 240\,s of data for 128 MWA tiles. 
The best triggering latencies are $<20$\,s between receiving an alert and beginning the observation, which are demonstrated by the GRB triggering programs \citep[e.g.][]{kaplan15,anderson21mwa,tian22a,tian22b,xu25}, with preparation for gravitational wave (GW) events \citep[][]{kaplan16,hancock19voevent,tian23b}. With the added benefit of dispersion delay at low frequencies, MWA can be on target in time to detect radio signals emitted at cosmological distances (see Section~\ref{sec:grb_prompt} for the GRB and GW science cases).

More recently, MWA can now trigger observations to capture transients associated with solar and heliospheric phenomena \citep{patra26}. For such observations, MWA is pointed at the Sun, which ensures that the voltage buffer is always filled with solar data, though the data is not recorded. Only when a solar trigger is received are the buffered voltages dumped to disk. 
The latter provides data prior to the arrival of the trigger and partially compensates for the latency of the trigger. 

\subsection{LOFAR}

The Low Frequency Array (LOFAR) has also been performing rapid-response triggering since the new mode was implemented in 2017. The early triggering strategies enable rapid response observations using the high band (110-240\,MHz) or the low band (30-80\,MHz). The triggered observations can use the full Dutch array, corresponding to 24 core stations and 14 remote stations \citep{rowlinson19}, and could either perform imaging or beamformed observations. With imaging observations, it is possible to use multiple beam pointing directions to tile out larger sky areas. 
The triggering software identifies the optimal calibrator source to be observed following the triggered observation and conducts various checks such as that the source remains above the horizon during the observation. Due to software and hardware limitations, the LOFAR rapid response mode has a response time of $4-6$\,minutes. Thus far LOFAR rapid-response triggering has targeted GRBs \citep[][see Section~\ref{sec:grb_prompt} for the science case]{rowlinson19,rowlinson21,rowlinson24,hennessy23,hennessy25}, with preparation for GW events \citep{gourdji23}. 

In addition to the rapid response mode, LOFAR has Transient Buffer Boards (TBBs) that can be triggered almost instantly giving up to 5 seconds of the full raw LOFAR data. These data give full sky coverage, individual antenna data, full time and frequency resolution, giving roughly 200\,GB per LOFAR station triggered. This mode is typically used for triggering on cosmic rays \citep[e.g.][]{schellart2013} and lightning \citep[e.g.][]{hare2020}.

LOFAR has undergone a significant upgrade to LOFAR2.0\footnote{https://www.lofar.eu/wp\-content/uploads/2023/04/LOFAR2\_0\_White\_Paper\_v2023.1.pdf}, which has provided significant improvements for the rapid response mode. Firstly, the new system aims to respond to triggers within 1 minute. Secondly LOFAR2.0 enables simultaneous use of the high and low band antennas or the use of more antennas to increase sensitivity. Thirdly, LOFAR2.0 enables simultaneous imaging and beamformed observations, giving the opportunity to simultaneously search for transient emission from millisecond duration to hours.

\subsection{ATCA}

ATCA began performing rapid-response triggering programs in 2017, with the operational mode outlined in \citet{anderson21}. If the target is in the ATCA observable sky, its repointing latency is slew-limited, allowing it to repoint within 10\,minutes. 
On receiving a trigger, an appropriate gain and flux calibrator is selected, with the gain calibrator observed before the GRB position. The resulting schedule file swaps between the gain calibrator and GRB position, with the appropriate time span automatically selected based on the current atmospheric conditions.
This mode can be operated using the ATCA 16\,cm, 4\,cm, 15\,mm and 7\,mm receivers, which covers a frequency range of $1-20$\,GHz. With the upgraded correlator BIGCAT,\footnote{https://www.atnf.csiro.au/projects/instrumentation/bigcat/} each receiver has an 8\,GHz bandwidth (except for the 16\,cm receiver, which continues to have a 2\,GHz bandwidth). The triggers can be sent to ATCA via the TRACE-T alert parsing tool (see Section~\ref{sec:trigg_service}), which are processed by the backend. 
Given the arcminute field of view, ATCA rapid-response observations are best triggered on well localised transients. ATCA rapid-response observations have primarily targeted GRBs detected by \swift{}  \citep{anderson21,chastain24}, resulting in the earliest radio detections of GRBs to date \citep[][see Section~\ref{sec:grb_synch} for the science case]{anderson24,anderson25,chastain26pp}.

\subsection{MeerKAT}

MeerKAT has a commensal system called MeerTRAP that searches for dispersed coherent transients, such as FRBs and pulsars. Since 2020, MeerTRAP has been piggybacking on the majority of observing programs, searching for dispersed coherent transient signals such as FRBs and pulsars. MeerTRAP operates on observations taken with any of the three receivers ($S$-band, $L$-band and the UHF), summing the incoming data from all MeerKAT dishes to form an incoherent beam that covers the whole field of view. Coherent beams from a subset of antennas that cover a smaller field of view, providing a factor of $\sim5$ increase in sensitivity to dispersed signals. MeerTRAP performs a real-time search for single pulses over a range of DMs, which are frequency- and buffer-size-dependent.
Transients are identified in near-real-time via a machine learning classifier, which generates a VOEvent (see Section~\ref{sec:auto}) that is sent to the South African Radio Astronomy Observatory (SARAO) broker. The broker then triggers the transient buffer within 45\,seconds of the transient detection, which dumps its 50\,seconds of data capacity \citep[see][for details on the instrumentation and software\footnote{https://github.com/fjankowsk/meertrig/}]{caleb20,rajwade20,rajwade22,rajwade24,jankowski22}.
As a result, MeerTRAP has successfully detected and localised new pulsars \citep{caleb22,smirnov24,turner25,tian25psr}, Galactic transients \citep{bezuidenhout22} and FRBs \citep{rajwade20,rajwade24,caleb23,jankowski23,driessen24,tian24,tian25frb}. 

Although the intention is for MeerTRAP VOEvents to be disseminated to the larger astronomical community in the future \citep{rajwade24}, the current false positive rate is too high. 
One issue that contributes to the high false positive rate is the inaccuracy of DMs in existing pulsar catalogues. This is particularly a problem with the MeerKAT $S$-band receiver, where the measured DMs are often outside the tolerance range and so not recognised as known sources. Another issue is that the removal of RFI at zero DM can distort the DM sweeps, causing them to be measured at a smaller value than they truly are. 
Such issues need to be considered for the SKAO and could be mitigated by creating an accurate DM pulsar catalogue and then using clustering algorithms on commensally measured DMs to prevent known sources from being misclassified.

\subsection{OVRO-LWA}

The most recent addition to radio rapid-response triggering facilities is the Owens Valley Radio Observatory Long Wavelength Array \citep[OVRO-LWA;][]{taylor12}, which is an all-sky ($\sim20,000$ field of view), low-frequency ($27$--$84$\,MHz) radio interferometer. 
On receiving transient alerts via the GCN, OVRO-LWA performs a buffer dump, which lead to searches for prompt radio emission. The original mode had a 13-second resolution, allowing for image dedispersion searches for coherent signals from a GRB \citep{anderson18lwa} and a GW event \citep{callister19}.
Their observational set-up now has a new integrated system called Time Machine that triggers voltage buffer and voltage capture observations on receiving transient alerts with $<10$\,s latency \citep{kosogorov25}. It involves a two-stage buffering system, with the first stage buffer providing a three-minute look-back time that is optimised for high-speed, low-latency storage, which is then transferred to the second stage that continues to record and store up to 30 minutes of data. Once the data is transferred to a long-term storage system, off-line beamforming is performed in the $55$--$85$\,MHz band with a temporal and spectral resolution of 1.3\,ms and 0.7 kHz, respectively. The transient pipeline then searches for dispersed signals up to a DM of 400\,pc\,cm$^{-3}$ with a fluence sensitivity of $\sim100$\,Jy\,ms. The first GW trigger with Time Machine is presented by \citet{kosogorov26}.

\section{Gamma-ray bursts and gravitational wave events: a rapid-response case study}

GRBs have historically been the main motivation for rapid-response triggering, likely due to the prompt nature of their gamma-ray emission.
GRB progenitors are thought to result from either massive stellar collapse or from the merger of a binary neutron star (BNS) or neutron star and black hole (NSBH). They are usually referred to as long or short GRBs, respectively, which is based on the duration of the gamma-ray emission \citep[][]{kouveliotou93}. Further confirmation of short GRBs originating from BNS mergers was obtained with the near-simultaneous detection of the GW-detected BNS merger GW170817 and the short GRB 170817A \citep{abbott17a,abbott17b}. 
The collapse or merger and subsequent accretion of material onto its central engine drives relativistic jets that interact with the surrounding medium and give rise to a synchrotron afterglow that is detected from radio up to TeV gamma-rays. 
See Chapter \citet{Colombo01.2026.SKA} for further details on GRB and GW event classification and radio properties and Chapter \citet{Castignani01.2026.SKA} for the links between radio and high-energy (TeV) gamma-rays. In this section, we motivate rapid-response triggering on GRBs to search for coherent, prompt radio signals associated with the initial event and central engine (Section~\ref{sec:grb_prompt}) and the early synchrotron emission generated by the outflows (Section~\ref{sec:grb_synch}).

\subsection{Coherent, prompt radio emission}\label{sec:grb_prompt}

\begin{figure}[h]
    \centering
	\includegraphics[width=0.7\textwidth]{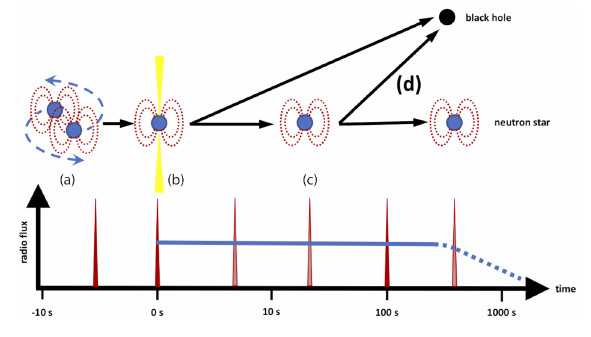}
    \caption{Figure from \citet{rowlinson19anderson} that depicts several different scenarios and their corresponding timescales of coherent emission emitted during a BNS merger. }
    \label{fig:grb_row19}
\end{figure}

There are several emission models that predict BNS mergers could emit coherent and prompt radio signals either just prior, during or following the merger. 
These signals would be analogous to FRBs \citep[see Chapter][]{Curtin01.2026.SKA} and indeed could represent a subset of the FRB population. 
For example, the interaction of magnetic fields between the binary neutron stars \citep[][]{lipunov96,2012Piro, Wang2016, metzger16,cooper23} or the excitation of the surrounding plasma by GWs \citep{moortgat03} could give rise to a prompt pulse. Similarly, radio pulses could be produced by interactions between the surrounding interstellar medium and accelerated winds \citep{Sridhar2021Wind} or the relativistic jet if it is strongly magnetised \citep[jet-ISM interaction;][]{usov00}. Following the merger a (quasi)stable, rapidly rotating, highly magnetised neutron star (magnetar) remnant could be formed that emits dipole radiation, which could be observed as pulses or persistent emission \citep{metzger11,metzger17,totani13}. If unstable, this magnetar would eventually collapse into a black hole, causing another prompt radio pulse due to magnetic reconnection \citep{zhang14}. 
Several of these potential coherent emission vectors are illustrated in Figure~\ref{fig:grb_row19}, taken from \citet{rowlinson19anderson}, which also provides a good overview of these prompt emission mechanisms. 
A subset may also apply to NSBH mergers that launch a jet \citep[see][for an overview]{clarke25} and long GRBs.
Relativistic outflows from long GRBs can also produce X-ray flares, which may also generate coherent radio emission if we again assume they are magnetically dominated \citep{starling20}.

\begin{figure}[h]
    \centering
	\includegraphics[width=0.45\textwidth]{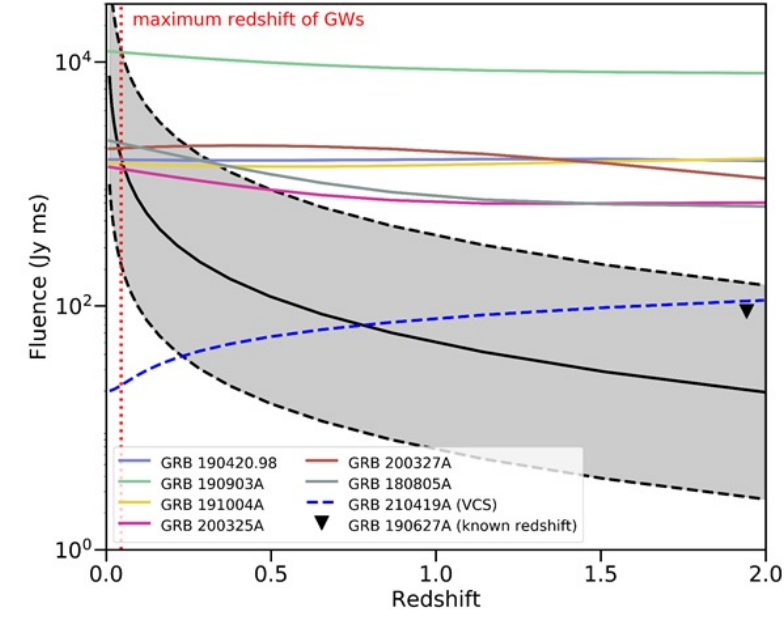}
    \includegraphics[width=0.45\textwidth]{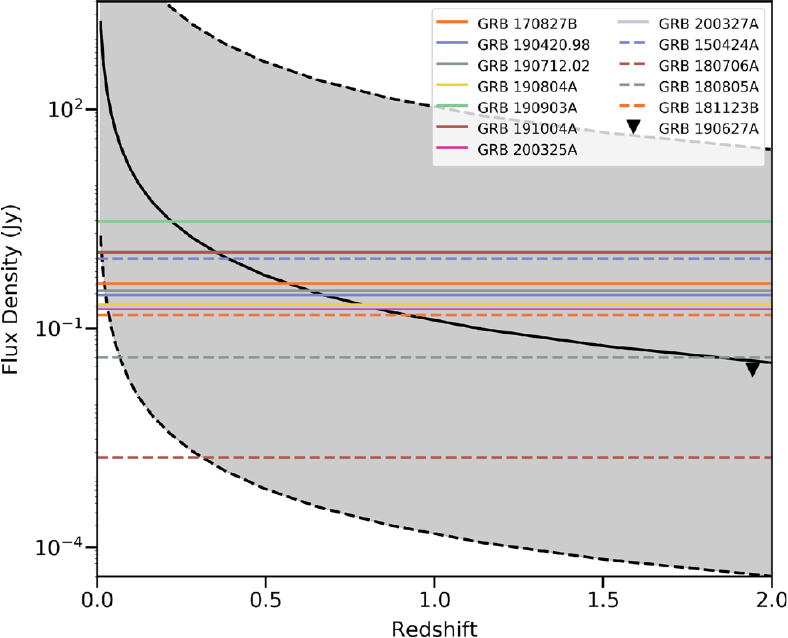}
    \caption{Constraints on coherent radio emission models from rapid-response observations of GRBs with MWA and LOFAR. Left: Fluence prediction of a pulsed radio signal produced by the jet interacting with the ISM assuming a magnetar remanant with average parameters is produce (grey region). Right: Flux density prediction of persistent radio emission produced by the spin down of a magnetar remnant (grey region). These Figures were adapted from \citet{tian22a}. }
    \label{fig:grb_models}
\end{figure}

Catching prompt, coherent signals emitted by GRBs are best targeted at low frequencies ($<350$\,MHz) as many of the emission mechanisms are brighter at lower frequencies. There is also the additional benefit that coherent signals experience dispersion delay, causing tens to hundreds of seconds delay in their arrival time at low frequencies if they are emitted at cosmological distances.
While this provides additional time for the telescope to repoint at the target, we still require the rapid dissemination of transient alerts from dedicated facilities (see Section~\ref{sec:tech}). 
The wide fields of view of low frequency instruments can also target GRBs with poor localisations like those from \fermi{} \citep[e.g. MWA,][]{tian22a}.
MWA and LOFAR have demonstrated the power of rapid-response observing modes by placing deep constraints on coherent emission mechanisms from GRBs \citep{anderson21mwa,rowlinson19,rowlinson21,tian22a,tian22b,hennessy23,hennessy25,xu25}.
In particular, Figure~\ref{fig:grb_models} shows the predicted radio emission from two models assuming an average magnetar remnant was produced \citep[see][]{rowlinson19anderson}, taken from \citet{tian22a}. They show the limits placed by rapid-response observations of GRBs triggered with both MWA and LOFAR for both the jet-ISM interaction (left) and persistent radio emission produced by spin-down radiation from a magnetar remnant (right). 
Indeed, recent results from LOFAR detected a potential prompt, coherent signal from a short GRB \citep{rowlinson24} that may have resulted from the collapse of a magnetar remnant into a black hole.

For SKA-Low to target coherent emission mechanism from GRBs at cosmological distances ($z>0.1$), the telescope must repoint within $20$\,s \citep[see figure 1 from][]{hancock19voevent} to catch any signals emitted just prior or during the GRB. Currently the target SKA-Low repointing speed is not defined in the SKAO baseline documents. However, if SKA-Low can repoint on this timeframe, \citet{cooper23} predict it will be sensitive to pre-merger, coherent radio pulses for 20 to 30 short GRBs per year. 
Given that a 5 minute observation with SKA-Low will provide a $3\sigma$ limit $<1\times 10^{-4}$\,Jy, it will be able to probe all coherent emission for a wide range of possible magnetar remnants up to $z=2$, and either finally detect these signals or influence the refinement of these models. 

\begin{figure}[h]
    \centering
	\includegraphics[width=0.7\textwidth]{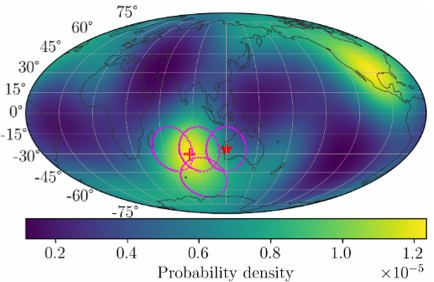}
    \caption{Figure adapted from \citet{tian23b} showing the proposed MWA primary beam coverage (magneta contours) if the array were split into 4 subarrays positioned to best cover the high sensitivity region of the LVK network as projected on Earth (colour scale). The position of the MWA is shown by a red star, with the position of the highest LVK sensitivity over the Indian Ocean marked by a red cross.}
    \label{fig:mwa_gw}
\end{figure}

Targeting this same prompt emission from GW-detected BNS mergers does afford significant challenges. Given GW-detected BNS are closer (within 325\,Mpc during O5,\footnote{https://www.ligo.caltech.edu/page/observing-plans} note that GW170817 was at 40 Mpc) we are unable to rely on dispersion delay to be on target. Even the MWA rapid-response theoretical lower limit of $<20$\,s is too slow for an GW alert transmitted within 10 seconds \citep{james19}.
GW alerts are also notoriously poorly localised (typically tens of square degrees for 3 detectors), and the mergers themselves may not have their jets beamed along our line-of-sight, which is the direction we expect most emission to be beamed.  
Targeting this predicted prompt radio emission has been explored for multiple SKA-Low precursors \citep{chu16,kaplan16,james19,tian23b}. 
For SKA-Low to detect coherent radio signals emitted at the moment of merger, it would need to depend on early-warning alerts generated by LVK through the detection of GWs for the BNS inspiral. 
With up to 30\,seconds warning from the inspiral, \citet{james19} demonstrated that observations up to 300\,MHz with the MWA rapid-response mode could be on target to detect prompt signals emitted at the time of merger.
Other options include whole sky transient monitoring \citep[e.g. AARTFAAC and OVRO-LWA;][]{kuiack21,shulevski22,callister19,kosogorov25,kosogorov26} or shadowing of the LVK network's highest probability sensitivity region \citep[e.g. MWA;][]{tian23b}, all of which may trigger a buffer dump on receiving a GW alert, obtaining up to several minutes of look-back time.
SKA-Low would also need to use subarraying and potentially apodising to create multiple beams to cover large sky area of the highest sensitivity regions provided by LVK, which has been explored for both MWA \citep[see Figure~\ref{fig:mwa_gw};][]{tian23b} and LOFAR \citep{gourdji23}. 

While lower frequencies offer the benefit of dispersion delay, higher frequency ($>350$\,MHz) rapid-response triggering has the advantage of probing radio emission at points when the environment may have been opaque to low frequency emission due to e.g., free-free absorption, synchrotron self-absorption, or induced Compton scattering \citep{zhang14,rowlinson19anderson}. 
Additionally, there are models that predict $\sim10$--$20$ GHz radio emission prior to compact object mergers due to electromagnetic flares interacting with the orbital current sheets  \citep{MostPhilippov2020, 2022MostPhilippov, 2023MostPhilippov}. 
Our current best limits on prompt emission at higher frequencies come from serendipitous simultaneous radio coverage of GRBs, GWs and FRBs. 
For example, the Canadian Hydrogen Intensity Mapping Experiment (CHIME) has placed constraints on radio emission between 400 to 800 MHz associated with GRBs \citep{curtin23,curtin24}, resulting in one of the best current constraints of less than $\sim$kJy on a coherent radio pulse from a short GRB \citep{curtin24}.
Several investigations have also looked for high-energy \citep{cunningham19,gourdji20} and gravitational wave \citep{abbott23} associations with known FRBs, placing limits on both types of counterparts. 
The only tentative association between a BNS merger and FRB was reported for GW190425 and CHIME FRB 20190425A by \citet{moroianu23}, however, further modelling suggests this is a chance coincidence \citep{bhardwaj24}.

SKA-Mid will probe orders of magnitude fainter radio emission, though with the key challenge of requiring a fast trigger time. At frequencies such as 10 GHz, sub-second triggering would be required to catch simultaneous radio emission, an unrealistic trigger time for SKA-Mid given the $\sim$minutes required to re-point the Australian SKA Pathfinder \citep[ASKAP;][]{Hotan2021} and MeerKAT. 
At more modest frequencies such as 1 GHz, re-pointing would need to occur on the timescale of second(s), which will likely only be possible for a small subset of sources. 
\citet{dobie19} and \citet{wang20} investigated the feasibility of using rapid-response triggering with ASKAP to catch coherent radio signals emitted at the moment of a BNS mergers at 888\,MHz. They suggested using a fly's eye mode capable of covering $\sim1000$\,deg$^{2}$ of the LVK BNS localisations, and relying on early-warning alerts from the inspiral, however, the detection rates would still be extremely low. 
Instead, SKA-Mid's strength will likely lie in detecting coherent emission associated with the merger remnant within a few hours post-burst \citep[e.g.][]{totani13,falcke14,zhang14,rowlinson19}.

\subsection{Synchrotron radio emission}\label{sec:grb_synch}

\begin{figure}[h]
    \centering
	\includegraphics[width=0.46\textwidth]{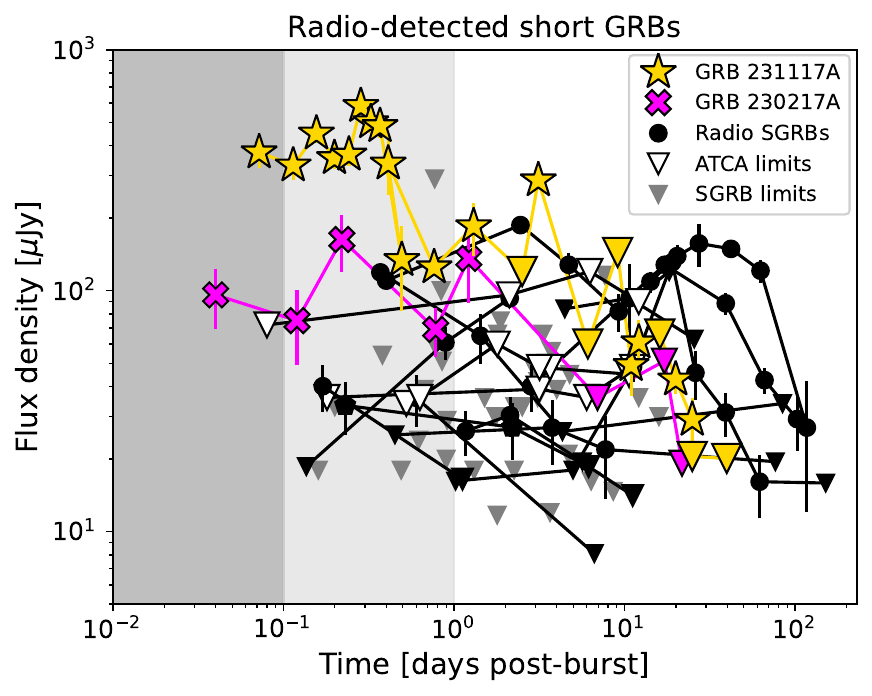}
    \includegraphics[width=0.532\textwidth]{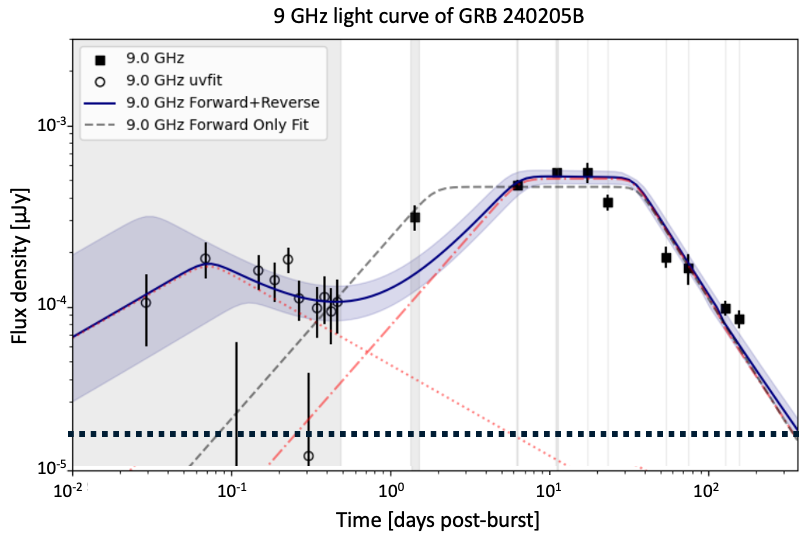}
    \caption{{\bf Left:} Adapted from \citet{anderson25} showing the $5$--$8$\,GHz light curves of radio detected short GRBs illustrating how the rapid-response observations are probing the previously under-sampled parameter space of $<0.1$\,days post-burst in the radio band.
    {\bf Right:} Adapted from \citet{chastain26pp} showing the 9\,GHz light curve of GRB 240205B. The duration of the ATCA rapid-response observation and all follow-up observations are indicated by the vertical gray shadowing. The fit approximates a reverse plus forward shock model, demonstrating that an entire synchrotron component would have been missed without the rapid-response observation. The SKA Band 5b 1 minute $3\sigma$ sensitivity is indicated by the dotted line.
    }
    \label{fig:sgrb_lc}
\end{figure}

When the relativistic jet is launched by a GRB, it produces a multi-wavelength afterglow as predicted by the Fireball model \citep{piran99}, which describes the generation of multiple synchrotron components caused by the outflow expanding into the circumburst medium. This includes a forward shock component, which takes weeks to peak in the GHz radio band. A reverse shock is also generated, which propagates back into the post-shocked ejecta, giving rise to a much faster evolving synchrotron component. The reverse shock emission usually fades below detectability within seconds to minutes at X-ray and optical wavelengths but takes hours to a few days to peak at radio GHz frequencies \citep[e.g.][]{anderson14,vanderhorst14}. This makes the radio band the most viable way of detecting the GRB reverse shock emission, which can provide direct insight into the outflow composition and magnetisation.
Rapid-response observations with AMI-LA \citep{anderson14,anderson18} and ATCA \citep[][]{anderson24,anderson25,chastain26pp} have enabled the earliest detections of GRBs, catching the reverse shock emission and early ($<0.1$\,days) evolution of the radio afterglow that would be missed if dependent on manually scheduled observations (e.g. left panel of Figure~\ref{fig:sgrb_lc}). 
The richness of science that can be extracted from early time and broadband radio observations of the reverse shock is exemplified by the rapid AMI-LA follow-up of GRB 221009A \citep[often referred to as the Brightest Of All Time or the BOAT;][]{burns23}. In this case the radio afterglow was so bright that the 12-18\,GHz AMI-LA bandwidth \citep{hickish18} could be split into 8 spectral windows on 15 minute timescales between $3-7$\,hr post-burst. This enabled \citep{bright23} to track the evolution of the outflow size, the bulk Lorentz factor, and the minimum total energy. 
This rapid radio follow-up and continued radio monitoring revealed multiple synchrotron components, including the reverse shock and two forward shocks from a structured jet \citep{rhodes24}, which can also help to reveal the origin of any associated TeV emission \citep[see Chapter][for more details]{Castignani01.2026.SKA}.
Implementing a rapid-response mode on SKA-Mid would enable the spectral and temporal details of the reverse shock to be tracked for much fainter GRB radio afterglows, and reveal previously hidden emission components. 

Unexpected early-time radio emission can also be probed with rapid-response observations as demonstrated by the detection of a radio flare that began 9 hours post-burst, and was likely caused by interstellar scintillation magnifying the GRB afterglow emission to above the ATCA minute timescale sensitivity \citep{anderson23}.
Rapid-response observations of GRB 231117A beginning 1 hour post-burst detected a radio plateau and flare likely caused by a violent collision between ejecta shells, probing central engine and outflow behavior \citep{anderson25}. 
SKA-Mid rapid-response observations of GRBs will be able to detect polarised radio emission in the reverse shock, which probes the magnetic field structure and strength within the outflow as has been done at mm wavelengths using rapid, mannually scheduled follow-up with the Atacama Large Millimeter Array \citep{laskar18,laskar19,laskar19grb181201a}.

The power of the rapid-response mode for probing GRB astrophysics is demonstrated in the right panel of Figure~\ref{fig:sgrb_lc}, which shows the 9\,GHz light curve of GRB 240205B \citep{chastain26pp}. The ATCA rapid-response mode triggered on the \swift{} detection of this event, and began observing 13\,minutes post-burst for 12 hours, detecting the radio counterpart. By splitting this rapid-response observation into 45 minute time blocks and fitting in the $uv$-plane, the reverse shock was revealed.
This demonstrates that without the rapid-response mode, an entire synchrotron emission component would have been missed. 
However, deeper sensitivity that will be afforded by SKA-Mid is necessary to properly constrain and model the evolution on minute timescales in order to derive the physical properties of the outflow. The right panel of Figure~\ref{fig:sgrb_lc} also shows the sensitivity that SKA-Mid AA4 will achieve on 1 minute timescales using Band 5b ($8.3$--$15.4$\,GHz, assuming a Briggs Robust weighting of 0), which will ensure a better characterisation of this early and fast evolving emission.

While similar synchrotron afterglow physics motivates the rapid follow-up of GW events at GHz frequencies, the large positional uncertainties and likely off-axis orientation of the associated jets make them difficult to locate, observe and detect at such early times. 
Even though GW170817 was located through an optical detection of its kilonova 11\,hr post-burst \citep{abbott17c}, the radio afterglow was not detected until 16\,days \citep{hallinan17} due to the event being viewed $\sim20$\,deg off-axis \citep{mooley18}.
In addition, multi-year searches for the radio counterparts associated with poorly localised GW events still have not yielded a detection \citep[e.g.][]{alexander21,dobie22,gulati25}.
However, the detection of the rapidly evolving reverse shock from an on-axis GW-detected BNS or NSBH merger could provide one of the earliest electromagnetic localizations. 
We therefore require wide-field follow-up that can cover large portions of the GW error regions \citep[e.g.][]{abbott16} as has been suggested for ASKAP, which can be operated in a fly's-eye mode to cover a significant portion of the LVK localisations \citep{dobie19,wang20}. 
Alas, even if SKA-Mid performed rapid-response follow-up that used multiple sub-arrays to instantaneously cover the highest probability positional region of a GW event, the $\sim$1\,deg field-of-view at 1\,GHz would not cover a significant portion of the GW uncertainty regions.
However, alternative rapid-response triggering scenarios could be utilised for SKA-Mid, preferencing Bands 5a and 5b given synchrotron afterglows peak more brightly at higher frequencies at earlier times. This could include triggering subarray pointings that target galaxies within the LVK positional uncertainty that fall within the GW-derived luminosity distance range \citep[e.g.][]{dobie19} or rapid-response follow-up of kilonova candidates identified within 1\,day post-merger when we still expect the reverse shock to be bright and rapidly evolving. 

\section{Transient science cases for rapid-response triggering}

\subsection{Flare stars}

Cool dwarf stars are capable of producing powerful flares that emit across the electromagnetic spectrum due to their strong magnetic activity. 
Radio observations probe the accelerated electron populations that produce these flares, which allow for the measurement of associated properties such as the brightness temperature, magnetic field strength and kinetic energy \citep[see Chapters][]{Driessen01.2026.SKA,Cavallaro01.2026.SKA}. 
The sudden release of energy that accelerates the electrons in magnetic loops give rise to gyrosynchrotron radio emission, which often dominate higher radio frequencies ($>5$\,GHz) with brightness temperatures $\leq10^{10}$\,K \citep{benz10}.
Giant radio flares vary in brightness, polarisation and spectral index on the timescale of seconds to hours \citep[e.g.][]{osten05}.

The low frequency radio flares detected from cool dwarf stars in precursor observations and surveys ($\lesssim1$\,GHz) are highly circularly polarised \citep{lenc18,pritchard21,driessen24} and show similar variability on seconds to hours timescales \citep[e.g.][see also the Variability of Radio Stars Chapter \citet{Driessen01.2026.SKA}]{lynch17,zic20, Rose23}. The emission is often elliptically polarised and generally has brightness temperatures in excess of $10^{12}$\,K \citep{Pritchard24}; as such these coherent bursts are thought to be generated by electron cyclotron maser emission \citep[ECME; ][]{dulk85, Hallinan08}.
These same SKA-Low precursor results illustrate the usefulness of Stokes V for detecting coherent radio emission, as low frequency circular polarisation observations have found sources at lower noise thresholds when compared to total intensity observations \citep[e.g.][]{lynch17,pritchard21}.
In addition, rapidly rotating massive main sequence stars (spectral type OBA) that have strong magnetic fields are also known to produce variable radio emission at lower frequencies \citep[$<5$\,GHz;][]{Das25a,Das25b} due to plasma break outs from their magnetospheres \citep[][]{schultz22,owocki22}.

\begin{figure}[h]
    \centering
    \includegraphics[width=0.47\textwidth]{Figures/fender15.pdf}
    \includegraphics[width=0.52\textwidth]{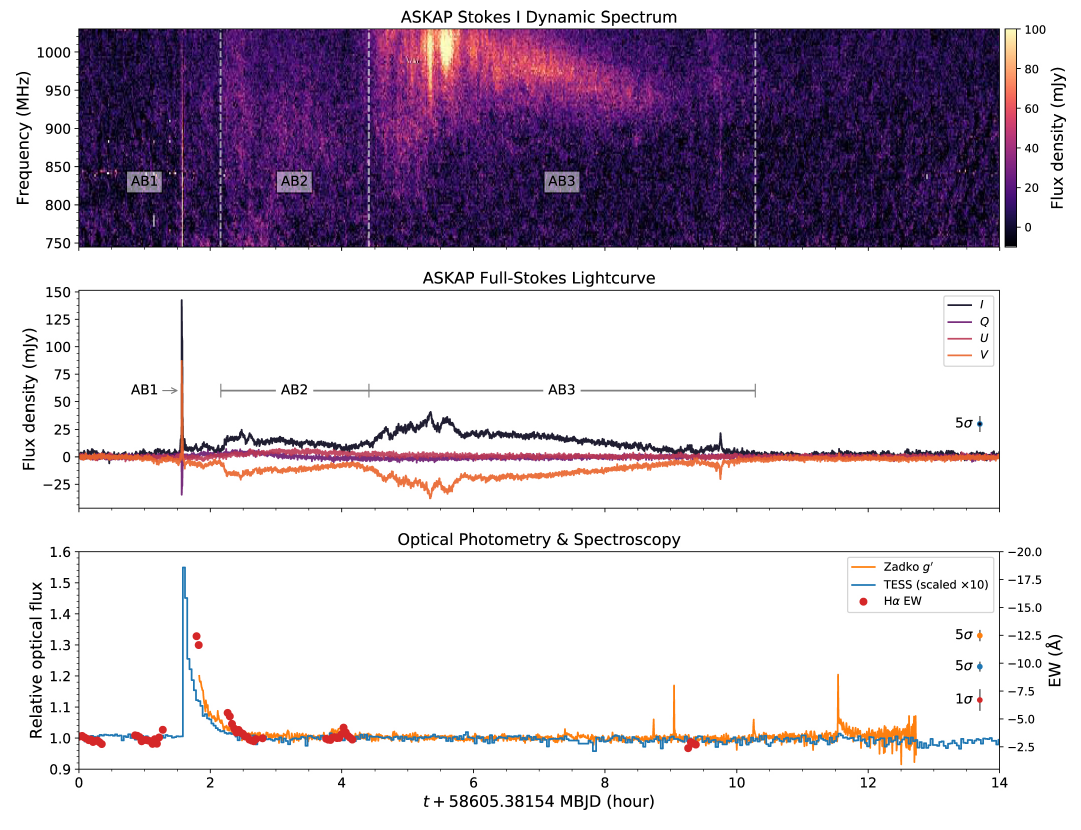}
    \caption{
    \textbf{Right:} Simultaneous radio and X-ray detection of a gyrosynchrotron flare from an M-dwarf binary, resulting from an AMI-LA rapid-response observation on the \swift\ detection of a high-energy X-ray/$\gamma$-ray superflare from DG CVn \citep{fender15}.
    \textbf{Left:} Dynamic radio spectrum and radio and optical light curves from simultaneous monitoring of dM5.5e star Proxima Centauri with ASKAP, the Transiting Exoplanet Survey Satellite \citep[TESS;][]{ricker15} and the Zadko Telescope \citep{coward17}. During 11 nights of monitoring, a bright, long-duration optical flare was accompanied by coherent radio bursts demonstrating both phenomena were caused by the same magnetic event \citep{zic20}. 
    }  
\label{fig:flarestars}
\end{figure}

While there some instances of rotationally-modulated emission \citep[e.g.][]{lynch17,Kao18,Rose23}, stochastic radio flares, whether gyrosynchrotron or ECME are impossible to predict, which means that radio monitoring campaigns may not yield a detection.
However, some extreme flaring events can result in high-energy X-ray/$\gamma$-ray superflares, which can trigger the detectors on instruments like \swift{} Burst Alert Telescope \citep[BAT;][]{barthelmy05}, the \textit{Monitor of All-sky X-ray Image} \citep[{\it MAXI};][]{matsuoka09} and the {\it Einstein Probe Wide-field X-ray Telescope} \citep[{\it EP-WXT};][]{yuan22ep}. These instruments transmit GCN allowing for rapid-response triggering and radio follow-up \citep[e.g.][]{rose25atel}. 
For example, AMI-LA triggered on the \swift-BAT detection of a X-ray/$\gamma$-ray superflare from the M dwarf system DG CVn, detecting a $>100$\,mJy gyrosynchrotron radio flare just 6 minutes post-burst \citep[see left panel of Figure~\ref{fig:flarestars} adapted from][]{fender15}. This observation was used to derive the kinetic energy of the flare, which ruled out a non-thermal emission mechanism for the associated X-ray flare \citep{osten16}.  
What is still unclear is whether such extreme magnetic events could also trigger the coherent ECME radio flares that dominate a lower frequencies ($<5$\,GHz). 

The SKAO rapid-response mode is in an excellent position to provide the broadest radio frequency coverage of superflares detected by high-energy satellites, as it can allow for simultaneous triggering of both SKA-Low and SKA-Mid. A commensal transient observing mode could also provide internal triggers of stellar radio flares for rapid follow-up with another observing mode or SKA subarray, which could help constrain the emission mechanism.
The close proximity of many of these cool dwarf systems means that associated radio flares reach tens to hundreds of mJy at frequencies $>5$\,GHz \citep[e.g.][]{osten05,fender15}. 
SKA-Mid will achieve unprecedented high time resolution monitoring of these flares on 1 minute timescale ($3\sigma$ sensitivity of $20\mu$Jy per beam), which can be used to create dynamic spectra of the flare as has been done with the Karl G. Jansky Very Large \citep[VLA;][]{perley2011} and ASKAP \citep[][see also right panel of Figure~\ref{fig:flarestars}]{villadsen19,zic20}. 

Simultaneous observations with SKA-Low will enable a search for coherent flares to determine whether the magnetic events that trigger the high-energy and gyrosynchrotron radio flares could trigger the coherent emission mechanism. 
In fact, simultaneous monitoring with ASKAP and optical facilities have detected a causal link between optical flares and coherent radio bursts, providing further insight into Solar-like bursts \citep[see right panel of Figure~\ref{fig:flarestars} adapted from][]{zic20}. This demonstrates the value that could be gained by triggering SKA-Low observations on optically detected stellar flares.
SKA-Low has a $3\sigma$ sensitivity of $150\mu$Jy per beam on one minute timescales, which will be sensitive to the sub-mJy radio detections of cool dwarfs with LOFAR \citep{vedantham20,callingham21} and the tens of mJy detections with MWA \citep{lynch17,lenc18}.

\subsection{Cosmic rays/neutrinos}


The SKAO will be able to detect high-energy cosmic particles both directly and indirectly. The direct detection of extensive air showers (EAS) from PeV--EeV cosmic ray interactions in the atmosphere above \skalow\ is outlined in Chapters \citet{Watanabe01.2026.SKA,Nelles01.2026.SKA,Buitink01.2026.SKA} and \citet{Corstanje01.2026.SKA}. Here, we highlight primarily that the trigger must be extremely quick: less than $\sim 4$ seconds, which is the expected size of the antenna-level voltage buffer. For cosmic ray triggering and read-out, it is intended to operate fully commensally with conventional observing so will not require any reconfiguration of observing parameters.

The SKAO will also be well-positioned to study high-energy astrophysical neutrinos, albeit indirectly \citep[see Chapter][]{Rosch01.2026.SKA}. Since their first identification by the IceCube detector \citep{2013Sci...342E...1I}, the flux has been studied in the $\sim$1\,TeV--10\,PeV range \citep{abbasi26}. In the $\sim$1--100\,TeV range, this flux can only be statistically distinguished from the background of neutrinos generated by cosmic ray interactions in the Earth's atmosphere; however, at higher energies, the majority of neutrinos are expected to be astrophysical.

\begin{figure}[h]
    \centering
    \includegraphics[width=0.9\textwidth]{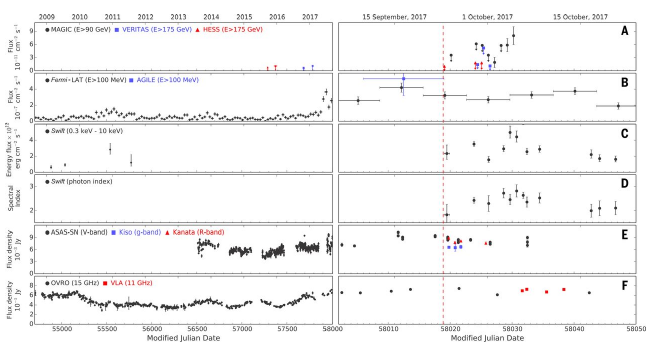}
    \caption{The gamma-ray, X-ray, optical and radio light curves before and after the high-energy neutrino IceCube-170922A event (vertical dashed red line) associated with blazar TXS 0506+056 \citep[adapted from][]{2018Sci...361.1378I}.}   
\label{fig:neutrino}
\end{figure}

Since neutrinos only weakly interact, they propagate directly from their sources to Earth, making them excellent probes of high-energy environments in which they are expected to be created. Yet, while the arrival directions of a small number of neutrinos point back to active blazars \citep[][see also Figure~\ref{fig:neutrino}]{2018Sci...361.1378I,2018Sci...361..147I}, the population as a whole does not have a significant association with blazars \citep{abbasi26b}, and searches for other potential point-sources have not found a highly significant candidate \citep{abbasi25}. 
While a potential source of cosmic rays and neutrinos could be GRBs due to hadronic loading in their jets \citep{1995PhRvL..75..386W}, no neutrino associations have been found \citep{2017MNRAS.469..906A,2022ApJ...939..116A,2022ApJ...941L..10M}, although choked bursts may be an alternative scenario \citep{2021JCAP...09..044F}. 
Thus, while neutrinos have long been thought to hold the key to identifying sites of cosmic ray acceleration \citep{2012PrPNP..67..651K}, current observations have raised more questions than answers.

SKAO is well timed to take full advantage of multi-messenger science with upcoming neutrino instruments, such as the Baikal-GVD \citep{2023PhRvD.107d2005A} and KM3NeT \citep{2016JPhG...43h4001A}, the latter having already detected the first ultra-high-energy neutrino \citep[but with no obvious astrophysical counterpart;][]{2025Natur.638..376K}. 
IceCube is also planning a large expansion \citep{2021JPhG...48f0501A}, while new detectors are planned in the Pacific Ocean \citep{2020NatAs...4..913A} and South China Sea \citep{2025APh...17103123Z}. 
Neutrino detectors can detect neutrinos from all 4$\pi$\,sr (albeit with complex energy and morphology dependence), and have a near-real-time alert system capable of sending alerts within seconds \citep[][]{2017APh....92...30A,2025EPJWC.31912010M}\footnote{https://icecube.wisc.edu/science/real-time-alerts/}.
For example, IceCube transmits alerts relating to track-like neutrino candidates via the GCN, which have $\lesssim1$\,deg localisations.\footnote{https://gcn.nasa.gov/missions/icecube} These localisations are also combined with correlated multi-messenger searches by the Astrophysical Multimessenger Observatory Network \citep[AMON;][]{smith13}, which searches photon, neutrino, cosmic ray, and gravitational wave data streams for associated sub-threshold events \citep[e.g.][]{ayalasolares23}, which are also transmitted via the GCN. 
AMON transmitted a GCN on the high-energy neutrino detected by IceCube\footnote{https://gcn.gsfc.nasa.gov/notices\_amon/50579430\_130033.amon} \citep{kopper17gcn}, where the source was later identified as the blazar TXS 0506+056 \citep{2018Sci...361.1378I,2018Sci...361..147I}, which resulted in follow-up from radio to TeV energies (see Figure~\ref{fig:neutrino}). Many instruments were already monitor this blazar, including the OVRO 40\,m telescope at 15\,GHz, however, the VLA did not begin observations until 2 weeks later \citep{tetarenko17atel}.
By enabling SKAO to trigger on neutrino events, we can be on target for events like TXS 0506+056 immediately, making it possible to search for rapidly brightening synchrotron transients that are expected to accelerate high-energy cosmic rays, and produce neutrinos through their interactions. Given that neutrinos can arrive from the distant Universe, the extra sensitivity provided by the SKAO will allow a much deeper probe of such scenarios.

\subsection{Long Period Transients} \label{sec:lpt}

Long-period radio transients (LPTs) are a phenomenological class of objects that produce periodic radio bursts at periodicities longer than expected for neutron star pulsars, i.e. $P\gtrsim1$\,min, up to several hours \citep[see Chapter][]{Qiu01.2026.SKA}. Their nature is not currently known, with magnetars and white dwarf pulsars in binaries being the two leading theories, but there may be more than one progenitor type; their mysterious nature means that more observations are desirable.

\begin{figure}[h]
    \centering
    \includegraphics[width=0.7\textwidth]{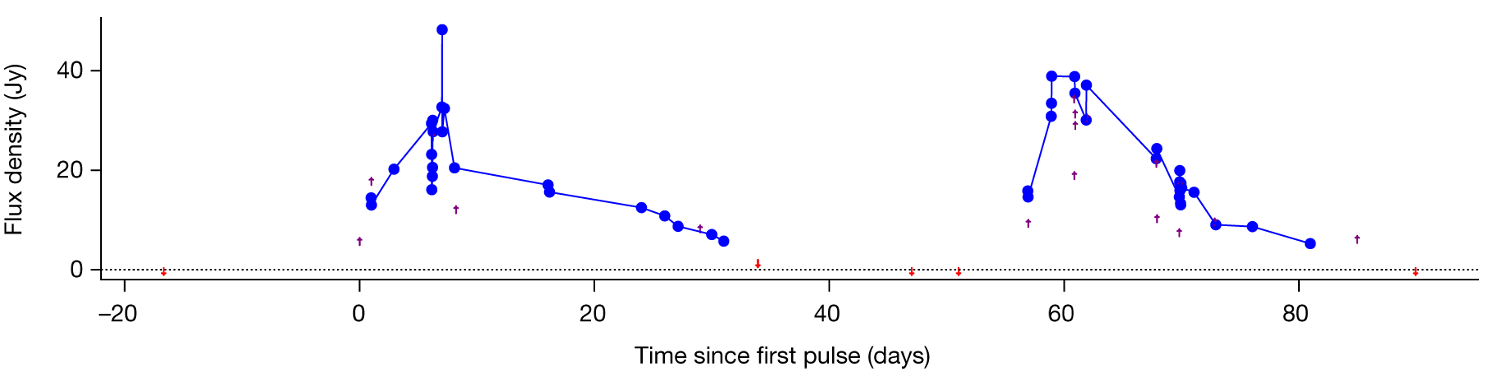}
    \includegraphics[width=0.75\textwidth]{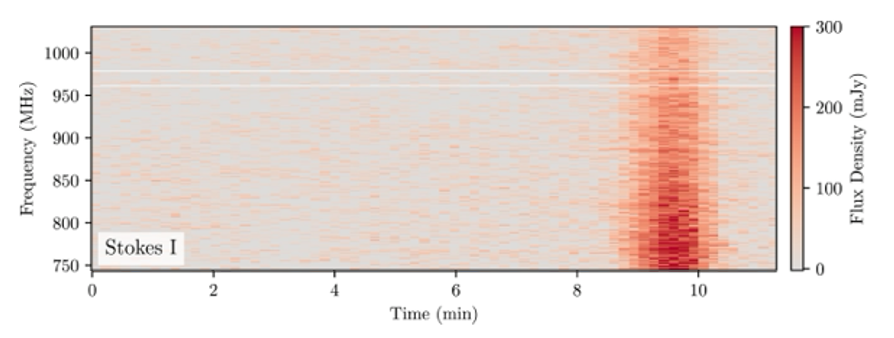}
    \caption{
    {\bf Top:} Radio light curve adapted from \citep{2022Natur.601..526H} showing the LPT GLEAM-X J162759.5-523504.3 was only active for 2 months over 8 years of MWA operation.
    {\bf Bottom:} Dynamic spectrum of LPT candidate ASKAP J175534.9-252749.1 adapted from \citep{Dobie24} that was first detected as a single 2\,minutes pulse in 60\,hours of observation with ASKAP.
    }    
\label{fig:lpt}
\end{figure}

At the time of writing, $\mathcal{O}\left(10\right)$ are known, with diverse characteristics, including the window over which they are active. Some LPTs have persisted for at least 10~years \citep{2023Natur.619..487H,2024ApJ...976L..21H} while others only switch on for weeks to months \citep[][see also the top panel of Figure~\ref{fig:lpt}]{2005Natur.434...50H,2022Natur.601..526H, Dobie24}. In these latter cases, rapid response is critical.
For instance, a $P\sim54$-minute LPT discovered with ASKAP \citep{2024NatAs...8.1159C} was followed up within days by MeerKAT, detecting dramatic changes in its pulse morphologies and polarisation, before the source faded from view over about three weeks.
In the case of ASKAP J162759.5-523504.3, only a single 2 minute burst was initially detected over 60 hours of ASKAP observations, which showed properties similar to both LPTs and pulsars \citep[][see also the bottom panel of Figure~\ref{fig:lpt}]{Dobie24}, demonstrating how quickly these events can disappear. Its eventual re-detection confirmed its LTP classification \citep{mcsweeney25}. 
Several other LPTs remain unpublished due to activity windows that were considerably less than a week, and a lack of rapid-response follow-up. Rapid-response follow-up will ensure observations of more pulses per target by catching the earliest pulse profiles to aid in understanding the mechanisms responsible for the sudden change in LPT radio states.

Simultaneous observations over multiple frequencies (often performed by different radio observatories) can determine a wideband radio spectrum, including potential cut-offs, which is expected for some emission mechanisms \citep{2023Natur.619..487H}.
Coordinated observations can reveal modulation of the pulse brightnesses that indicates binary activity
\citep{horvath26}.
Currently, such follow-up is handled by either target-of-opportunity programmes or director discretionary time requests, and due to strict timing requirements to catch pulsations, can be challenging to schedule. 

The SKAO therefore presents a tremendous opportunity for discovering new or newly active LPTs (via fast imaging or monitoring programs).
Some LPTs have relatively steep spectra and are brighter at low frequencies; objects like this will be most easily discovered by SKA-Low, with its large field-of-view (likely widened further with substations). On the other hand, other LPTs have a steep turnover and are invisible at low frequencies -- it is not currently known whether this is intrinsic or extrinsic to the sources \citep[][]{2025Natur.642..583W,2025NatAs...9..393L}. The luminosity function is currently unknown, and there may be a population of faint LPTs; due to high scattering ($\propto\nu^{-4}$) in the Milky Way, these are more likely to be detected by SKA-Mid.

Taking these factors together, the detection of a newly active LPTs  
by either SKA-Mid or SKA-Low should generate an automatic internal alert \citep[see Chapter][on commensal image plane searches for transients]{AlexAndersson01.2026.SKA} that triggers immediate automated follow-up with both instruments at the highest possible temporal resolution. The aim is to capture as many pulses as possible over the ensuing days to weeks, with as much simultaneity as possible. 
A triggered rapid-response observation with SKA-Mid band 5 would also provide the earliest highest-precision source position, which will enable optical cross-identification.
Sub-arrays could be used if the source is of sufficient brightness. 
Given that most LPTs have pulses that are only $\sim$ seconds to minutes long, once the rotational period is established, sufficiently clever scheduling software could capture a single pulse without being too detrimental to other projects. 
An automated alert system that could trigger coordinated follow-up in the Western sky (e.g. ATCA and VLA or their successors) would ensure continuous monitoring within the first 24 hours of discovery to accumulate as many pulses as possible before a potential disappearance. Once an ephemeris is established, the spacing could be decreased to e.g. logarithmic, or in the case of a persistent LPT, added to a timing program.

\subsection{Fast Radio Bursts}

Fast radio bursts are enigmatic extragalactic radio transients lasting of order milliseconds \citep{lorimer07}. 
FRBs can be either repeating or (apparently) non-repeating events \citep[][see also Chapter \citet{Curtin01.2026.SKA}]{spitler16,CHIME_cat1_morphology_2021,chime_repeaters23}.
Both \skalow\ and \skamid\ will detect FRBs by running real-time detection algorithms on tied-array beams produced by the pulsar beamformer, and generate triggers to read out antenna-level/station-level buffers of coarse channelised data for offline analysis. The science cases to use FRBs as probes of the ionised matter distribution of the Universe, and the extreme astrophysics of their progenitors, are described in accompanying Chapters \citet{Curtin01.2026.SKA,Caleb01.2026.SKA,Caleb02.2026.SKA}. Here, we discuss FRB science cases for rapid-response triggering.

\begin{figure}[h]
    \centering
    \includegraphics[width=0.7\textwidth]{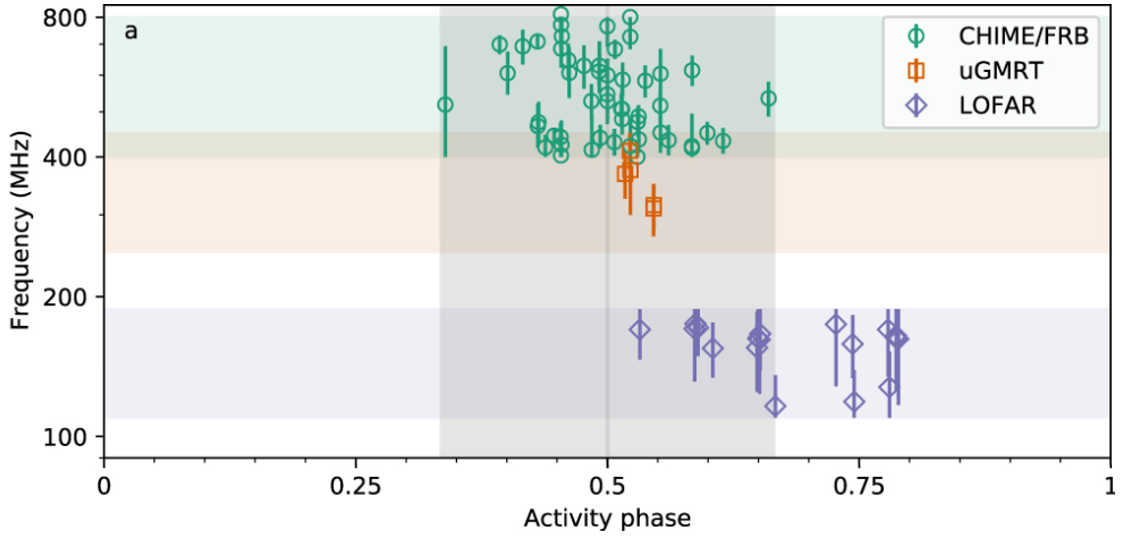}
    \caption{Figure adapted from \citet{pleunis21} showing bursts detected by LOFAR, uGMRT and CHIME/FRB folded on the 16.33\,day activity period of repeater FRB 20180916B. The error bars indicate the spectral width of each individual burst. 
    }    
\label{fig:frb}
\end{figure}

The high dispersion measures of FRBs means that the time delay between a signal arriving at \skamid\ and at \skalow\ frequencies will be significant (at least 10 seconds for a 100\,pc\,cm$^{-3}$ FRB between 1\,GHz and 200\,MHz). 
This may be sufficient to allow an FRB detected by \skamid\ to trigger observations with \skalow, providing broad-band coverage of the FRB spectrum. 
Such observations are necessary as the properties of FRBs emission over a broad bandwidth are not well understood, with a wide variety of potential spectral properties \citep[][]{CHIME_cat1_morphology_2021}. Only one FRB is known to emit down to 110\,MHz \citep{pleunis21,pastor-marazuela21}, and another up to 8\,GHz \citep{gajjar18}, both of which are repeaters. 
In the case of repeater FRB 20180916B, simultaneous monitoring with multiple radio telescopes, including the LOFAR high-band antenna (110–188 MHz), the upgraded Giant Metre Wavelength Radio Telescope (uGMRT; 200–450 MHz), and CHIME/FRB (400–800 MHz) showed that while bursts are emitted across this frequency window, their individual bandwidths are narrow, with no simultaneous detections between CHIME/FRB and LOFAR \citep[see Figure~\ref{fig:frb} adapted from][]{pleunis21}. Triggering \skalow\ observations of \skamid-detected FRBs would investigate if this phenomenon is common among all repeaters and also non-repeating events.

Meanwhile, non-detections of low-frequency ($<350$\,MHz) emission from FRBs detected at higher frequencies have been attributed to plasma lensing, free-free absorption, or the intrinsic emission mechanism \citep{2018ApJ...867L..12S}, or in the case of repeaters, to the low duty cycles of their activity windows \citep{tian23}. The wide burst-to-burst variation between FRBs makes studying such effects with ensemble samples very difficult, which is a problem that can be overcome by studying exactly the same bursts with \skamid\ and \skalow.

Recently both CHIME/FRB\footnote{https://gcn.nasa.gov/missions/chime} and the Deep Synoptic Survey - 110 \citep[DSA-110;][]{kocz19}\footnote{https://gcn.nasa.gov/missions/dsa110} have started broadcasting FRB alerts over GCN Kafka, providing the opportunity for fast responding low frequency instruments like MWA to be on target before the signal arrives at <200\,MHz.
During the SKA era, other instruments will also be detecting FRBs and transmitting FRB transient alerts. In particular, the Canadian Hydrogen Observatory and Radio-transient Detector \citep[CHORD; ][]{CHORD_white_paper_2019} will share a small (albeit, not simultaneous) sky overlap with the SKA, while the Bustling Universe Radio Survey Telescope in Taiwan \citep[BURSTT; ][]{BURSTT2022} will view down to low latitudes, with simultaneous sky overlap with \skalow. These wide field-of-view instruments will detect the nearest and brightest coherent transients, such as FRB-like events from galactic magnetars \citep{bochenek20,chime_magnetar20} or FRBs from nearby globular clusters \citep{kirsten22}.
They will search for extremely active repeating FRBs undergoing `burst storms' \citep[e.g.][]{2022MNRAS.515.3577H,2023MNRAS.520.2281N} and periodic windows of burst activity \citep[e.g.][]{pleunis21}.
Given the rarity of such events and the non-Poissonion nature of burst emission from repeating sources, simultaneous rapid response follow-up with SKA-Low and SKA-Mid will provide the most comprehensive simultaneous frequency coverage (from 50\,MHz to 15\,GHz) of FRBs to date. Such observations would allow for comprehensive sampling of the spectral and energy distributions of bursts across repeater activity windows. It would also provide instantaneous spectral coverage of non-repeating bursts to determine what components of their complex spectra are intrinsic or influenced by extrinsic effects.

A small subset of FRBs have been localized to $<100$ milliarcsecond precision. When matched with high resolution optical imaging, these sources can be studied in the context of their position within the host galaxy \citep[][]{marcote17} providing advances toward understanding their progenitors. Very long baseline interferometry (VLBI) is required to achieve such angular scales. 
Given the above-mentioned non-Poissonion nature of repeaters, it can be challenging to detect bursts by the time a VLBI network is on target. 
Extension of a rapid response capability to the SKA-Mid VLBI mode (and SKA-Low for FRBs that emit at lower frequencies) and similarly enabled at other capable antennas would increase the feasibility of SKA VLBI localization of repeating FRBs.

\subsection{Novae}

Galactic binary systems comprising white dwarf stars in compact orbits are known to produce recurrent thermonuclear explosions \citep{chomiuk2021}. Two different classes of systems are usually considered: cataclysmic variables (those systems where a white dwarf accretes from a Roche-lobe–filling main-sequence or slightly evolved stellar companion) and symbiotic stars (there is a Roche-lobe overflow or wind accretion from an evolved stellar companion instead). 
In both cases, the white dwarf undergoes explosive ejections due to thermonuclear reactions produced by the material accreted onto its surface, which are known as classical novae \citep[see e.g.][and Chapter \citet{Lico01.2026.SKA}]{Starrfield2016}.

Radio emission is produced by free-free processes from the ionized gas and non-thermal synchrotron processes from the accelerated particles via shocks \citep{Chomiuk2021b}. The emission evolves on timescales from minutes to weeks, with the first hours to days being particularly sensitive for mapping the transition from optically thick to thin ejecta, revealing how the expanding material interacts with the environment and accelerates particles detecting shock signatures, early jet formation, and initial mass ejection structures \citep{Munari2022,deRuiter2023,nyamai23,Lico2024}.
The fast ($\gtrsim1000~\text{km s}^{-1}$) outflows launched in these events allow us to trace such emission on timescales ranging from hours to tens of days \citep[see e.g.][]{Giroletti2020,Lico2024}. For much closer systems like $\uptau$~CrB (expected to explode in the coming years, \citealt{Schaefer2023}), the evolution would be much faster and hence the first hours will be critical to trace the outflow. 

\begin{figure}[h]
    \centering
    \includegraphics[width=0.48\textwidth]{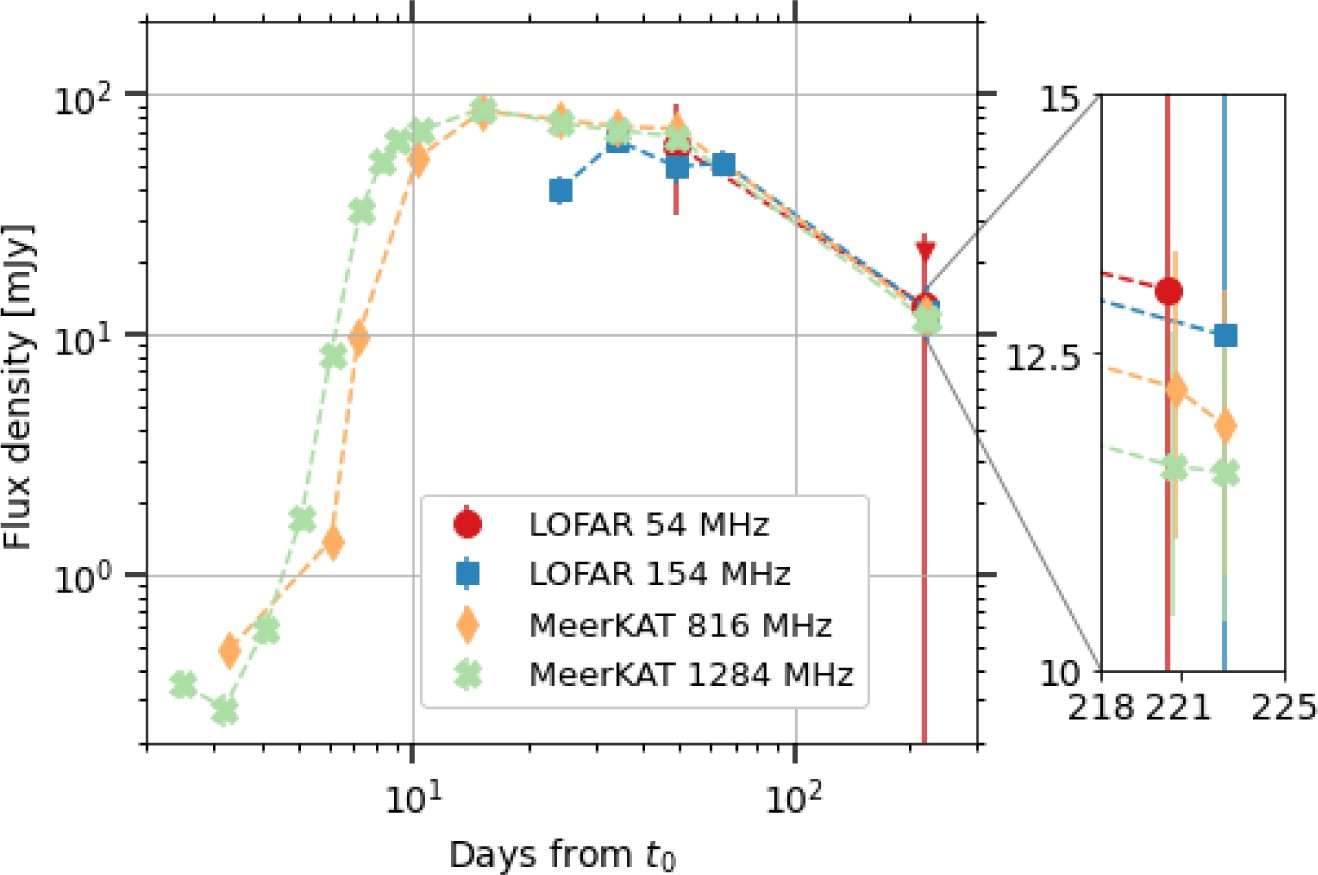}
    \includegraphics[width=0.5\textwidth]{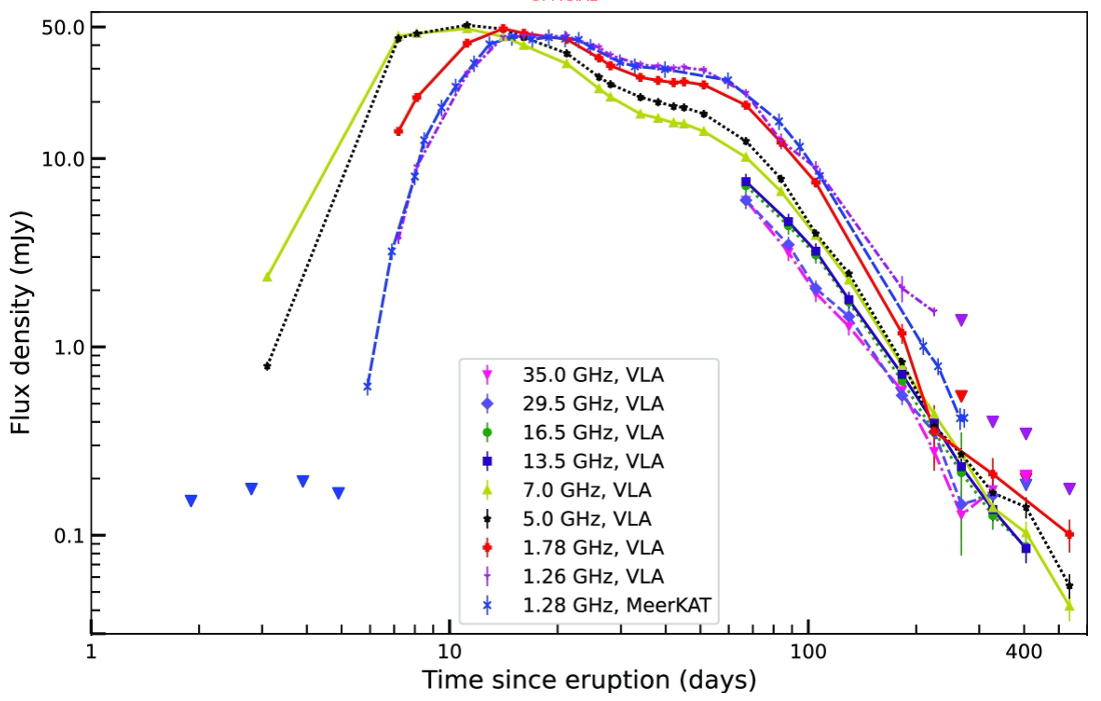}
    \caption{Radio light curves of the RS Ophiuchi 2021 eruption \citep[left;][]{deRuiter2023} and the V3890 Sagitarii 2019 August eruption \citep[right;][]{nyamai23}, which detected radio emission within $2-3$\,days of the outburst beginning. 
    }    
\label{fig:nova}
\end{figure}

Rapid-response multi-frequency radio interferometric monitoring can thus probe the evolving emission from both thermal material and accelerated non-thermal particles in the expanding ejecta, providing critical insight into the nature and location of energetic processes.
For example, Figure~\ref{fig:nova} shows that radio observations began within $2-3$ days of the outburst starting for both RS Ophiuchi \citep[left panel;][]{deRuiter2023} and V3890 Sagitarii \citep[right panel;][]{nyamai23}. Rapid-response observations may have detected the radio counterpart within 1 day to track the optically thick to optically transition from very early times.
Rapid triggering of SKAO observations following a nova eruption is therefore essential for capturing the rapid evolution of physical conditions that occur in the immediate aftermath of the explosion to differentiate both components \citep[][see also Figure~\ref{fig:nova}]{deRuiter2023,nyamai23}, establishing constraints on the efficient particle acceleration \citep{Giroletti2020} resulting in the production of gamma-ray emission \citep{Acciari2022}.

The hourly timescale variability of the radio emission and its intricate extended components forces the requirement of high-sensitivity snapshot observations to obtain precise images that trace the evolution of the outflow. 
SKAO will stand out in this regard, offering high instantaneous sensitivity and exceptional image fidelity. At lower frequencies, there are limited observations of novae \citep[e.g.][]{deRuiter2023,nyamai23,nayana24}, primarily due to triggering and sensitivity constraints. SKA-Low would provide a novel opportunity to obtain accurate light curves that when correlated with SKA-Mid observations, would enable us to trace the optically-thick to optically-thin transition along the outflow.

\subsection{Galactic X-ray Binaries}

X-ray binaries (XRBs) are binary star systems in which a compact object (either a black hole or neutron star) accretes matter from a companion star, emitting from radio to X-rays or even gamma rays. These systems exhibit complex and dynamic behavior, including state transitions and rapid flaring activity associated with accretion phenomena and transient ejections via relativistic jets \citep{fender04}. 
Notable examples, including Cygnus X-1, V404~Cygni or GRS~1915+105, show intermittent radio flares coinciding with hard-to-soft X-ray state transitions. These flares last from minutes to hours and flux density variations of tens to hundreds of mJy at GHz frequencies \citep[see e.g.][and Chapter \citet{Beri01.2026.SKA}]{Fender2006,Trushkin2017,MillerJones2019,Gandhi2025}.

\begin{figure}[h]
    \centering
    \includegraphics[width=0.63\textwidth]{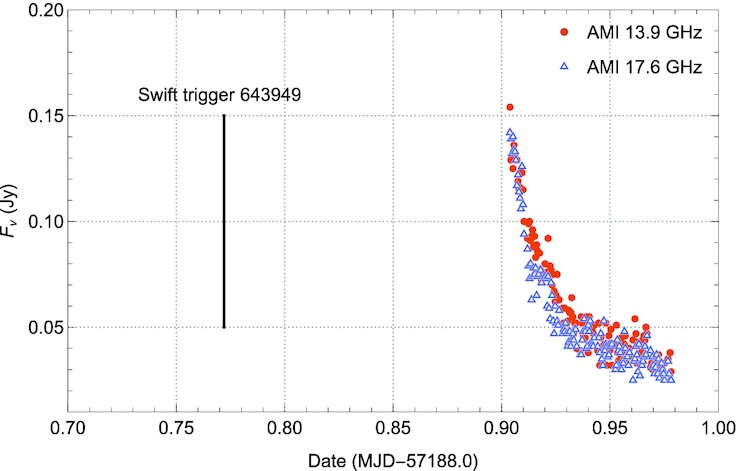}
    \includegraphics[width=0.36\textwidth]{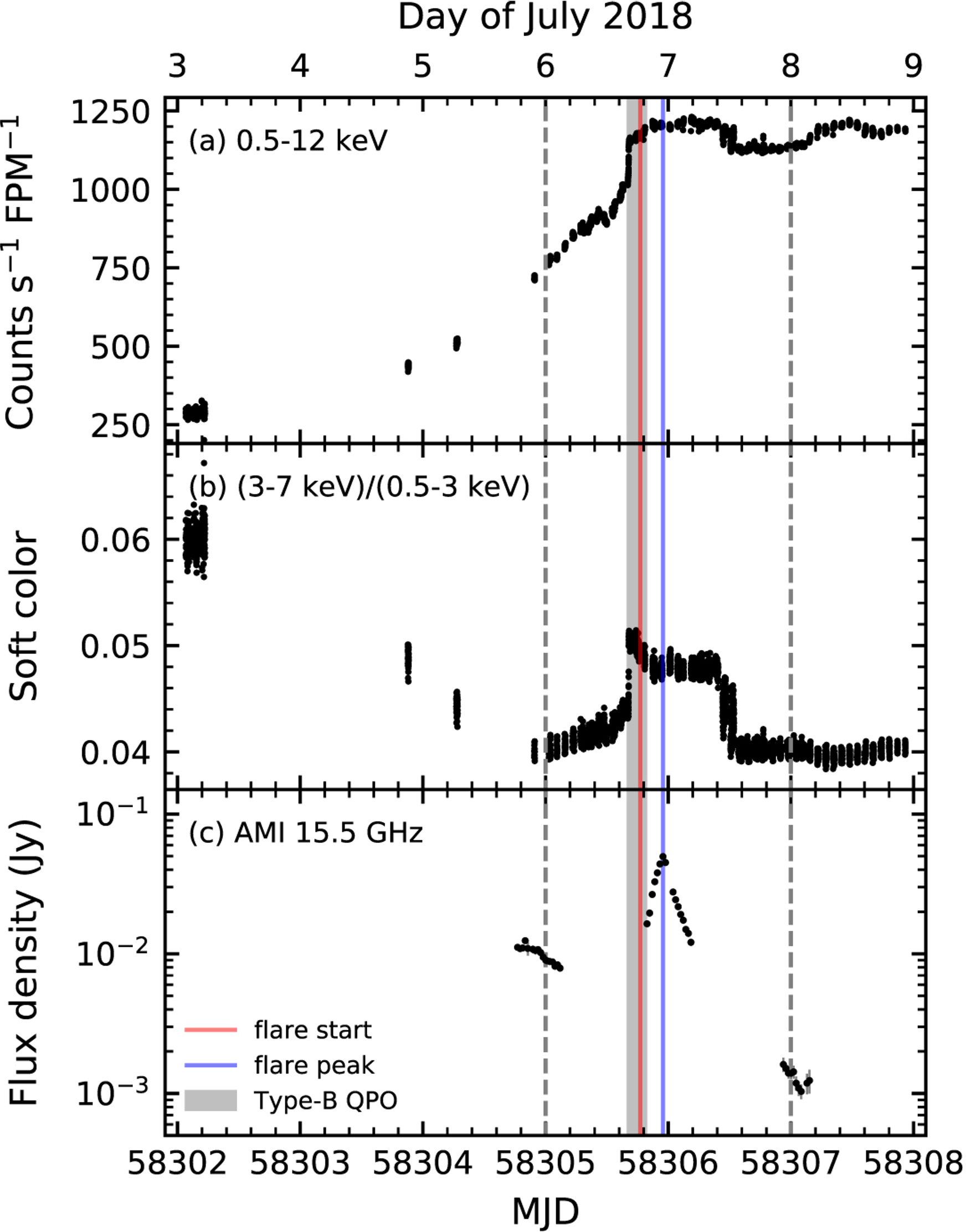}
    \caption{{\bf Left:} AMI-LA rapid-response observations of the V404 Cygni 2015 outburst from \citet[][]{fender23}. {\bf Right:} Figure adapted from \citet{homan20} showing the state transition in MAXI J1820+070, including the X-ray light curve (a), X-ray hardness ratio (b), and radio light curve (c). The start (red line) and peak (blue line) of the radio flare directly follows the QPO transition (grey region).
    }    
\label{fig:xrb}
\end{figure}

XRB outbursts are occasionally detected by all-sky X-ray and gamma-ray monitors such as \swift, which have resulted in triggering rapid-response observations with the AMI-LA as part of the ALARRM program. For example,  ALARRM triggered on the \swift{} detection of the V404 Cygni 2015 outburst, beginning observations within $3$\,hours when the target had risen to an elevation of $30^{\circ}$ \citep{fender23}. The left panel of Figure~\ref{fig:xrb} shows AMI-LA detected V404 Cyngi decaying from a flare that was $>1000$ times more luminous than its quiescent level. Ongoing radio and X-ray monitoring showed the flaring behaviour to be correlated on all timescales, and required extended periods of particle acceleration from the central engine rather than discrete, impulsive injection. 
Similarly, AMI-LA also triggered on the \swift{} detection of the MAXI J1820+070 2018 outburst, resulting in the earliest radio detection of a new black hole XRB just 90\,minutes later \citep{bright20}. Continued AMI-LA monitoring during the hard to soft state transition, detected a radio flare directly following a switch between two types of quasi-periodic oscillations (QPOs) in the X-ray band \citep[see right panel of Figure~\ref{fig:xrb} adapted from][]{homan20}. This demonstrated a link between QPO transitions and discrete jet ejections, providing a temporal signature that could be used to trigger SKA-VLBI observations to track the associated radio emission.

The unprecedented temporal and spectral sensitivity of SKAO will revolutionize rapid radio follow-up of XRBs flaring activity, allowing for the detection and monitoring of faint ejecta or short-duration flares.
While quasi-simultaneous multi-frequency radio observations have been obtained through the coordination of multiple radio facilities \citep[e.g.][]{chauhan21}, simultaneous SKA-Low and SKA-Mid observations would enable real time broadband imaging of the jet dynamics on (sub)hour-to-day timescales, constraining the jet evolution, acceleration and collimation at a new level. 
Rapid and sensitive radio follow-up, would enable the tracking of propagation velocities, jet geometry changes, and correlations with X-ray states, ultimately constraining the coupling between accretion/ejection processes and jet formation \citep{Fender2025}. 
In addition, several systems show jet features moving significantly in intra-hour observations \citep{MillerJones2019}. The compelling high instantaneous sensitivity of SKA-Mid, in combination with VLBI would enable the creation of snapshots on minute exposure times while capturing the moving emission, which is limited to hour timescales with current instruments \citep{MillerJones2019,wood21,wood23,wood24,wood25}. On the detection of bright radio flaring from an XRB, SKA-Mid could triggering rapid-response follow-up with SKA-VLBI to capture these geometric changes, shedding light on the launching timescales and mechanism \citep[also see Chapter][]{TaoAn01.2026.SKA}.

\subsection{Neutron Stars}

\subsubsection{Pulsars}

\begin{figure}[h]
    \centering
    \includegraphics[width=0.9\textwidth]{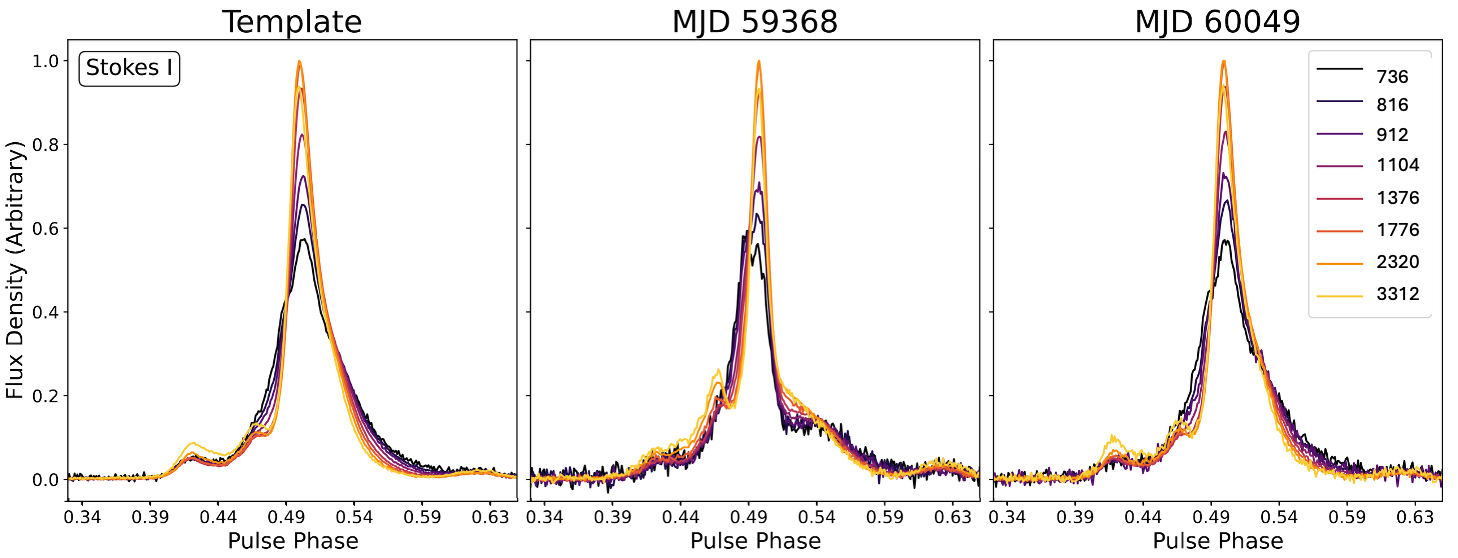}
    \caption{Figure adapted from \citet{mandow25} showing the stokes I profiles of PSR J1713+0747 at 47 days (middle panel) and 2 years (right panel) following the event compared to the template profile from before the event.
    }    
\label{fig:msp}
\end{figure}

Time-integrated pulse profiles of millisecond pulsar (MSPs) exhibit exceptional long-term stability, surpassing that observed in the canonical pulsar population \citep{2009MNRAS.400..951V}. It is this stability that is exploited for precision timing experiments \citep{2021PhRvX..11d1050K,2024ApJ...971L..18R}, such as the usage of MSPs in pulsar timing arrays \citep[][see also Chapter \citet{Shannon01.2026.SKA}]{2023ApJ...951L...8A,2023A&A...678A..50E,2023ApJ...951L...6R,2025MNRAS.536.1467M}. However, a small subset of the MSP population have exhibited temporal profile variability, including long-term discrete profile changes, which are rare and not yet well understood.
Long-term discrete profile changes are characterised as infrequent events and exhibit a sudden emergence of alteration to profile components, followed by a slow recovery towards the pre-event morphology. To date, only three MSPs have been observed to exhibit such behaviour: PSR J0437$-$4715 \citep{2021MNRAS.502..478G}; PSR J1643$-$1224 \citep{2016ApJ...828L...1S}; and PSR J1713$+$0747 \citep{mandow25}. 
For example, Figure~\ref{fig:msp} adapted from \citet{mandow25} show pulse profiles of PSR J1713+0747 following the event compared to its template as observed by the Ultra-Wideband Low-frequency receiver \citep[UWL;][]{hobbs20} on Murriyang, CSIRO's Parkes Radio Telescope.
In some of these cases, these events have also exhibited accompanying changes to the polarisation profiles, offering unique insights into the evolving, dynamic and complex magnetospheric environment, which remains poorly understood.
Currently, the International Pulsar Timing Array 
\citep{2024ApJ...966..105A} regularly monitors a large number of MSPs on cadences ranging between days to a month for discrete profile change events, which initiate internal alerts for multi-telescope, broad-frequency follow-up campaigns.

Due to the abrupt onset and transient nature of discrete profile changes in MSPs, it is important to conduct follow up observations as soon as detections are announced. 
The profile change is most significant when the transient event is first discovered, motivating the need for rapid follow-up within the first 12 to 24 hours, which can be best accommodated by a rapid-response mode.
Currently, the Parkes Pulsar Timing Array \citep{manchester13} collaboration operates a rapid follow-up observing programme with Murriyang, CSIRO's Parkes Radio Telescope.
Given the similar sky coverage, the same approach is well-suited for implementation for any SKAO follow-up observations using both \skalow\ and \skamid. SKAO monitoring of MSP could also provide internal alerts, enabling the fastest follow-up of sources showing discrete profile changes. This would be followed by daily monitoring for at least seven days, enabling the tracking of temporal evolution and the recovery of profile morphology across a broad range of frequencies.

\subsubsection{Magnetars}


Magnetars are a sub-class of the neutron star population \citep[see Chapter][]{Levin01.2026.SKA} characterized by bright emission typically seen at high energies \citep[X-rays/gamma-rays, for a review see][]{kaspi17}. They exhibit episodic outbursts that are seen across the electromagnetic spectrum. A small fraction of them are observed as radio pulsars with the onset of radio emission correlated with the outburst \citep{halpern05,camilo06,camilo07,levin10,anderson12,torne15,dai19}. It is still unclear when the radio emission mechanism in magnetars turns on at the start of the outburst phase \citep{rea12,rajwade22magnetar}. Finding the exact epoch can provide clues to the elusive radio emission physics of neutron stars. 

\begin{figure}[h]
    \centering
    \includegraphics[width=0.75\textwidth]{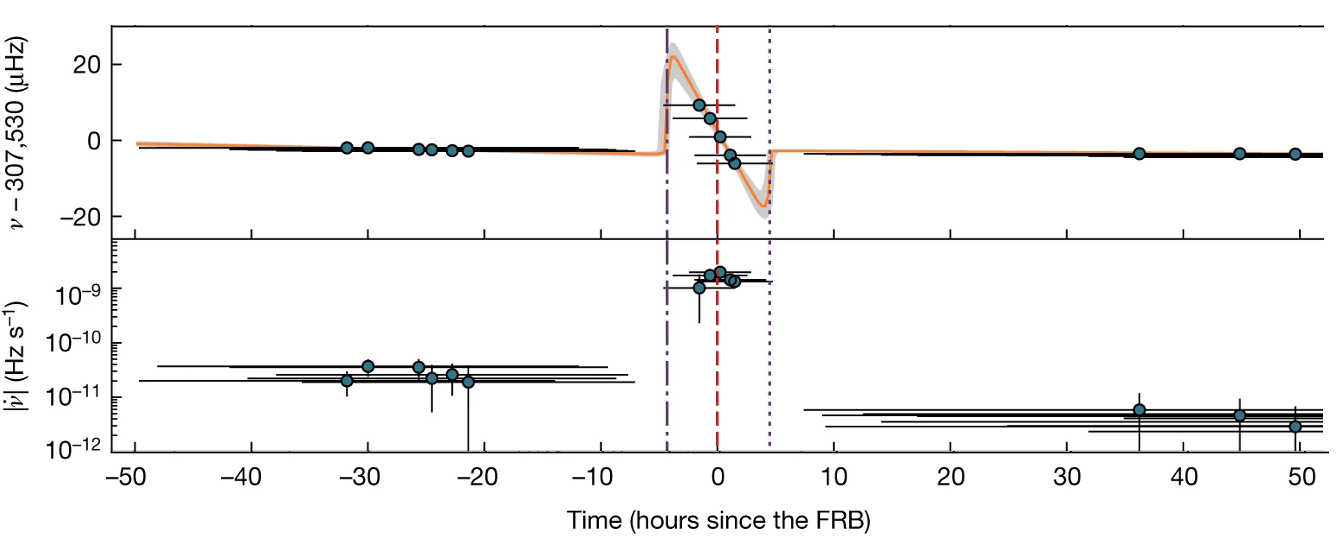}
    \caption{Figure adapted from \citet{hu24} showing the change in SGR 1935+2154's spin frequency (top) and spin down rate (bottom) before, during and after the first and second X-ray detected glitches (vertical dashed-dotted and dotted lines), which bracket the CHIME/FRB-detected burst (vertical red dashed line).
    }    
\label{fig:mag}
\end{figure}

Interest in magnetars grew further when a bright, MJy burst was observed from the magnetar SGR 1935+2148 \citep{bochenek20,chime_magnetar20} showing a connection with FRBs; which still have an unknown origin. 
Several FRB-like bursts have been observed from SGR 1935+2154 that seem to be correlated with the onset of high bursting activity at X-ray wavelengths~\citep[e.g.][]{bochenek20, kirsten2021}. 
Indeed, \citet{hu24} reported the X-ray detection of two glitches bracketing an FRB that occurred on 2022 October 14 \citep{maan22atel,giri23pp}, which they attribute to the strong magnetospheric wind providing a torque that rapidly slows the star's rotation \citep[see Figure~\ref{fig:mag} adapted from][]{hu24}.
Additionally, magnetars are also a proposed progenitor for LPTs \citep[][see also Section~\ref{sec:lpt}]{2022Natur.601..526H,hurley-walker23}, which rotate too slowly for the associated radio emission to be attributed to dipole spin down radiation \citep{rea22,rea24,cooper24}. 

Although the FRB-like bursts from SGRB 1935+2154 were associated with high-energy outbursts reported via the GCN, their detection was not due to rapid-response radio follow-up. Rather, they resulted either serendipitously through regular monitoring with, for example, CHIME-FRB or from radio observations scheduled days to weeks following the outbursts \citep[e.g.][]{kirsten2021,zhang20ATel}. 
In addition, associations between FRB-bursts and high-energy outbursts are quite rare \citep[e.g.][]{younes17,lin20}, demonstrating that rapid-response follow-up is crucial for catching these unusual events.
Understanding this association and exploring the potential link with LTPs requires simultaneous SKA-Mid and SKA-Low rapid-response observations of magnetars experiencing high-energy activity for 
1) detecting the emergence of coherent pulsed radio emission  
and 2) detecting FRB-like bursts during a high bursting activity phase.
SKA-Mid observations will provide valuable insight into the radio emission mechanism in the immediate aftermath of the outburst when one can investigate how the radio pulse profile, polarisation and single burst properties depend on time and the temporal and spectral properties of the X-ray emission. 
SKA-Low observations will constrain the low frequency flux density spectrum of radio-loud magnetars, which have been difficult to obtain due to the high degree of scattering and dispersion smearing. Since the radio spectral energy distribution for magnetars differs from canonical pulsars, constraining the spectral behaviour across a wide bandwidth would enable a deeper understanding of the radio emission mechanism. Furthermore, identifying the epoch of FRB-like bursts and coherent radio emission will be critical to establish a physical link between FRBs and magnetars.

\subsection{Solar and heliospheric observations}

\subsubsection{Solar triggering}

Solar emission is highly variable along all observable axes -- brightness temperature, time, frequency, variability, morphology and polarization, providing valuable insights into solar physics and space weather. 
Capturing all aspects of this variation simultaneously requires snapshot spectro-polarimetric imaging with high dynamic range and fidelity while maintaining high temporal and spectral resolution. 
However, solar activity is inherently unpredictable, especially from the perspective of scheduling observations. 
As a result, the need to capture solar activity has resulted in the construction of dedicated solar instruments (both single elements and interferometers),
but their capabilities fall short of what is needed to capture the highly varied, structured and variable solar emission. 
Fortunately, these needs can be met by state-of-the-art versatile interferometers, such as the SKAO and its precursors \citep[see Chapter][for an overview of Solar and  Heliospheric physics that can be explored with SKAO]{Zucca01.2026.SKA}.
The scientific merits of using these instruments for solar and heliospheric observations are already well established \citep[e.g.][]{Oberoi2023,Zhang2024-LOFAR-Uburst, Normo2025-LOFAR-typeII, Mondal2025-uGMRT, Dey2025-LinPol, Kansabanik2025-MeerKAT}.

\begin{figure}[h]
    \centering
    \includegraphics[width=0.75\textwidth]{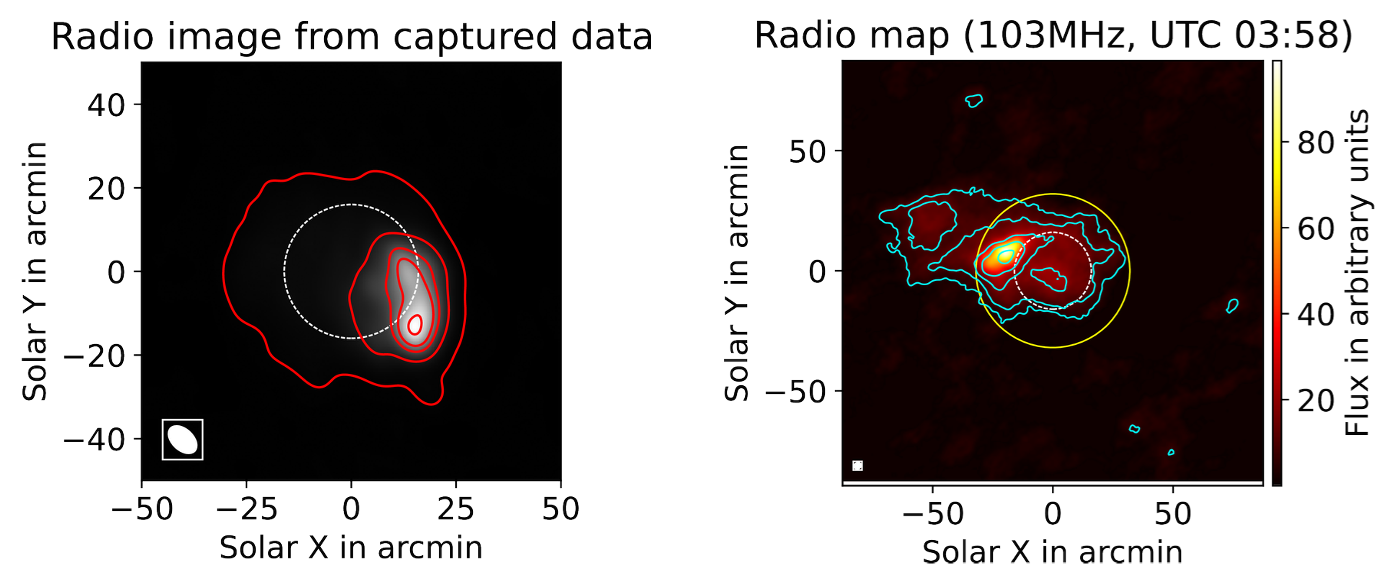}
    \caption{Figure adapted from \citet{patra26} showing MWA rapid-response images of solar bursts, where the observations were triggered by alerts from the Yamagawa heliospectograph in Japan. The dotted white circle shows the optical disc of the Sun.
    {\bf Left:} MWA observation at 126\,MHz on 2024 August 1. Red contours are 1, 10, 30, 50, 90\% of the peak.
    {\bf Right:} MWA observation at 103\,MHz on 2024 November 4. The cyan contours are 5, 10, 20, 40, 80\% of the peak.
    }    
\label{fig:sol}
\end{figure}

SKAO observing time will be highly oversubscribed, with limited time for solar observations. 
Solar radio transients are also infrequent and unpredictable, making blind scheduling efforts inefficient. 
This approach would risk missing major episodes of solar activity and delay the collection of a statistically significant sample. 
Implementing a robust, automated near-real-time triggering system will capture these events with an efficient use of telescope time. 
The first radio telescope to perform rapid-response observations based on external solar triggers is MWA, which has demonstrated its efficacy by capturing $\sim$110 bursts over 6 months \citep[see Figure~\ref{fig:sol} adapted from][]{patra26}.
This system uses unallocated observing time during the day in conjunction with the MWA voltage buffer system \citep{morrison2023mwax}, to record data up to 160\,s before the trigger arrival. This partially mitigates the $\sim4$\,minutes of latency between the trigger notice and the associated radio emission arriving at Earth.
The bulk of this latency is due to relying on external triggers, which could be substantially reduced by using an internal trigger. 

Multiple approaches for catching solar events can be implemented with the SKAO telescopes, including responding to external triggers from co-located (e.g. solar spectrometer) or remote instruments (e.g. satellite solar observatories), as well as internal SKAO triggers.
Using SKAO subarraying and substation capabilities, the smallest substation \citep[a footprint of 6\,m for $\sim$60 SKA-Low stations within a 10\,km diameter adding up to $<$0.5\% of the collecting area;][]{SKAO-Low-SubStation-Templates} could be used for solar monitoring when the Sun is above the elevation limit, with minimal sensitivity impact on other observations.
\skalow's flexible spectral sampling \citep{SKAO-YearInLife} would enable observations with just 100 channels spread across 50--350\,MHz, each with a bandwidth of 5.4\,kHz (total 0.5\,MHz), leaving a large fraction of the telescope resources available for the commensal observation \citep{patra26}.
This solar monitoring station could directly trigger the \skalow{} buffer, which has a substantial 900\,s capacity \citep{SKAO-YearInLife}, directly compensating for any latency in the trigger and therefore collecting data of the full solar event.

An \skamid{} subarray of 4 telescopes could also be used for solar monitoring using a configuration optimised to match the size of the radio solar disk \citep{SKAO-subarray-templates2025}. 
If monitoring with three subarrays to simultaneously cover SKA Bands 1, 2 and 5a, near-real-time analysis of the dynamic spectra could identify solar activity, which then generates a trigger to initiate solar observations with a larger array.  
Automated alerts could also be generated for triggering other solar observatories.

\subsubsection{Heliospheric triggering}

Coronal mass ejections (CMEs) are potent drivers of space weather, which can impact power grids and satellite based assets, astronaut safety, outages in radio communications, and Global Navigation Satellite Systems (GNSS) based services.
The largest factor driving the geo-effectiveness of CMEs is the strength and orientation of their vector magnetic field.
Our best information of CME entrained magnetic fields comes from in-situ measurements from the spacecraft located at the L1 Lagrange point, about 1.5 million km ($\sim$1\% of an AU) towards the Sun on the Earth-Sun line. 
These measurements provide only 30\,min warning for the fastest CMEs traveling at speeds in excess of a 1000\,km\ s$^{-1}$ and only a factor of two more for the more usual CMEs.
This is grossly insufficient for implementing geo-magnetic storm mitigation steps.
The only method to estimate the CME magnetic field while it is close to the Sun is by measuring Faraday rotation changes in the background linearly polarized sources as their emission passes through the magnetized CME plasma \citep[e.g.][]{Oberoi2004}. 
This promising approach has been successfully demonstrated \citep{Kooi2022-FR-review}. However, due to limitations in sensitivity and image quality, has been limited to a handful of suitable sources located in the ecliptic \citep{Kooi2021-CME-FR} or radio beacon experiments \citep{Bird2007}. 
The sensitivity of SKAO telescopes will enable more robust experiments through observations of a larger number of discreet Galactic and extra-galactic sources and the polarised Galactic background.

While CMEs are unpredictable, space based Extreme-UV observations and low latency coronagraph observations can identify them lifting off from the Sun and estimate their direction of propagation, speed and angular width.
Using community standard CME propagation models, it is possible to estimate its trajectory in the plane of the sky, identify suitable background polarized sources and schedule observations. 
The vast distances needed to be traversed by the CME imply that the observations need to be scheduled on a time scale of a few hours.
The first such observing programs have already been implemented at MWA, ASKAP and most recently MeerKAT, and have yielded multiple successful observations \citep{Devojyoti2024-heliopolarimetry}.

\section{Triggering Technology and Landscape} \label{sec:tech}

Rapid response to transient phenomena requires seamless integration into a complex global ecosystem of observatories, surveys, and communication platforms. The landscape of these tools has significantly evolved over the last years to handle the increasing volume and rapidity of alerts from multi-wavelength and multimessenger facilities, moving towards a highly automated, machine (and increasingly AI) driven paradigm, while also allowing for human-in-the-loop decision making.

\subsection{The Modern Alert Ecosystem} \label{sec:alerts}
The coordination of follow-up observations is managed by a variety of specialized services, each providing rapid access to different types of astronomical events.

\begin{description}
    \item[General Coordinates Network (GCN):]
    For decades, the GCN has been the cornerstone for disseminating alerts on high-energy phenomena like GRBs and, more recently, multimessenger events such as GWs and high-energy neutrinos. It provides low-latency alerts (‘notices’) that are essential for capturing the earliest electromagnetic counterparts as well as human-written reports (‘circulars’) that provide results on follow-up observations. Both can be accessed via alert streams sent via the Apache Kafka\footnote{https://kafka.apache.org/} streaming platform.
    \item[Radio Alert Systems:] In the radio domain, dedicated systems have been established to announce extremely rapid phenomena. A key example is the CHIME/FRB \texttt{frb-voe}\footnote{https://www.chime-frb.ca/voevents} \citep{2025AJ....169...39A}, which is a VOEvent broker that provides real-time announcements of FRB detections (see Section~\ref{sec:auto} for a description of VOEvents). These alerts use a specific VOEvent schema for FRBs~\citep{2017arXiv171008155P}, a standardized, machine-readable format for communicating essential event parameters, such as the burst time, position, and dispersion measure, to automatically trigger follow-up at other wavelengths.
    \item[Transient Name Server\footnote{https://www.wis-tns.org/} (TNS):]  
    As the official International Astronomical Union (IAU) mechanism, the TNS is the central service for reporting and naming new extragalactic transients, primarily supernovae and Tidal Disruption Events. While not designed for the same sub-minute latency as GCN, it is a crucial tool for access to transient classification information. Data can be accessed via an API.
    \item[Vera C. Rubin Observatory\footnote{https://rubinobservatory.org/for-scientists/data-products/alerts-and-brokers} (LSST) Alert Brokers:]
    The upcoming Legacy Survey of Space and Time (LSST) will generate an unprecedented flood of alerts, up to 10 million per night. This data volume makes manual inspection impossible. 
    Consequently, there is a network of alert brokers, including ALeRCE, AMPEL, ANTARES, Babamul, Fink, Lasair, Pitt-Google, SNAPS, POI Broker \citep{forster21,nordin19,matheson21,jegoudulaz26,moller21,williams24,wood-vasey24zen,trilling23}.
    These brokers depend on in-house and community-driven 
    algorithms to filter, classify, and prioritize the LSST alert stream in real-time.
    Several of these systems also ingest other transient alert notices (e.g. the GCN) and/or cross-matches with multi-wavelength catalogues or transient surveys (e.g. the ZTF) for transient aggregation and classification (for example, the Bursts and Outbursts Observation Monitor or BOOM on which Babamul is built).\footnote{https://gcn.nasa.gov/missions/boom}
    The resulting Kafka streams are made available to the scientific community \citep[e.g. Fink;][]{moller21}. 
\end{description}
 
\subsection{Aggregation and Accessibility: Astro-COLIBRI}
While services like the GCN and LSST brokers are built primarily for high-speed, machine-to-machine communication, they create a significant challenge of information overload for the individual researcher. The Astro-COLIBRI platform~\citep{2021ApJS..256....5R} is designed specifically to address this by keeping the expert human-in-the-loop. It provides a crucial synthesis layer on top of the raw alert streams, aggregating data from the GCN, TNS, LSST brokers, and other sources into a single, intuitive interface. By offering powerful filtering capabilities via modern interfaces, including a web portal, smartphone applications, and a programmable API, Astro-COLIBRI transforms the flood of machine-readable data into actionable information. 
This enables scientists to quickly assess the relevance of an event and make informed decisions about triggering follow-up observations, bridging the gap between automated detection and human-driven discovery.

\subsection{Telescope triggering services}\label{sec:trigg_service}

Software is being developed that take alert notices from the services and brokers to trigger automated observations with target telescopes.

\begin{description}
    \item [TRACE-T] The Transient Rapid-response using Coordinated Event Triggering (TRACE-T)\footnote{https://github.com/ADACS-Australia/TraceT2} is a web application that automatically monitors and filters VOEvents broadcast via Kafka alert streams (e.g. the GCN or Fink) to trigger observations with radio telescopes based on user defined parameters. Developed by the Australian Astronomy Data and Computing Services, TRACE-T currently provides the modules for triggering rapid-response observations with both ATCA \citep{anderson21} and MWA \citep{hancock19voevent}, allowing the user to specify the observing set up. The TRACE-T software package has been modularised so that it can be expanded for use with other telescope facilities. 
    \item[TOM Toolkit:]  Target and Observation Manager software for powerful, programmable control of astronomical observing programs (TOM Toolkit)\footnote{https://lco.global/tomtoolkit/} allows users to build Target and Observation Managers (TOMs) 
    to manage large astronomical programs and data volumes \citep{street18}. It enables users to submit requests directly to networked optical telescopes with highly flexible queue-scheduling for rapid transient follow-up. TOM Toolkit is used to combine large telescope networks including AEON+\footnote{https://aeonplus.github.io/} \citep{street20} and the Siding Spring Observatory Alert System.\footnote{https://indico.in2p3.fr/event/37156/contributions/164880/attachments/97729/150477/ozfink\_sssoalertsystem.pdf}
\end{description}

\subsection{A Suggested Framework for SKAO Triggering}
For the SKAO to operate effectively, its triggering framework must support two complementary modes: rapid, fully automated, machine-to-machine communication and manual, expert-driven observations. The system must be capable of both automatically responding to external alerts and disseminating its own discoveries, relying on standardized machine-readable formats and modern distribution protocols. Simultaneously, it must be flexible enough to allow astronomers to initiate follow-up observations on compelling targets identified via interactive platforms.

\subsubsection{Automated Triggering} \label{sec:auto}
Alerts are increasingly distributed via high-throughput, low-latency messaging systems like Apache Kafka. This ensures observatories receive information with minimal delay. The format of these notices is currently evolving. The traditional format is the VOEvent, an XML-based schema established by the International Virtual Observatory Alliance (IVOA). While robust, XML is verbose. To improve efficiency, the community is increasingly moving towards JSON (JavaScript Object Notation) schemas. GCN has already introduced a novel JSON-based schema for its notices, and the IVOA is actively developing a JSON serialization for the VOEvent standard. The SKAO trigger system should be built upon Kafka streams and be able to parse both the legacy XML and the modern JSON formats to ensure comprehensive coverage and future-proofing. This infrastructure will allow the SKAO to automatically ingest an alert, verify its visibility and scientific priority, and command the observatory to repoint/reconfigure, all within seconds of the initial discovery.

\subsubsection{Manual and Expert-in-the-Loop Triggering}
While automated systems are essential for handling the sheer volume of alerts, the expertise of astronomers remains invaluable for identifying unique or unexpected phenomena. Modern platforms like Astro-COLIBRI are not just for monitoring; they are designed to be interactive, allowing for direct triggering of follow-up observations. Once an interesting multi-wavelength or multi-messenger transient is identified, an authenticated researcher can request SKAO observations with a simple action, such as pressing a button in a smartphone app. This direct-triggering paradigm is already well-established and has been successfully implemented with current observatories like H.E.S.S.~\citep{2006A&A...457..899A, 2022A&A...666A.119H}, MAGIC~\citep{aleksic_2016}, VERITAS~\citep{Holder2006}, and the Cherenkov Telescope Array Observatory \citep[CTAO;][]{2019scta.book.....C} operating in the high-energy gamma-ray domain, as well as in the optical domain through the Las Cumbres Observatory (LCO) Telescope Observation Manager (TOM) framework. We note that this framework is already been prototyped for radio facilities like the European VLBI Network (EVN). Providing human interfaces for triggering or exploiting existing platforms like Astro-COLIBRI will empower the community to react quickly and flexibly to the most scientifically compelling transients.

\section*{Acknowledgments}

We acknowledge the Whadjuk Nyungar as the traditional owners of the land where the majority of this chapter was written. 
A.P.C. is a Canadian SKA Scientist and is funded by the Government of Canada / est financ\'e par le gouvernement du Canada.
AR acknowledges funding from the European Research Council (ERC) under the European Union’s Horizon research and innovation programme (‘QuickBlitz’; grant agreement number 101170284), funding from the NWO Aspasia grant (number 015.016.033), and support through the project CORTEX (project number NWA.1160.18.316) of the research programme NWA-ORC, which is (partly) financed by the Dutch Research Council (NWO). Views and opinions expressed are however those of the author(s) only and do not necessarily reflect those of the European Union or the European Research Council. Neither the European Union nor the granting authority can be held responsible for them.
DO acknowledges the support of the Department of Atomic Energy, Government of India, under project no. 12-R\&D-TFR-5.02-0700.
FS acknowledges ANR (French National Research Agency) for its support of the project "Multi-messenger observations of the Transient Sky (MOTS)" under grant no. ANR-22-CE31-0012.

\bibliographystyle{abbrvnat-maxbibnames4.bst}
\bibliography{chapter} 

@incollection{Colombo01.2026.SKA, author = {Alberto Colombo and author2 and author3 and author4 and author5},title = {},year = {2026},publisher = {},note = {arXiv search: Report number AASKAII/Colombo01},booktitle = {Advancing Astrophysics with the SKA -- II (AASKAII)}}

@incollection{Curtin01.2026.SKA, author = {Alice P. Curtin and author2 and author3 and author4 and author5},title = {},year = {2026},publisher = {},note = {arXiv search: Report number AASKAII/Curtin01},booktitle = {Advancing Astrophysics with the SKA -- II (AASKAII)}}

@incollection{Caleb01.2026.SKA, author = {Manisha Caleb and author2 and author3 and author4 and author5},title = {},year = {2026},publisher = {},note = {arXiv search: Report number AASKAII/Caleb01},booktitle = {Advancing Astrophysics with the SKA -- II (AASKAII)}}

@incollection{Caleb02.2026.SKA, author = {Manisha Caleb and author2 and author3 and author4 and author5},title = {},year = {2026},publisher = {},note = {arXiv search: Report number AASKAII/Caleb02},booktitle = {Advancing Astrophysics with the SKA -- II (AASKAII)}}

@incollection{Castignani01.2026.SKA, author = {Gianluca Castignani and author2 and author3 and author4 and author5},title = {},year = {2026},publisher = {},note = {arXiv search: Report number AASKAII/Castignani01},booktitle = {Advancing Astrophysics with the SKA -- II (AASKAII)}}

@incollection{Driessen01.2026.SKA, author = {Laura N. Driessen and author2 and author3 and author4 and author5},title = {},year = {2026},publisher = {},note = {arXiv search: Report number AASKAII/Driessen01},booktitle = {Advancing Astrophysics with the SKA -- II (AASKAII)}}

@incollection{Cavallaro01.2026.SKA, author = {Francesco Cavallaro and author2 and author3 and author4 and author5},title = {},year = {2026},publisher = {},note = {arXiv search: Report number AASKAII/Cavallaro01},booktitle = {Advancing Astrophysics with the SKA -- II (AASKAII)}}

@incollection{Watanabe01.2026.SKA, author = {Keito Watanabe and author2 and author3 and author4 and author5},title = {},year = {2026},publisher = {},note = {arXiv search: Report number AASKAII/Watanabe01},booktitle = {Advancing Astrophysics with the SKA -- II (AASKAII)}}

@incollection{Nelles01.2026.SKA, author = {Anna Nelles and author2 and author3 and author4 and author5},title = {},year = {2026},publisher = {},note = {arXiv search: Report number AASKAII/Nelles01},booktitle = {Advancing Astrophysics with the SKA -- II (AASKAII)}}

@incollection{Buitink01.2026.SKA, author = {Stijn Buitink and author2 and author3 and author4 and author5},title = {},year = {2026},publisher = {},note = {arXiv search: Report number AASKAII/Buitink01},booktitle = {Advancing Astrophysics with the SKA -- II (AASKAII)}}

@incollection{Corstanje01.2026.SKA, author = {Arthur Corstanje and author2 and author3 and author4 and author5},title = {},year = {2026},publisher = {},note = {arXiv search: Report number AASKAII/Corstanje01},booktitle = {Advancing Astrophysics with the SKA -- II (AASKAII)}}

@incollection{Rosch01.2026.SKA, author = {Florian Rösch and author2 and author3 and author4 and author5},title = {},year = {2026},publisher = {},note = {arXiv search: Report number AASKAII/Rosch01},booktitle = {Advancing Astrophysics with the SKA -- II (AASKAII)}}

@incollection{Qiu01.2026.SKA, author = {Hao Qiu and author2 and author3 and author4 and author5},title = {},year = {2026},publisher = {},note = {arXiv search: Report number AASKAII/Qiu01},booktitle = {Advancing Astrophysics with the SKA -- II (AASKAII)}}

@incollection{Beri01.2026.SKA, author = {Aru Beri and author2 and author3 and author4 and author5},title = {},year = {2026},publisher = {},note = {arXiv search: Report number AASKAII/Beri01},booktitle = {Advancing Astrophysics with the SKA -- II (AASKAII)}}

@incollection{TaoAn01.2026.SKA, author = {Tao An and author2 and author3 and author4 and author5},title = {},year = {2026},publisher = {},note = {arXiv search: Report number AASKAII/TaoAn01},booktitle = {Advancing Astrophysics with the SKA -- II (AASKAII)}}

@incollection{Lico01.2026.SKA, author = {Rocco Lico and author2 and author3 and author4 and author5},title = {},year = {2026},publisher = {},note = {arXiv search: Report number AASKAII/Lico01},booktitle = {Advancing Astrophysics with the SKA -- II (AASKAII)}}

@incollection{AlexAndersson01.2026.SKA, author = {Alex Andersson and author2 and author3 and author4 and author5},title = {},year = {2026},publisher = {},note = {arXiv search: Report number AASKAII/AlexAndersson01},booktitle = {Advancing Astrophysics with the SKA -- II (AASKAII)}}

@incollection{Levin01.2026.SKA, author = {Lina Levin and author2 and author3 and author4 and author5},title = {},year = {2026},publisher = {},note = {arXiv search: Report number AASKAII/Levin01},booktitle = {Advancing Astrophysics with the SKA -- II (AASKAII)}}

@incollection{Shannon01.2026.SKA, author = {Ryan M. Shannon and author2 and author3 and author4 and author5},title = {},year = {2026},publisher = {},note = {arXiv search: Report number AASKAII/Shannon01},booktitle = {Advancing Astrophysics with the SKA -- II (AASKAII)}}

@incollection{Zucca01.2026.SKA, author = {Zucca Pietro and author2 and author3 and author4 and author5},title = {},year = {2026},publisher = {},note = {arXiv search: Report number AASKAII/Zucca01},booktitle = {Advancing Astrophysics with the SKA -- II (AASKAII)}}

@ARTICLE{hare2020,
       author = {{Hare}, B.~M. and {Scholten}, O. and {Dwyer}, J. and {Ebert}, U. and {Nijdam}, S. and {Bonardi}, A. and {Buitink}, S. and {Corstanje}, A. and {Falcke}, H. and {Huege}, T. and {H{\"o}randel}, J.~R. and {Krampah}, G.~K. and {Mitra}, P. and {Mulrey}, K. and {Neijzen}, B. and {Nelles}, A. and {Pandya}, H. and {Rachen}, J.~P. and {Rossetto}, L. and {Trinh}, T.~N.~G. and {ter Veen}, S. and {Winchen}, T.},
        title = "{Radio Emission Reveals Inner Meter-Scale Structure of Negative Lightning Leader Steps}",
      journal = {\prl},
         year = 2020,
        month = mar,
       volume = {124},
       number = {10},
          eid = {105101},
        pages = {105101},
          doi = {10.1103/PhysRevLett.124.105101},
archivePrefix = {arXiv},
       eprint = {2007.03231},
 primaryClass = {physics.ao-ph},
       adsurl = {https://ui.adsabs.harvard.edu/abs/2020PhRvL.124j5101H}
}

@ARTICLE{kirsten2021,
       author = {{Kirsten}, F. and {Snelders}, M.~P. and {Jenkins}, M. and {Nimmo}, K. and {van den Eijnden}, J. and {Hessels}, J.~W.~T. and {Gawro{\'n}ski}, M.~P. and {Yang}, J.},
        title = "{Detection of two bright radio bursts from magnetar SGR 1935 + 2154}",
      journal = {Nature Astronomy},
         year = 2021,
        month = apr,
       volume = {5},
        pages = {414-422},
          doi = {10.1038/s41550-020-01246-3},
archivePrefix = {arXiv},
       eprint = {2007.05101},
 primaryClass = {astro-ph.HE},
       adsurl = {https://ui.adsabs.harvard.edu/abs/2021NatAs...5..414K}
}

@ARTICLE{marcote17,
       author = {{Marcote}, B. and {Paragi}, Z. and {Hessels}, J.~W.~T. and {Keimpema}, A. and {van Langevelde}, H.~J. and {Huang}, Y. and {Bassa}, C.~G. and {Bogdanov}, S. and {Bower}, G.~C. and {Burke-Spolaor}, S. and {Butler}, B.~J. and {Campbell}, R.~M. and {Chatterjee}, S. and {Cordes}, J.~M. and {Demorest}, P. and {Garrett}, M.~A. and {Ghosh}, T. and {Kaspi}, V.~M. and {Law}, C.~J. and {Lazio}, T.~J.~W. and {McLaughlin}, M.~A. and {Ransom}, S.~M. and {Salter}, C.~J. and {Scholz}, P. and {Seymour}, A. and {Siemion}, A. and {Spitler}, L.~G. and {Tendulkar}, S.~P. and {Wharton}, R.~S.},
        title = "{The Repeating Fast Radio Burst FRB 121102 as Seen on Milliarcsecond Angular Scales}",
      journal = {\apjl},
         year = 2017,
        month = jan,
       volume = {834},
       number = {2},
          eid = {L8},
        pages = {L8},
          doi = {10.3847/2041-8213/834/2/L8},
archivePrefix = {arXiv},
       eprint = {1701.01099},
 primaryClass = {astro-ph.HE},
       adsurl = {https://ui.adsabs.harvard.edu/abs/2017ApJ...834L...8M}
}

@ARTICLE{pastor-marazuela21,
       author = {{Pastor-Marazuela}, In{\'e}s and {Connor}, Liam and {van Leeuwen}, Joeri and {Maan}, Yogesh and {ter Veen}, Sander and {Bilous}, Anna and {Oostrum}, Leon and {Petroff}, Emily and {Straal}, Samayra and {Vohl}, Dany and {Attema}, Jisk and {Boersma}, Oliver M. and {Kooistra}, Eric and {van der Schuur}, Daniel and {Sclocco}, Alessio and {Smits}, Roy and {Adams}, Elizabeth A.~K. and {Adebahr}, Bj{\"o}rn and {de Blok}, W.~J.~G. and {Coolen}, Arthur H.~W.~M. and {Damstra}, Sieds and {D{\'e}nes}, Helga and {Hess}, Kelley M. and {van der Hulst}, Thijs and {Hut}, Boudewijn and {Ivashina}, V. Marianna and {Kutkin}, Alexander and {Loose}, G. Marcel and {Lucero}, Danielle M. and {Mika}, {\'A}gnes and {Moss}, Vanessa A. and {Mulder}, Henk and {Norden}, Menno J. and {Oosterloo}, Tom and {Orr{\'u}}, Emanuela and {Ruiter}, Mark and {Wijnholds}, Stefan J.},
        title = "{Chromatic periodic activity down to 120 megahertz in a fast radio burst}",
      journal = {\nat},
         year = 2021,
        month = aug,
       volume = {596},
       number = {7873},
        pages = {505-508},
          doi = {10.1038/s41586-021-03724-8},
archivePrefix = {arXiv},
       eprint = {2012.08348},
 primaryClass = {astro-ph.HE},
       adsurl = {https://ui.adsabs.harvard.edu/abs/2021Natur.596..505P}
}

@ARTICLE{pleunis21,
       author = {{Pleunis}, Z. and {Michilli}, D. and {Bassa}, C.~G. and {Hessels}, J.~W.~T. and {Naidu}, A. and {Andersen}, B.~C. and {Chawla}, P. and {Fonseca}, E. and {Gopinath}, A. and {Kaspi}, V.~M. and {Kondratiev}, V.~I. and {Li}, D.~Z. and {Bhardwaj}, M. and {Boyle}, P.~J. and {Brar}, C. and {Cassanelli}, T. and {Gupta}, Y. and {Josephy}, A. and {Karuppusamy}, R. and {Keimpema}, A. and {Kirsten}, F. and {Leung}, C. and {Marcote}, B. and {Masui}, K.~W. and {Mckinven}, R. and {Meyers}, B.~W. and {Ng}, C. and {Nimmo}, K. and {Paragi}, Z. and {Rahman}, M. and {Scholz}, P. and {Shin}, K. and {Smith}, K.~M. and {Stairs}, I.~H. and {Tendulkar}, S.~P.},
        title = "{LOFAR Detection of 110-188 MHz Emission and Frequency-dependent Activity from FRB 20180916B}",
      journal = {\apjl},
         year = 2021,
        month = apr,
       volume = {911},
       number = {1},
          eid = {L3},
        pages = {L3},
          doi = {10.3847/2041-8213/abec72},
archivePrefix = {arXiv},
       eprint = {2012.08372},
 primaryClass = {astro-ph.HE},
       adsurl = {https://ui.adsabs.harvard.edu/abs/2021ApJ...911L...3P}
}

@ARTICLE{schellart2013,
       author = {{Schellart}, P. and {Nelles}, A. and {Buitink}, S. and {Corstanje}, A. and {Enriquez}, J.~E. and {Falcke}, H. and {Frieswijk}, W. and {H{\"o}randel}, J.~R. and {Horneffer}, A. and {James}, C.~W. and {Krause}, M. and {Mevius}, M. and {Scholten}, O. and {ter Veen}, S. and {Thoudam}, S. and {van den Akker}, M. and {Alexov}, A. and {Anderson}, J. and {Avruch}, I.~M. and {B{\"a}hren}, L. and {Beck}, R. and {Bell}, M.~E. and {Bennema}, P. and {Bentum}, M.~J. and {Bernardi}, G. and {Best}, P. and {Bregman}, J. and {Breitling}, F. and {Brentjens}, M. and {Broderick}, J. and {Br{\"u}ggen}, M. and {Ciardi}, B. and {Coolen}, A. and {de Gasperin}, F. and {de Geus}, E. and {de Jong}, A. and {de Vos}, M. and {Duscha}, S. and {Eisl{\"o}ffel}, J. and {Fallows}, R.~A. and {Ferrari}, C. and {Garrett}, M.~A. and {Grie{\ss}meier}, J. and {Grit}, T. and {Hamaker}, J.~P. and {Hassall}, T.~E. and {Heald}, G. and {Hessels}, J.~W.~T. and {Hoeft}, M. and {Holties}, H.~A. and {Iacobelli}, M. and {Juette}, E. and {Karastergiou}, A. and {Klijn}, W. and {Kohler}, J. and {Kondratiev}, V.~I. and {Kramer}, M. and {Kuniyoshi}, M. and {Kuper}, G. and {Maat}, P. and {Macario}, G. and {Mann}, G. and {Markoff}, S. and {McKay-Bukowski}, D. and {McKean}, J.~P. and {Miller-Jones}, J.~C.~A. and {Mol}, J.~D. and {Mulcahy}, D.~D. and {Munk}, H. and {Nijboer}, R. and {Norden}, M.~J. and {Orru}, E. and {Overeem}, R. and {Paas}, H. and {Pandey-Pommier}, M. and {Pizzo}, R. and {Polatidis}, A.~G. and {Renting}, A. and {Romein}, J.~W. and {R{\"o}ttgering}, H. and {Schoenmakers}, A. and {Schwarz}, D. and {Sluman}, J. and {Smirnov}, O. and {Sobey}, C. and {Stappers}, B.~W. and {Steinmetz}, M. and {Swinbank}, J. and {Tang}, Y. and {Tasse}, C. and {Toribio}, C. and {van Leeuwen}, J. and {van Nieuwpoort}, R. and {van Weeren}, R.~J. and {Vermaas}, N. and {Vermeulen}, R. and {Vocks}, C. and {Vogt}, C. and {Wijers}, R.~A.~M.~J. and {Wijnholds}, S.~J. and {Wise}, M.~W. and {Wucknitz}, O. and {Yatawatta}, S. and {Zarka}, P. and {Zensus}, A.},
        title = "{Detecting cosmic rays with the LOFAR radio telescope}",
      journal = {\aap},
         year = 2013,
        month = dec,
       volume = {560},
          eid = {A98},
        pages = {A98},
          doi = {10.1051/0004-6361/201322683},
archivePrefix = {arXiv},
       eprint = {1311.1399},
 primaryClass = {astro-ph.IM},
       adsurl = {https://ui.adsabs.harvard.edu/abs/2013A&A...560A..98S}
}

@ARTICLE{MostPhilippov2020,
       author = {{Most}, Elias R. and {Philippov}, Alexander A.},
        title = "{Electromagnetic Precursors to Gravitational-wave Events: Numerical Simulations of Flaring in Pre-merger Binary Neutron Star Magnetospheres}",
      journal = {\apjl},
         year = 2020,
        month = apr,
       volume = {893},
       number = {1},
          eid = {L6},
        pages = {L6},
          doi = {10.3847/2041-8213/ab8196},
archivePrefix = {arXiv},
       eprint = {2001.06037},
 primaryClass = {astro-ph.HE},
       adsurl = {https://ui.adsabs.harvard.edu/abs/2020ApJ...893L...6M}
}

@ARTICLE{2022MostPhilippov,
       author = {{Most}, Elias R. and {Philippov}, Alexander A.},
        title = "{Electromagnetic precursor flares from the late inspiral of neutron star binaries}",
      journal = {\mnras},
         year = 2022,
        month = sep,
       volume = {515},
       number = {2},
        pages = {2710-2724},
          doi = {10.1093/mnras/stac1909},
archivePrefix = {arXiv},
       eprint = {2205.09643},
 primaryClass = {astro-ph.HE},
       adsurl = {https://ui.adsabs.harvard.edu/abs/2022MNRAS.515.2710M}
}

@ARTICLE{2012Piro,
       author = {{Piro}, Anthony L.},
        title = "{Magnetic Interactions in Coalescing Neutron Star Binaries}",
      journal = {\apj},
         year = 2012,
        month = aug,
       volume = {755},
       number = {1},
          eid = {80},
        pages = {80},
          doi = {10.1088/0004-637X/755/1/80},
archivePrefix = {arXiv},
       eprint = {1205.6482},
 primaryClass = {astro-ph.HE},
       adsurl = {https://ui.adsabs.harvard.edu/abs/2012ApJ...755...80P}
}

@ARTICLE{Wang2016,
       author = {{Wang}, Jie-Shuang and {Yang}, Yuan-Pei and {Wu}, Xue-Feng and {Dai}, Zi-Gao and {Wang}, Fa-Yin},
        title = "{Fast Radio Bursts from the Inspiral of Double Neutron Stars}",
      journal = {\apjl},
         year = 2016,
        month = may,
       volume = {822},
       number = {1},
          eid = {L7},
        pages = {L7},
          doi = {10.3847/2041-8205/822/1/L7},
archivePrefix = {arXiv},
       eprint = {1603.02014},
 primaryClass = {astro-ph.HE},
       adsurl = {https://ui.adsabs.harvard.edu/abs/2016ApJ...822L...7W}
}

@ARTICLE{Sridhar2021Wind,
       author = {{Sridhar}, Navin and {Zrake}, Jonathan and {Metzger}, Brian D. and {Sironi}, Lorenzo and {Giannios}, Dimitrios},
        title = "{Shock-powered radio precursors of neutron star mergers from accelerating relativistic binary winds}",
      journal = {\mnras},
         year = 2021,
        month = mar,
       volume = {501},
       number = {3},
        pages = {3184-3202},
          doi = {10.1093/mnras/staa3794},
archivePrefix = {arXiv},
       eprint = {2010.09214},
 primaryClass = {astro-ph.HE},
       adsurl = {https://ui.adsabs.harvard.edu/abs/2021MNRAS.501.3184S}
}

@ARTICLE{2023MostPhilippov,
       author = {{Most}, Elias R. and {Philippov}, Alexander A.},
        title = "{Reconnection-Powered Fast Radio Transients from Coalescing Neutron Star Binaries}",
      journal = {\prl},
         year = 2023,
        month = jun,
       volume = {130},
       number = {24},
          eid = {245201},
        pages = {245201},
          doi = {10.1103/PhysRevLett.130.245201},
archivePrefix = {arXiv},
       eprint = {2207.14435},
 primaryClass = {astro-ph.HE},
       adsurl = {https://ui.adsabs.harvard.edu/abs/2023PhRvL.130x5201M}
}

@ARTICLE{Oberoi2023,
       author = {{Oberoi}, Divya and {Bisoi}, Susanta Kumar and {Sasikumar Raja}, K. and {Kansabanik}, Devojyoti and {Mohan}, Atul and {Mondal}, Surajit and {Sharma}, Rohit},
        title = "{Preparing for solar and heliospheric science with the SKAO: An Indian perspective}",
      journal = {Journal of Astrophysics and Astronomy},
         year = 2023,
        month = jun,
       volume = {44},
       number = {1},
          eid = {40},
        pages = {40},
          doi = {10.1007/s12036-023-09917-z},
archivePrefix = {arXiv},
       eprint = {2211.03791},
 primaryClass = {astro-ph.IM},
       adsurl = {https://ui.adsabs.harvard.edu/abs/2023JApA...44...40O}
}

@ARTICLE{aartsen17,
       author = {{Aartsen}, M.~G. and {Ackermann}, M. and {Adams}, J. and {Aguilar}, J.~A. and {Ahlers}, M. and {Ahrens}, M. and {Altmann}, D. and {Andeen}, K. and {Anderson}, T. and {Ansseau}, I. and {Anton}, G. and {Archinger}, M. and {Arg{\"u}elles}, C. and {Auer}, R. and {Auffenberg}, J. and {Axani}, S. and {Baccus}, J. and {Bai}, X. and {Barnet}, S. and {Barwick}, S.~W. and {Baum}, V. and {Bay}, R. and {Beattie}, K. and {Beatty}, J.~J. and {Becker Tjus}, J. and {Becker}, K.-H. and {Bendfelt}, T. and {BenZvi}, S. and {Berley}, D. and {Bernardini}, E. and {Bernhard}, A. and {Besson}, D.~Z. and {Binder}, G. and {Bindig}, D. and {Bissok}, M. and {Blaufuss}, E. and {Blot}, S. and {Boersma}, D. and {Bohm}, C. and {B{\"o}rner}, M. and {Bos}, F. and {Bose}, D. and {B{\"o}ser}, S. and {Botner}, O. and {Bouchta}, A. and {Braun}, J. and {Brayeur}, L. and {Bretz}, H.-P. and {Bron}, S. and {Burgman}, A. and {Burreson}, C. and {Carver}, T. and {Casier}, M. and {Cheung}, E. and {Chirkin}, D. and {Christov}, A. and {Clark}, K. and {Classen}, L. and {Coenders}, S. and {Collin}, G.~H. and {Conrad}, J.~M. and {Cowen}, D.~F. and {Cross}, R. and {Day}, C. and {Day}, M. and {de Andr{\'e}}, J.~P.~A.~M. and {De Clercq}, C. and {del Pino Rosendo}, E. and {Dembinski}, H. and {De Ridder}, S. and {Descamps}, F. and {Desiati}, P. and {de Vries}, K.~D. and {de Wasseige}, G. and {de With}, M. and {DeYoung}, T. and {D{\'\i}az-V{\'e}lez}, J.~C. and {di Lorenzo}, V. and {Dujmovic}, H. and {Dumm}, J.~P. and {Dunkman}, M. and {Eberhardt}, B. and {Edwards}, W.~R. and {Ehrhardt}, T. and {Eichmann}, B. and {Eller}, P. and {Euler}, S. and {Evenson}, P.~A. and {Fahey}, S. and {Fazely}, A.~R. and {Feintzeig}, J. and {Felde}, J. and {Filimonov}, K. and {Finley}, C. and {Flis}, S. and {F{\"o}sig}, C.-C. and {Franckowiak}, A. and {Fr{\`e}re}, M. and {Friedman}, E. and {Fuchs}, T. and {Gaisser}, T.~K. and {Gallagher}, J. and {Gerhardt}, L. and {Ghorbani}, K. and {Giang}, W. and {Gladstone}, L. and {Glauch}, T. and {Glowacki}, D. and {Gl{\"u}senkamp}, T. and {Goldschmidt}, A. and {Gonzalez}, J.~G. and {Grant}, D. and {Griffith}, Z. and {Gustafsson}, L. and {Haack}, C. and {Hallgren}, A. and {Halzen}, F. and {Hansen}, E. and {Hansmann}, T. and {Hanson}, K. and {Haugen}, J. and {Hebecker}, D. and {Heereman}, D. and {Helbing}, K. and {Hellauer}, R. and {Heller}, R. and {Hickford}, S. and {Hignight}, J. and {Hill}, G.~C. and {Hoffman}, K.~D. and {Hoffmann}, R. and {Hoshina}, K. and {Huang}, F. and {Huber}, M. and {Hulth}, P.~O. and {Hultqvist}, K. and {In}, S. and {Inaba}, M. and {Ishihara}, A. and {Jacobi}, E. and {Jacobsen}, J. and {Japaridze}, G.~S. and {Jeong}, M. and {Jero}, K. and {Jones}, A. and {Jones}, B.~J.~P. and {Joseph}, J. and {Kang}, W. and {Kappes}, A. and {Karg}, T. and {Karle}, A. and {Katz}, U. and {Kauer}, M. and {Keivani}, A. and {Kelley}, J.~L. and {Kemp}, J. and {Kheirandish}, A. and {Kim}, J. and {Kim}, M. and {Kintscher}, T. and {Kiryluk}, J. and {Kitamura}, N. and {Kittler}, T. and {Klein}, S.~R. and {Kleinfelder}, S. and {Kleist}, M. and {Kohnen}, G. and {Koirala}, R. and {Kolanoski}, H. and {Konietz}, R. and {K{\"o}pke}, L. and {Kopper}, C. and {Kopper}, S. and {Koskinen}, D.~J. and {Kowalski}, M. and {Krasberg}, M. and {Krings}, K. and {Kroll}, M. and {Kr{\"u}ckl}, G. and {Kr{\"u}ger}, C. and {Kunnen}, J. and {Kunwar}, S. and {Kurahashi}, N. and {Kuwabara}, T. and {Labare}, M. and {Laihem}, K. and {Landsman}, H. and {Lanfranchi}, J.~L. and {Larson}, M.~J. and {Lauber}, F. and {Laundrie}, A. and {Lennarz}, D. and {Leich}, H. and {Lesiak-Bzdak}, M. and {Leuermann}, M. and {Lu}, L. and {Ludwig}, J. and {L{\"u}nemann}, J. and {Mackenzie}, C. and {Madsen}, J.},
        title = "{The IceCube Neutrino Observatory: instrumentation and online systems}",
      journal = {Journal of Instrumentation},
         year = 2017,
        month = mar,
       volume = {12},
       number = {3},
        pages = {P03012},
          doi = {10.1088/1748-0221/12/03/P03012},
archivePrefix = {arXiv},
       eprint = {1612.05093},
 primaryClass = {astro-ph.IM},
       adsurl = {https://ui.adsabs.harvard.edu/abs/2017JInst..12P3012A}
}

@ARTICLE{abbott16,
       author = {{Abbott}, B.~P. and {Abbott}, R. and {Abbott}, T.~D. and {Abernathy}, M.~R. and {Acernese}, F. and {Ackley}, K. and {Adams}, C. and {Adams}, T. and {Addesso}, P. and {Adhikari}, R.~X. and {Adya}, V.~B. and {Affeldt}, C. and {Agathos}, M. and {Agatsuma}, K. and {Aggarwal}, N. and {Aguiar}, O.~D. and {Aiello}, L. and {Ain}, A. and {Ajith}, P. and {Allen}, B. and {Allocca}, A. and {Altin}, P.~A. and {Anderson}, S.~B. and {Anderson}, W.~G. and {Arai}, K. and {Araya}, M.~C. and {Arceneaux}, C.~C. and {Areeda}, J.~S. and {Arnaud}, N. and {Arun}, K.~G. and {Ascenzi}, S. and {Ashton}, G. and {Ast}, M. and {Aston}, S.~M. and {Astone}, P. and {Aufmuth}, P. and {Aulbert}, C. and {Babak}, S. and {Bacon}, P. and {Bader}, M.~K.~M. and {Baker}, P.~T. and {Baldaccini}, F. and {Ballardin}, G. and {Ballmer}, S.~W. and {Barayoga}, J.~C. and {Barclay}, S.~E. and {Barish}, B.~C. and {Barker}, D. and {Barone}, F. and {Barr}, B. and {Barsotti}, L. and {Barsuglia}, M. and {Barta}, D. and {Barthelmy}, S. and {Bartlett}, J. and {Bartos}, I. and {Bassiri}, R. and {Basti}, A. and {Batch}, J.~C. and {Baune}, C. and {Bavigadda}, V. and {Bazzan}, M. and {Behnke}, B. and {Bejger}, M. and {Bell}, A.~S. and {Bell}, C.~J. and {Berger}, B.~K. and {Bergman}, J. and {Bergmann}, G. and {Berry}, C.~P.~L. and {Bersanetti}, D. and {Bertolini}, A. and {Betzwieser}, J. and {Bhagwat}, S. and {Bhandare}, R. and {Bilenko}, I.~A. and {Billingsley}, G. and {Birch}, J. and {Birney}, R. and {Biscans}, S. and {Bisht}, A. and {Bitossi}, M. and {Biwer}, C. and {Bizouard}, M.~A. and {Blackburn}, J.~K. and {Blair}, C.~D. and {Blair}, D.~G. and {Blair}, R.~M. and {Bloemen}, S. and {Bock}, O. and {Bodiya}, T.~P. and {Boer}, M. and {Bogaert}, G. and {Bogan}, C. and {Bohe}, A. and {Bojtos}, P. and {Bond}, C. and {Bondu}, F. and {Bonnand}, R. and {Boom}, B.~A. and {Bork}, R. and {Boschi}, V. and {Bose}, S. and {Bouffanais}, Y. and {Bozzi}, A. and {Bradaschia}, C. and {Brady}, P.~R. and {Braginsky}, V.~B. and {Branchesi}, M. and {Brau}, J.~E. and {Briant}, T. and {Brillet}, A. and {Brinkmann}, M. and {Brisson}, V. and {Brockill}, P. and {Brooks}, A.~F. and {Brown}, D.~A. and {Brown}, D.~D. and {Brown}, N.~M. and {Buchanan}, C.~C. and {Buikema}, A. and {Bulik}, T. and {Bulten}, H.~J. and {Buonanno}, A. and {Buskulic}, D. and {Buy}, C. and {Byer}, R.~L. and {Cadonati}, L. and {Cagnoli}, G. and {Cahillane}, C. and {Bustillo}, J.~C. and {Callister}, T. and {Calloni}, E. and {Camp}, J.~B. and {Cannon}, K.~C. and {Cao}, J. and {Capano}, C.~D. and {Capocasa}, E. and {Carbognani}, F. and {Caride}, S. and {Diaz}, J.~C. and {Casentini}, C. and {Caudill}, S. and {Cavagli{\'a}}, M. and {Cavalier}, F. and {Cavalieri}, R. and {Cella}, G. and {Cepeda}, C.~B. and {Baiardi}, L.~C. and {Cerretani}, G. and {Cesarini}, E. and {Chakraborty}, R. and {Chalermsongsak}, T. and {Chamberlin}, S.~J. and {Chan}, M. and {Chao}, S. and {Charlton}, P. and {Chassande-Mottin}, E. and {Chen}, H.~Y. and {Chen}, Y. and {Cheng}, C. and {Chincarini}, A. and {Chiummo}, A. and {Cho}, H.~S. and {Cho}, M. and {Chow}, J.~H. and {Christensen}, N. and {Chu}, Q. and {Chua}, S. and {Chung}, S. and {Ciani}, G. and {Clara}, F. and {Clark}, J.~A. and {Cleva}, F. and {Coccia}, E. and {Cohadon}, P.-F. and {Colla}, A. and {Collette}, C.~G. and {Cominsky}, L. and {Constancio}, Jr., M. and {Conte}, A. and {Conti}, L. and {Cook}, D. and {Corbitt}, T.~R. and {Cornish}, N. and {Corsi}, A. and {Cortese}, S. and {Costa}, C.~A. and {Coughlin}, M.~W. and {Coughlin}, S.~B. and {Coulon}, J.-P. and {Countryman}, S.~T. and {Couvares}, P. and {Cowan}, E.~E. and {Coward}, D.~M. and {Cowart}, M.~J. and {Coyne}, D.~C. and {Coyne}, R. and {Craig}, K. and {Creighton}, J.~D.~E.},
        title = "{Localization and Broadband Follow-up of the Gravitational-wave Transient GW150914}",
      journal = {\apjl},
         year = 2016,
        month = jul,
       volume = {826},
       number = {1},
          eid = {L13},
        pages = {L13},
          doi = {10.3847/2041-8205/826/1/L13},
archivePrefix = {arXiv},
       eprint = {1602.08492},
 primaryClass = {astro-ph.HE},
       adsurl = {https://ui.adsabs.harvard.edu/abs/2016ApJ...826L..13A}
}

@ARTICLE{abbott17a,
       author = {{Abbott}, B.~P. and {LIGO Scientific Collaboration} and {Virgo Collaboration}},
        title = "{GW170817: Observation of Gravitational Waves from a Binary Neutron Star Inspiral}",
      journal = {\prl},
         year = 2017,
        month = oct,
       volume = {119},
       number = {16},
          eid = {161101},
        pages = {161101},
          doi = {10.1103/PhysRevLett.119.161101},
archivePrefix = {arXiv},
       eprint = {1710.05832},
 primaryClass = {gr-qc},
       adsurl = {https://ui.adsabs.harvard.edu/abs/2017PhRvL.119p1101A}
}

@ARTICLE{abbott17b,
       author = {{Abbott}, B.~P. and {Abbott}, R. and {Abbott}, T.~D. and {Acernese}, F. and {Ackley}, K. and {Adams}, C. and {Adams}, T. and {Addesso}, P. and {Adhikari}, R.~X. and {Adya}, V.~B. and {Affeldt}, C. and {Afrough}, M. and {Agarwal}, B. and {Agathos}, M. and {Agatsuma}, K. and {Aggarwal}, N. and {Aguiar}, O.~D. and {Aiello}, L. and {Ain}, A. and {Ajith}, P. and {Allen}, B. and {Allen}, G. and {Allocca}, A. and {Aloy}, M.~A. and {Altin}, P.~A. and {Amato}, A. and {Ananyeva}, A. and {Anderson}, S.~B. and {Anderson}, W.~G. and {Angelova}, S.~V. and {Antier}, S.},
        title = "{Gravitational Waves and Gamma-Rays from a Binary Neutron Star Merger: GW170817 and GRB 170817A}",
      journal = {\apjl},
         year = 2017,
        month = oct,
       volume = {848},
       number = {2},
          eid = {L13},
        pages = {L13},
          doi = {10.3847/2041-8213/aa920c},
archivePrefix = {arXiv},
       eprint = {1710.05834},
 primaryClass = {astro-ph.HE},
       adsurl = {https://ui.adsabs.harvard.edu/abs/2017ApJ...848L..13A}
}

@ARTICLE{abbott17c,
       author = {{Abbott}, B.~P. and {Abbott}, R. and {Abbott}, T.~D. and {Acernese}, F. and {Ackley}, K. and {Adams}, C. and {Adams}, T. and {Addesso}, P. and {Adhikari}, R.~X. and {Adya}, V.~B. and {Affeldt}, C. and {Afrough}, M. and {Agarwal}, B. and {Agathos}, M. and {Agatsuma}, K. and {Aggarwal}, N. and {Aguiar}, O.~D. and {Aiello}, L. and {Ain}, A. and {Ajith}, P. and {Allen}, B. and {Allen}, G. and {Allocca}, A. and {Altin}, P.~A. and {Amato}, A. and {Ananyeva}, A. and {Anderson}, S.~B. and {Anderson}, W.~G. and {Angelova}, S.~V. and {Antier}, S. and {Appert}, S. and {Arai}, K. and {Araya}, M.~C. and {Areeda}, J.~S. and {Arnaud}, N. and {Arun}, K.~G. and {Ascenzi}, S. and {Ashton}, G. and {Ast}, M. and {Aston}, S.~M. and {Astone}, P. and {Atallah}, D.~V. and {Aufmuth}, P. and {Aulbert}, C. and {AultONeal}, K. and {Austin}, C. and {Avila-Alvarez}, A. and {Babak}, S. and {Bacon}, P. and {Bader}, M.~K.~M. and {Bae}, S. and {Baker}, P.~T. and {Baldaccini}, F. and {Ballardin}, G. and {Ballmer}, S.~W. and {Banagiri}, S. and {Barayoga}, J.~C. and {Barclay}, S.~E. and {Barish}, B.~C. and {Barker}, D. and {Barkett}, K. and {Barone}, F. and {Barr}, B. and {Barsotti}, L. and {Barsuglia}, M. and {Barta}, D. and {Barthelmy}, S.~D. and {Bartlett}, J. and {Bartos}, I. and {Bassiri}, R. and {Basti}, A. and {Batch}, J.~C. and {Bawaj}, M. and {Bayley}, J.~C. and {Bazzan}, M. and {B{\'e}csy}, B. and {Beer}, C. and {Bejger}, M. and {Belahcene}, I. and {Bell}, A.~S. and {Berger}, B.~K. and {Bergmann}, G. and {Bero}, J.~J. and {Berry}, C.~P.~L. and {Bersanetti}, D. and {Bertolini}, A. and {Betzwieser}, J. and {Bhagwat}, S. and {Bhandare}, R. and {Bilenko}, I.~A. and {Billingsley}, G. and {Billman}, C.~R. and {Birch}, J. and {Birney}, R. and {Birnholtz}, O. and {Biscans}, S. and {Biscoveanu}, S. and {Bisht}, A. and {Bitossi}, M. and {Biwer}, C. and {Bizouard}, M.~A. and {Blackburn}, J.~K. and {Blackman}, J. and {Blair}, C.~D. and {Blair}, D.~G. and {Blair}, R.~M. and {Bloemen}, S. and {Bock}, O. and {Bode}, N. and {Boer}, M. and {Bogaert}, G. and {Bohe}, A. and {Bondu}, F. and {Bonilla}, E. and {Bonnand}, R. and {Boom}, B.~A. and {Bork}, R. and {Boschi}, V. and {Bose}, S. and {Bossie}, K. and {Bouffanais}, Y. and {Bozzi}, A. and {Bradaschia}, C. and {Brady}, P.~R. and {Branchesi}, M. and {Brau}, J.~E. and {Briant}, T. and {Brillet}, A. and {Brinkmann}, M. and {Brisson}, V. and {Brockill}, P. and {Broida}, J.~E. and {Brooks}, A.~F. and {Brown}, D.~A. and {Brown}, D.~D. and {Brunett}, S. and {Buchanan}, C.~C. and {Buikema}, A. and {Bulik}, T. and {Bulten}, H.~J. and {Buonanno}, A. and {Buskulic}, D. and {Buy}, C. and {Byer}, R.~L. and {Cabero}, M. and {Cadonati}, L. and {Cagnoli}, G. and {Cahillane}, C. and {Calder{\'o}n Bustillo}, J. and {Callister}, T.~A. and {Calloni}, E. and {Camp}, J.~B. and {Canepa}, M. and {Canizares}, P. and {Cannon}, K.~C. and {Cao}, H. and {Cao}, J. and {Capano}, C.~D. and {Capocasa}, E. and {Carbognani}, F. and {Caride}, S. and {Carney}, M.~F. and {Casanueva Diaz}, J. and {Casentini}, C. and {Caudill}, S. and {Cavagli{\`a}}, M. and {Cavalier}, F. and {Cavalieri}, R. and {Cella}, G. and {Cepeda}, C.~B. and {Cerd{\'a}-Dur{\'a}n}, P. and {Cerretani}, G. and {Cesarini}, E. and {Chamberlin}, S.~J. and {Chan}, M. and {Chao}, S. and {Charlton}, P. and {Chase}, E. and {Chassande-Mottin}, E. and {Chatterjee}, D. and {Chatziioannou}, K. and {Cheeseboro}, B.~D. and {Chen}, H.~Y. and {Chen}, X. and {Chen}, Y. and {Cheng}, H.-P. and {Chia}, H. and {Chincarini}, A. and {Chiummo}, A. and {Chmiel}, T. and {Cho}, H.~S. and {Cho}, M. and {Chow}, J.~H. and {Christensen}, N. and {Chu}, Q. and {Chua}, A.~J.~K. and {Chua}, S. and {Chung}, A.~K.~W. and {Chung}, S. and {Ciani}, G.},
        title = "{Multi-messenger Observations of a Binary Neutron Star Merger}",
      journal = {\apjl},
         year = 2017,
        month = oct,
       volume = {848},
       number = {2},
          eid = {L12},
        pages = {L12},
          doi = {10.3847/2041-8213/aa91c9},
archivePrefix = {arXiv},
       eprint = {1710.05833},
 primaryClass = {astro-ph.HE},
       adsurl = {https://ui.adsabs.harvard.edu/abs/2017ApJ...848L..12A}
}

@ARTICLE{abbott20,
       author = {{Abbott}, B.~P. and {Abbott}, R. and {Abbott}, T.~D. and {Abraham}, S. and {Acernese}, F. and {Ackley}, K. and {Adams}, C. and {Adya}, V.~B. and {Affeldt}, C. and {Agathos}, M. and {Agatsuma}, K. and {Aggarwal}, N. and {Aguiar}, O.~D. and {Aiello}, L. and {Ain}, A. and {Ajith}, P. and {Akutsu}, T. and {Allen}, G. and {Allocca}, A. and {Aloy}, M.~A. and {Altin}, P.~A. and {Amato}, A. and {Ananyeva}, A. and {Anderson}, S.~B. and {Anderson}, W.~G. and {Ando}, M. and {Angelova}, S.~V. and {Antier}, S. and {Appert}, S. and {Arai}, K. and {Arai}, Koya and {Arai}, Y. and {Araki}, S. and {Araya}, A. and {Araya}, M.~C. and {Areeda}, J.~S. and {Ar{\`e}ne}, M. and {Aritomi}, N. and {Arnaud}, N. and {Arun}, K.~G. and {Ascenzi}, S. and {Ashton}, G. and {Aso}, Y. and {Aston}, S.~M. and {Astone}, P. and {Aubin}, F. and {Aufmuth}, P. and {Aultoneal}, K. and {Austin}, C. and {Avendano}, V. and {Avila-Alvarez}, A. and {Babak}, S. and {Bacon}, P. and {Badaracco}, F. and {Bader}, M.~K.~M. and {Bae}, S.~W. and {Bae}, Y.~B. and {Baiotti}, L. and {Bajpai}, R. and {Baker}, P.~T. and {Baldaccini}, F. and {Ballardin}, G. and {Ballmer}, S.~W. and {Banagiri}, S. and {Barayoga}, J.~C. and {Barclay}, S.~E. and {Barish}, B.~C. and {Barker}, D. and {Barkett}, K. and {Barnum}, S. and {Barone}, F. and {Barr}, B. and {Barsotti}, L. and {Barsuglia}, M. and {Barta}, D. and {Bartlett}, J. and {Barton}, M.~A. and {Bartos}, I. and {Bassiri}, R. and {Basti}, A. and {Bawaj}, M. and {Bayley}, J.~C. and {Bazzan}, M. and {B{\'e}csy}, B. and {Bejger}, M. and {Belahcene}, I. and {Bell}, A.~S. and {Beniwal}, D. and {Berger}, B.~K. and {Bergmann}, G. and {Bernuzzi}, S. and {Bero}, J.~J. and {Berry}, C.~P.~L. and {Bersanetti}, D. and {Bertolini}, A. and {Betzwieser}, J. and {Bhandare}, R. and {Bidler}, J. and {Bilenko}, I.~A. and {Bilgili}, S.~A. and {Billingsley}, G. and {Birch}, J. and {Birney}, R. and {Birnholtz}, O. and {Biscans}, S. and {Biscoveanu}, S. and {Bisht}, A. and {Bitossi}, M. and {Bizouard}, M.~A. and {Blackburn}, J.~K. and {Blair}, C.~D. and {Blair}, D.~G. and {Blair}, R.~M. and {Bloemen}, S. and {Bode}, N. and {Boer}, M. and {Boetzel}, Y. and {Bogaert}, G. and {Bondu}, F. and {Bonilla}, E. and {Bonnand}, R. and {Booker}, P. and {Boom}, B.~A. and {Booth}, C.~D. and {Bork}, R. and {Boschi}, V. and {Bose}, S. and {Bossie}, K. and {Bossilkov}, V. and {Bosveld}, J. and {Bouffanais}, Y. and {Bozzi}, A. and {Bradaschia}, C. and {Brady}, P.~R. and {Bramley}, A. and {Branchesi}, M. and {Brau}, J.~E. and {Briant}, T. and {Briggs}, J.~H. and {Brighenti}, F. and {Brillet}, A. and {Brinkmann}, M. and {Brisson}, V. and {Brockill}, P. and {Brooks}, A.~F. and {Brown}, D.~A. and {Brown}, D.~D. and {Brunett}, S. and {Buikema}, A. and {Bulik}, T. and {Bulten}, H.~J. and {Buonanno}, A. and {Buskulic}, D. and {Buy}, C. and {Byer}, R.~L. and {Cabero}, M. and {Cadonati}, L. and {Cagnoli}, G. and {Cahillane}, C. and {Bustillo}, J. Calder{\'o}n and {Callister}, T.~A. and {Calloni}, E. and {Camp}, J.~B. and {Campbell}, W.~A. and {Canepa}, M. and {Cannon}, K. and {Cannon}, K.~C. and {Cao}, H. and {Cao}, J. and {Capocasa}, E. and {Carbognani}, F. and {Caride}, S. and {Carney}, M.~F. and {Carullo}, G. and {Diaz}, J. Casanueva and {Casentini}, C. and {Caudill}, S. and {Cavagli{\`a}}, M. and {Cavalier}, F. and {Cavalieri}, R. and {Cella}, G. and {Cerd{\'a}-Dur{\'a}n}, P. and {Cerretani}, G. and {Cesarini}, E. and {Chaibi}, O. and {Chakravarti}, K. and {Chamberlin}, S.~J. and {Chan}, M. and {Chan}, M.~L. and {Chao}, S. and {Charlton}, P. and {Chase}, E.~A. and {Chassande-Mottin}, E. and {Chatterjee}, D. and {Chaturvedi}, M. and {Chatziioannou}, K. and {Cheeseboro}, B.~D. and {Chen}, C.~S. and {Chen}, H.~Y. and {Chen}, K.~H.},
        title = "{Prospects for observing and localizing gravitational-wave transients with Advanced LIGO, Advanced Virgo and KAGRA}",
      journal = {Living Reviews in Relativity},
         year = 2020,
        month = dec,
       volume = {23},
       number = {1},
          eid = {3},
        pages = {3},
          doi = {10.1007/s41114-020-00026-9},
       adsurl = {https://ui.adsabs.harvard.edu/abs/2020LRR....23....3A}
}

@ARTICLE{abbott23,
       author = {{Abbott}, R. and {Abbott}, T.~D. and {Acernese}, F. and {Ackley}, K. and {Adams}, C. and {Adhikari}, N. and {Adhikari}, R.~X. and {Adya}, V.~B. and {Affeldt}, C. and {Agarwal}, D. and {Agathos}, M. and {Agatsuma}, K. and {Aggarwal}, N. and {Aguiar}, O.~D. and {Aiello}, L. and {Ain}, A. and {Ajith}, P. and {Akutsu}, T. and {Albanesi}, S. and {Allocca}, A. and {Altin}, P.~A. and {Amato}, A. and {Anand}, C. and {Anand}, S. and {Ananyeva}, A. and {Anderson}, S.~B. and {Anderson}, W.~G. and {Ando}, M. and {Andrade}, T. and {Andres}, N. and {Andri{\'c}}, T. and {Angelova}, S.~V. and {Ansoldi}, S. and {Antelis}, J.~M. and {Antier}, S. and {Appert}, S. and {Arai}, Koji and {Arai}, Koya and {Arai}, Y. and {Araki}, S. and {Araya}, A. and {Araya}, M.~C. and {Areeda}, J.~S. and {Ar{\`e}ne}, M. and {Aritomi}, N. and {Arnaud}, N. and {Aronson}, S.~M. and {Arun}, K.~G. and {Asada}, H. and {Asali}, Y. and {Ashton}, G. and {Aso}, Y. and {Assiduo}, M. and {Aston}, S.~M. and {Astone}, P. and {Aubin}, F. and {Austin}, C. and {Babak}, S. and {Badaracco}, F. and {Bader}, M.~K.~M. and {Badger}, C. and {Bae}, S. and {Bae}, Y. and {Baer}, A.~M. and {Bagnasco}, S. and {Bai}, Y. and {Baiotti}, L. and {Baird}, J. and {Bajpai}, R. and {Ball}, M. and {Ballardin}, G. and {Ballmer}, S.~W. and {Balsamo}, A. and {Baltus}, G. and {Banagiri}, S. and {Bankar}, D. and {Barayoga}, J.~C. and {Barbieri}, C. and {Barish}, B.~C. and {Barker}, D. and {Barneo}, P. and {Barone}, F. and {Barr}, B. and {Barsotti}, L. and {Barsuglia}, M. and {Barta}, D. and {Bartlett}, J. and {Barton}, M.~A. and {Bartos}, I. and {Bassiri}, R. and {Basti}, A. and {Bawaj}, M. and {Bayley}, J.~C. and {Baylor}, A.~C. and {Bazzan}, M. and {B{\'e}csy}, B. and {Bedakihale}, V.~M. and {Bejger}, M. and {Belahcene}, I. and {Benedetto}, V. and {Beniwal}, D. and {Bennett}, T.~F. and {Bentley}, J.~D. and {Benyaala}, M. and {Bergamin}, F. and {Berger}, B.~K. and {Bernuzzi}, S. and {Berry}, C.~P.~L. and {Bersanetti}, D. and {Bertolini}, A. and {Betzwieser}, J. and {Beveridge}, D. and {Bhandare}, R. and {Bhardwaj}, U. and {Bhattacharjee}, D. and {Bhaumik}, S. and {Bilenko}, I.~A. and {Billingsley}, G. and {Bini}, S. and {Birney}, I.~A. and {Birnholtz}, O. and {Biscans}, S. and {Bischi}, M. and {Biscoveanu}, S. and {Bisht}, A. and {Biswas}, B. and {Bitossi}, M. and {Bizouard}, M.-A. and {Blackburn}, J.~K. and {Blair}, C.~D. and {Blair}, D.~G. and {Blair}, R.~M. and {Bobba}, F. and {Bode}, N. and {Boer}, M. and {Bogaert}, G. and {Boldrini}, M. and {Bonavena}, L.~D. and {Bondu}, F. and {Bonilla}, E. and {Bonnand}, R. and {Booker}, P. and {Boom}, B.~A. and {Bork}, R. and {Boschi}, V. and {Bose}, N. and {Bose}, S. and {Bossilkov}, V. and {Boudart}, V. and {Bouffanais}, Y. and {Boumerdassi}, A. and {Bozzi}, A. and {Bradaschia}, C. and {Brady}, P.~R. and {Bramley}, A. and {Branch}, A. and {Branchesi}, M. and {Brau}, J.~E. and {Breschi}, M. and {Briant}, T. and {Briggs}, J.~H. and {Brillet}, A. and {Brinkmann}, M. and {Brockill}, P. and {Brooks}, A.~F. and {Brooks}, J. and {Brown}, D.~D. and {Brunett}, S. and {Bruno}, G. and {Bruntz}, R. and {Bryant}, J. and {Buchanan}, J. and {Bulik}, T. and {Bulten}, H.~J. and {Buonanno}, A. and {Buscicchio}, R. and {Buskulic}, D. and {Buy}, C. and {Byer}, R.~L. and {Cadonati}, L. and {Cagnoli}, G. and {Cahillane}, C. and {Bustillo}, J. Calder{\'o}n and {Callaghan}, J.~D. and {Callister}, T.~A. and {Calloni}, E. and {Cameron}, J. and {Camp}, J.~B. and {Canepa}, M. and {Canevarolo}, S. and {Cannavacciuolo}, M. and {Cannon}, K.~C. and {Cao}, H. and {Cao}, Z. and {Capocasa}, E. and {Capote}, E. and {Carapella}, G. and {Carbognani}, F. and {Carlin}, J.~B. and {Carney}, M.~F.},
        title = "{Search for Gravitational Waves Associated with Fast Radio Bursts Detected by CHIME/FRB during the LIGO{\textendash}Virgo Observing Run O3a}",
      journal = {\apj},
         year = 2023,
        month = oct,
       volume = {955},
       number = {2},
          eid = {155},
        pages = {155},
          doi = {10.3847/1538-4357/acd770},
archivePrefix = {arXiv},
       eprint = {2203.12038},
 primaryClass = {astro-ph.HE},
       adsurl = {https://ui.adsabs.harvard.edu/abs/2023ApJ...955..155A}
}

@ARTICLE{acciari2022,
       author = {{Acciari}, V.~A. and {Ansoldi}, S. and {Antonelli}, L.~A. and {Arbet Engels}, A. and {Artero}, M. and {Asano}, K. and {Baack}, D. and {Babi{\'c}}, A. and {Baquero}, A. and {Barres de Almeida}, U. and {Barrio}, J.~A. and {Batkovi{\'c}}, I. and {Becerra Gonz{\'a}lez}, J. and {Bednarek}, W. and {Bellizzi}, L. and {Bernardini}, E. and {Bernardos}, M. and {Berti}, A. and {Besenrieder}, J. and {Bhattacharyya}, W. and {Bigongiari}, C. and {Biland}, A. and {Blanch}, O. and {B{\"o}kenkamp}, H. and {Bonnoli}, G. and {Bo{\v{s}}njak}, {\v{Z}}. and {Busetto}, G. and {Carosi}, R. and {Ceribella}, G. and {Cerruti}, M. and {Chai}, Y. and {Chilingarian}, A. and {Cikota}, S. and {Colak}, S.~M. and {Colombo}, E. and {Contreras}, J.~L. and {Cortina}, J. and {Covino}, S. and {D'Amico}, G. and {D'Elia}, V. and {Da Vela}, P. and {Dazzi}, F. and {De Angelis}, A. and {De Lotto}, B. and {Del Popolo}, A. and {Delfino}, M. and {Delgado}, J. and {Delgado Mendez}, C. and {Depaoli}, D. and {Di Pierro}, F. and {Di Venere}, L. and {Do Souto Espi{\~n}eira}, E. and {Prester}, D. Dominis and {Donini}, A. and {Dorner}, D. and {Doro}, M. and {Elsaesser}, D. and {Fallah Ramazani}, V. and {Fari{\~n}a Alonso}, L. and {Fattorini}, A. and {Fonseca}, M.~V. and {Font}, L. and {Fruck}, C. and {Fukami}, S. and {Fukazawa}, Y. and {Garc{\'\i}a L{\'o}pez}, R.~J. and {Garczarczyk}, M. and {Gasparyan}, S. and {Gaug}, M. and {Giglietto}, N. and {Giordano}, F. and {Gliwny}, P. and {Godinovi{\'c}}, N. and {Green}, J.~G. and {Green}, D. and {Hadasch}, D. and {Hahn}, A. and {Hassan}, T. and {Heckmann}, L. and {Herrera}, J. and {Hoang}, J. and {Hrupec}, D. and {H{\"u}tten}, M. and {Inada}, T. and {Ishio}, K. and {Iwamura}, Y. and {Jim{\'e}nez Mart{\'\i}nez}, I. and {Jormanainen}, J. and {Jouvin}, L. and {Kerszberg}, D. and {Kobayashi}, Y. and {Kubo}, H. and {Kushida}, J. and {Lamastra}, A. and {Lelas}, D. and {Leone}, F. and {Lindfors}, E. and {Linhoff}, L. and {Lombardi}, S. and {Longo}, F. and {L{\'o}pez-Coto}, R. and {L{\'o}pez-Moya}, M. and {L{\'o}pez-Oramas}, A. and {Loporchio}, S. and {Machado de Oliveira Fraga}, B. and {Maggio}, C. and {Majumdar}, P. and {Makariev}, M. and {Mallamaci}, M. and {Maneva}, G. and {Manganaro}, M. and {Mannheim}, K. and {Maraschi}, L. and {Mariotti}, M. and {Mart{\'\i}nez}, M. and {Mas Aguilar}, A. and {Mazin}, D. and {Menchiari}, S. and {Mender}, S. and {Mi{\'c}anovi{\'c}}, S. and {Miceli}, D. and {Miener}, T. and {Miranda}, J.~M. and {Mirzoyan}, R. and {Molina}, E. and {Moralejo}, A. and {Morcuende}, D. and {Moreno}, V. and {Moretti}, E. and {Nakamori}, T. and {Nava}, L. and {Neustroev}, V. and {Nievas Rosillo}, M. and {Nigro}, C. and {Nilsson}, K. and {Nishijima}, K. and {Noda}, K. and {Nozaki}, S. and {Ohtani}, Y. and {Oka}, T. and {Otero-Santos}, J. and {Paiano}, S. and {Palatiello}, M. and {Paneque}, D. and {Paoletti}, R. and {Paredes}, J.~M. and {Pavleti{\'c}}, L. and {Pe{\~n}il}, P. and {Persic}, M. and {Pihet}, M. and {Prada Moroni}, P.~G. and {Prandini}, E. and {Priyadarshi}, C. and {Puljak}, I. and {Rhode}, W. and {Rib{\'o}}, M. and {Rico}, J. and {Righi}, C. and {Rugliancich}, A. and {Sahakyan}, N. and {Saito}, T. and {Sakurai}, S. and {Satalecka}, K. and {Saturni}, F.~G. and {Schleicher}, B. and {Schmidt}, K. and {Schweizer}, T. and {Sitarek}, J. and {{\v{S}}nidari{\'c}}, I. and {Sobczynska}, D. and {Spolon}, A. and {Stamerra}, A. and {Stri{\v{s}}kovi{\'c}}, J. and {Strom}, D. and {Strzys}, M. and {Suda}, Y. and {Suri{\'c}}, T. and {Takahashi}, M. and {Takeishi}, R. and {Tavecchio}, F. and {Temnikov}, P. and {Terzi{\'c}}, T. and {Teshima}, M. and {Tosti}, L. and {Truzzi}, S. and {Tutone}, A. and {Ubach}, S. and {van Scherpenberg}, J. and {Vanzo}, G. and {Vazquez Acosta}, M. and {Ventura}, S. and {Verguilov}, V. and {Vigorito}, C.~F. and {Vitale}, V. and {Vovk}, I. and {Will}, M. and {Wunderlich}, C. and {Yamamoto}, T. and {Zari{\'c}}, D. and {Ambrosino}, F.},
        title = "{Proton acceleration in thermonuclear nova explosions revealed by gamma rays}",
      journal = {Nature Astronomy},
         year = 2022,
        month = apr,
       volume = {6},
        pages = {689-697},
          doi = {10.1038/s41550-022-01640-z},
archivePrefix = {arXiv},
       eprint = {2202.07681},
 primaryClass = {astro-ph.HE},
       adsurl = {https://ui.adsabs.harvard.edu/abs/2022NatAs...6..689A}
}

@ARTICLE{alexander21,
       author = {{Alexander}, K.~D. and {Schroeder}, G. and {Paterson}, K. and {Fong}, W. and {Cowperthwaite}, P. and {Gomez}, S. and {Margalit}, B. and {Margutti}, R. and {Berger}, E. and {Blanchard}, P. and {Chornock}, R. and {Eftekhari}, T. and {Laskar}, T. and {Metzger}, B.~D. and {Nicholl}, M. and {Villar}, V.~A. and {Williams}, P.~K.~G.},
        title = "{A Late-time Galaxy-targeted Search for the Radio Counterpart of GW190814}",
      journal = {\apj},
         year = 2021,
        month = dec,
       volume = {923},
       number = {1},
          eid = {66},
        pages = {66},
          doi = {10.3847/1538-4357/ac281a},
archivePrefix = {arXiv},
       eprint = {2102.08957},
 primaryClass = {astro-ph.HE},
       adsurl = {https://ui.adsabs.harvard.edu/abs/2021ApJ...923...66A}
}

@ARTICLE{anderson12,
       author = {{Anderson}, Gemma E. and {Gaensler}, B.~M. and {Slane}, Patrick O. and {Rea}, Nanda and {Kaplan}, David L. and {Posselt}, Bettina and {Levin}, Lina and {Johnston}, Simon and {Murray}, Stephen S. and {Brogan}, Crystal L. and {Bailes}, Matthew and {Bates}, Samuel and {Benjamin}, Robert A. and {Bhat}, N.~D. Ramesh and {Burgay}, Marta and {Burke-Spolaor}, Sarah and {Chakrabarty}, Deepto and {D'Amico}, Nichi and {Drake}, Jeremy J. and {Esposito}, Paolo and {Grindlay}, Jonathan E. and {Hong}, Jaesub and {Israel}, G.~L. and {Keith}, Michael J. and {Kramer}, Michael and {Lazio}, T. Joseph W. and {Lee}, Julia C. and {Mauerhan}, Jon C. and {Milia}, Sabrina and {Possenti}, Andrea and {Stappers}, Ben and {Steeghs}, Danny T.~H.},
        title = "{Multi-wavelength Observations of the Radio Magnetar PSR J1622-4950 and Discovery of Its Possibly Associated Supernova Remnant}",
      journal = {\apj},
         year = 2012,
        month = may,
       volume = {751},
       number = {1},
          eid = {53},
        pages = {53},
          doi = {10.1088/0004-637X/751/1/53},
archivePrefix = {arXiv},
       eprint = {1203.2719},
 primaryClass = {astro-ph.HE},
       adsurl = {https://ui.adsabs.harvard.edu/abs/2012ApJ...751...53A}
}

@ARTICLE{anderson14,
       author = {{Anderson}, G.~E. and {van der Horst}, A.~J. and {Staley}, T.~D. and {Fender}, R.~P. and {Wijers}, R.~A.~M.~J. and {Scaife}, A.~M.~M. and {Rumsey}, C. and {Titterington}, D.~J. and {Rowlinson}, A. and {Saunders}, R.~D.~E.},
        title = "{Probing the bright radio flare and afterglow of GRB 130427A with the Arcminute Microkelvin Imager}",
      journal = {\mnras},
         year = 2014,
        month = may,
       volume = {440},
       number = {3},
        pages = {2059-2065},
          doi = {10.1093/mnras/stu478},
archivePrefix = {arXiv},
       eprint = {1403.2217},
 primaryClass = {astro-ph.HE},
       adsurl = {https://ui.adsabs.harvard.edu/abs/2014MNRAS.440.2059A}
}

@ARTICLE{anderson18,
       author = {{Anderson}, G.~E. and {Staley}, T.~D. and {van der Horst}, A.~J. and {Fender}, R.~P. and {Rowlinson}, A. and {Mooley}, K.~P. and {Broderick}, J.~W. and {Wijers}, R.~A.~M.~J. and {Rumsey}, C. and {Titterington}, D.~J.},
        title = "{The Arcminute Microkelvin Imager catalogue of gamma-ray burst afterglows at 15.7 GHz}",
      journal = {\mnras},
         year = 2018,
        month = jan,
       volume = {473},
       number = {2},
        pages = {1512-1536},
          doi = {10.1093/mnras/stx2407},
archivePrefix = {arXiv},
       eprint = {1709.06232},
 primaryClass = {astro-ph.HE},
       adsurl = {https://ui.adsabs.harvard.edu/abs/2018MNRAS.473.1512A}
}

@ARTICLE{anderson18lwa,
       author = {{Anderson}, Marin M. and {Hallinan}, Gregg and {Eastwood}, Michael W. and {Monroe}, Ryan M. and {Vedantham}, Harish K. and {Bourke}, Stephen and {Greenhill}, Lincoln J. and {Kocz}, Jonathon and {Lazio}, T. Joseph W. and {Price}, Danny C. and {Schinzel}, Frank K. and {Wang}, Yuankun and {Woody}, David P.},
        title = "{A Simultaneous Search for Prompt Radio Emission Associated with the Short GRB 170112A Using the All-sky Imaging Capability of the OVRO-LWA}",
      journal = {\apj},
         year = 2018,
        month = sep,
       volume = {864},
       number = {1},
          eid = {22},
        pages = {22},
          doi = {10.3847/1538-4357/aad2d7},
archivePrefix = {arXiv},
       eprint = {1711.06665},
 primaryClass = {astro-ph.HE},
       adsurl = {https://ui.adsabs.harvard.edu/abs/2018ApJ...864...22A}
}

@ARTICLE{anderson21,
       author = {{Anderson}, G.~E. and {Bell}, M.~E. and {Stevens}, J. and {Aksulu}, M.~D. and {Miller-Jones}, J.~C.~A. and {van der Horst}, A.~J. and {Wijers}, R.~A.~M.~J. and {Rowlinson}, A. and {Bahramian}, A. and {Hancock}, P.~J. and {Macquart}, J. -P. and {Ryder}, S.~D. and {Plotkin}, R.~M.},
        title = "{Rapid-response radio observations of short GRB 181123B with the Australia Telescope Compact Array}",
      journal = {\mnras},
         year = 2021,
        month = may,
       volume = {503},
       number = {3},
        pages = {4372-4386},
          doi = {10.1093/mnras/stab727},
archivePrefix = {arXiv},
       eprint = {2103.05209},
 primaryClass = {astro-ph.HE},
       adsurl = {https://ui.adsabs.harvard.edu/abs/2021MNRAS.503.4372A}
}

@ARTICLE{anderson21mwa,
       author = {{Anderson}, G.~E. and {Hancock}, P.~J. and {Rowlinson}, A. and {Sokolowski}, M. and {Williams}, A. and {Tian}, J. and {Miller-Jones}, J.~C.~A. and {Hurley-Walker}, N. and {Bannister}, K.~W. and {Bell}, M.~E. and {James}, C.~W. and {Kaplan}, D.~L. and {Murphy}, Tara and {Tingay}, S.~J. and {Meyers}, B.~W. and {Johnston-Hollitt}, M. and {Wayth}, R.~B.},
        title = "{Murchison Widefield Array rapid-response observations of the short GRB 180805A}",
      journal = {\pasa},
         year = 2021,
        month = jun,
       volume = {38},
          eid = {e026},
        pages = {e026},
          doi = {10.1017/pasa.2021.15},
archivePrefix = {arXiv},
       eprint = {2104.14758},
 primaryClass = {astro-ph.HE},
       adsurl = {https://ui.adsabs.harvard.edu/abs/2021PASA...38...26A}
}

@ARTICLE{anderson23,
       author = {{Anderson}, G.~E. and {Russell}, T.~D. and {Fausey}, H.~M. and {van der Horst}, A.~J. and {Hancock}, P.~J. and {Bahramian}, A. and {Bell}, M.~E. and {Miller-Jones}, J.~C.~A. and {Rowell}, G. and {Sammons}, M.~W. and {Wijers}, R.~A.~M.~J. and {Galvin}, T.~J. and {Goodwin}, A.~J. and {Konno}, R. and {Rowlinson}, A. and {Ryder}, S.~D. and {Sch{\"u}ssler}, F. and {Wagner}, S.~J. and {Zhu}, S.~J.},
        title = "{Rapid radio brightening of GRB 210702A}",
      journal = {\mnras},
         year = 2023,
        month = aug,
       volume = {523},
       number = {4},
        pages = {4992-5005},
          doi = {10.1093/mnras/stad1635},
archivePrefix = {arXiv},
       eprint = {2211.11212},
 primaryClass = {astro-ph.HE},
       adsurl = {https://ui.adsabs.harvard.edu/abs/2023MNRAS.523.4992A}
}

@ARTICLE{anderson24,
       author = {{Anderson}, G.~E. and {Schroeder}, G. and {van der Horst}, A.~J. and {Rhodes}, L. and {Rowlinson}, A. and {Bahramian}, A. and {Chastain}, S.~I. and {Gompertz}, B.~P. and {Hancock}, P.~J. and {Laskar}, T. and {Leung}, J.~K. and {Wijers}, R.~A.~M.~J.},
        title = "{The Early Radio Afterglow of Short GRB 230217A}",
      journal = {\apjl},
         year = 2024,
        month = nov,
       volume = {975},
       number = {1},
          eid = {L13},
        pages = {L13},
          doi = {10.3847/2041-8213/ad85e9},
archivePrefix = {arXiv},
       eprint = {2409.07686},
 primaryClass = {astro-ph.HE},
       adsurl = {https://ui.adsabs.harvard.edu/abs/2024ApJ...975L..13A}
}

@ARTICLE{anderson25,
       author = {{Anderson}, G.~E. and {Lamb}, G.~P. and {Gompertz}, B.~P. and {Rhodes}, L. and {Martin-Carrillo}, A. and {van der Horst}, A.~J. and {Rowlinson}, A. and {Bell}, M.~E. and {Chen}, T.-W. and {Fausey}, H.~M. and {Ferro}, M. and {Hancock}, P.~J. and {Oates}, S.~R. and {Schulze}, S. and {Starling}, R.~L.~C. and {Yang}, S. and {Ackley}, K. and {Anderson}, J.~P. and {Andersson}, A. and {Fern{\'a}ndez}, J.~F. Ag{\"u}{\'\i} and {Brivio}, R. and {Burns}, E. and {Chambers}, K.~C. and {de Boer}, T. and {D'Elia}, V. and {De Pasquale}, M. and {de Ugarte Postigo}, A. and {Dimple} and {Fender}, R. and {Fulton}, M.~D. and {Gao}, H. and {Gillanders}, J.~H. and {Green}, D.~A. and {Gromadzki}, M. and {Gulati}, A. and {Hartmann}, D.~H. and {Huber}, M.~E. and {Klingler}, N.~J. and {Kuin}, N.~P.~M. and {Leung}, J.~K. and {Levan}, A.~J. and {Lin}, C.-C. and {Magnier}, E. and {Malesani}, D.~B. and {Minguez}, P. and {Mooley}, K.~P. and {Mukherjee}, T. and {Nicholl}, M. and {O'Brien}, P.~T. and {Pugliese}, G. and {Rossi}, A. and {Ryder}, S.~D. and {Sbarufatti}, B. and {Schneider}, B. and {Sch{\"u}ssler}, F. and {Smartt}, S.~J. and {Smith}, K.~W. and {Srivastav}, S. and {Steeghs}, D. and {Tanvir}, N.~R. and {Thoene}, C.~C. and {Vergani}, S.~D. and {Wainscoat}, R.~J. and {Wang}, Z.-N. and {Wijers}, R.~A.~M.~J. and {Williams-Baldwin}, D. and {Worssam}, I. and {Zafar}, T.},
        title = "{The Radio Flare and Multiwavelength Afterglow of the Short GRB 231117A: Energy Injection from a Violent Shell Collision}",
      journal = {\apj},
         year = 2025,
        month = nov,
       volume = {994},
       number = {1},
          eid = {5},
        pages = {5},
          doi = {10.3847/1538-4357/adfed7},
archivePrefix = {arXiv},
       eprint = {2508.14650},
 primaryClass = {astro-ph.HE},
       adsurl = {https://ui.adsabs.harvard.edu/abs/2025ApJ...994....5A}
}

@ARTICLE{atwood09,
       author = {{Atwood}, W.~B. and {Abdo}, A.~A. and {Ackermann}, M. and {Althouse}, W. and {Anderson}, B. and {Axelsson}, M. and {Baldini}, L. and {Ballet}, J. and {Band}, D.~L. and {Barbiellini}, G. and {Bartelt}, J. and {Bastieri}, D. and {Baughman}, B.~M. and {Bechtol}, K. and {B{\'e}d{\'e}r{\`e}de}, D. and {Bellardi}, F. and {Bellazzini}, R. and {Berenji}, B. and {Bignami}, G.~F. and {Bisello}, D. and {Bissaldi}, E. and {Blandford}, R.~D. and {Bloom}, E.~D. and {Bogart}, J.~R. and {Bonamente}, E. and {Bonnell}, J. and {Borgland}, A.~W. and {Bouvier}, A. and {Bregeon}, J. and {Brez}, A. and {Brigida}, M. and {Bruel}, P. and {Burnett}, T.~H. and {Busetto}, G. and {Caliandro}, G.~A. and {Cameron}, R.~A. and {Caraveo}, P.~A. and {Carius}, S. and {Carlson}, P. and {Casandjian}, J.~M. and {Cavazzuti}, E. and {Ceccanti}, M. and {Cecchi}, C. and {Charles}, E. and {Chekhtman}, A. and {Cheung}, C.~C. and {Chiang}, J. and {Chipaux}, R. and {Cillis}, A.~N. and {Ciprini}, S. and {Claus}, R. and {Cohen-Tanugi}, J. and {Condamoor}, S. and {Conrad}, J. and {Corbet}, R. and {Corucci}, L. and {Costamante}, L. and {Cutini}, S. and {Davis}, D.~S. and {Decotigny}, D. and {DeKlotz}, M. and {Dermer}, C.~D. and {de Angelis}, A. and {Digel}, S.~W. and {do Couto e Silva}, E. and {Drell}, P.~S. and {Dubois}, R. and {Dumora}, D. and {Edmonds}, Y. and {Fabiani}, D. and {Farnier}, C. and {Favuzzi}, C. and {Flath}, D.~L. and {Fleury}, P. and {Focke}, W.~B. and {Funk}, S. and {Fusco}, P. and {Gargano}, F. and {Gasparrini}, D. and {Gehrels}, N. and {Gentit}, F.-X. and {Germani}, S. and {Giebels}, B. and {Giglietto}, N. and {Giommi}, P. and {Giordano}, F. and {Glanzman}, T. and {Godfrey}, G. and {Grenier}, I.~A. and {Grondin}, M.-H. and {Grove}, J.~E. and {Guillemot}, L. and {Guiriec}, S. and {Haller}, G. and {Harding}, A.~K. and {Hart}, P.~A. and {Hays}, E. and {Healey}, S.~E. and {Hirayama}, M. and {Hjalmarsdotter}, L. and {Horn}, R. and {Hughes}, R.~E. and {J{\'o}hannesson}, G. and {Johansson}, G. and {Johnson}, A.~S. and {Johnson}, R.~P. and {Johnson}, T.~J. and {Johnson}, W.~N. and {Kamae}, T. and {Katagiri}, H. and {Kataoka}, J. and {Kavelaars}, A. and {Kawai}, N. and {Kelly}, H. and {Kerr}, M. and {Klamra}, W. and {Kn{\"o}dlseder}, J. and {Kocian}, M.~L. and {Komin}, N. and {Kuehn}, F. and {Kuss}, M. and {Landriu}, D. and {Latronico}, L. and {Lee}, B. and {Lee}, S.-H. and {Lemoine-Goumard}, M. and {Lionetto}, A.~M. and {Longo}, F. and {Loparco}, F. and {Lott}, B. and {Lovellette}, M.~N. and {Lubrano}, P. and {Madejski}, G.~M. and {Makeev}, A. and {Marangelli}, B. and {Massai}, M.~M. and {Mazziotta}, M.~N. and {McEnery}, J.~E. and {Menon}, N. and {Meurer}, C. and {Michelson}, P.~F. and {Minuti}, M. and {Mirizzi}, N. and {Mitthumsiri}, W. and {Mizuno}, T. and {Moiseev}, A.~A. and {Monte}, C. and {Monzani}, M.~E. and {Moretti}, E. and {Morselli}, A. and {Moskalenko}, I.~V. and {Murgia}, S. and {Nakamori}, T. and {Nishino}, S. and {Nolan}, P.~L. and {Norris}, J.~P. and {Nuss}, E. and {Ohno}, M. and {Ohsugi}, T. and {Omodei}, N. and {Orlando}, E. and {Ormes}, J.~F. and {Paccagnella}, A. and {Paneque}, D. and {Panetta}, J.~H. and {Parent}, D. and {Pearce}, M. and {Pepe}, M. and {Perazzo}, A. and {Pesce-Rollins}, M. and {Picozza}, P. and {Pieri}, L. and {Pinchera}, M. and {Piron}, F. and {Porter}, T.~A. and {Poupard}, L. and {Rain{\`o}}, S. and {Rando}, R. and {Rapposelli}, E. and {Razzano}, M. and {Reimer}, A. and {Reimer}, O. and {Reposeur}, T. and {Reyes}, L.~C. and {Ritz}, S. and {Rochester}, L.~S. and {Rodriguez}, A.~Y. and {Romani}, R.~W. and {Roth}, M. and {Russell}, J.~J. and {Ryde}, F. and {Sabatini}, S. and {Sadrozinski}, H.~F.-W. and {Sanchez}, D. and {Sander}, A. and {Sapozhnikov}, L. and {Parkinson}, P.~M. Saz and {Scargle}, J.~D. and {Schalk}, T.~L. and {Scolieri}, G.},
        title = "{The Large Area Telescope on the Fermi Gamma-Ray Space Telescope Mission}",
      journal = {\apj},
         year = 2009,
        month = jun,
       volume = {697},
       number = {2},
        pages = {1071-1102},
          doi = {10.1088/0004-637X/697/2/1071},
archivePrefix = {arXiv},
       eprint = {0902.1089},
 primaryClass = {astro-ph.IM},
       adsurl = {https://ui.adsabs.harvard.edu/abs/2009ApJ...697.1071A}
}

@ARTICLE{ayalasolares23,
       author = {{Ayala Solares}, H.~A. and {Coutu}, S. and {Cowen}, D. and {Fox}, D.~B. and {Gr{\'e}goire}, T. and {McBride}, F. and {Mostaf{\'a}}, M. and {Murase}, K. and {Wissel}, S. and {Albert}, A. and {Alves}, S. and {Andr{\'e}}, M. and {Ardid}, M. and {Ardid}, S. and {Aubert}, J.-J. and {Aublin}, J. and {Baret}, B. and {Basa}, S. and {Belhorma}, B. and {Bendahman}, M. and {Benfenati}, F. and {Bertin}, V. and {Biagi}, S. and {Bissinger}, M. and {Boumaaza}, J. and {Bouta}, M. and {Bouwhuis}, M.~C. and {Br{\^a}nza{\c{s}}}, H. and {Bruijn}, R. and {Brunner}, J. and {Busto}, J. and {Caiffi}, B. and {Calvo}, D. and {Capone}, A. and {Caramete}, L. and {Carr}, J. and {Carretero}, V. and {Celli}, S. and {Chabab}, M. and {Chau}, T.~N. and {El Moursli}, R. Cherkaoui and {Chiarusi}, T. and {Circella}, M. and {Coelho}, J.~A.~B. and {Coleiro}, A. and {Coniglione}, R. and {Coyle}, P. and {Creusot}, A. and {D{\'\i}az}, A.~F. and {de Wasseige}, G. and {De Martino}, B. and {Distefano}, C. and {Di Palma}, I. and {Domi}, A. and {Donzaud}, C. and {Dornic}, D. and {Drouhin}, D. and {Eberl}, T. and {van Eeden}, T. and {van Eijk}, D. and {El Khayati}, N. and {Enzenh{\"o}fer}, A. and {Fermani}, P. and {Ferrara}, G. and {Filippini}, F. and {Fusco}, L. and {Garc{\'\i}a}, J. and {Gay}, P. and {Glotin}, H. and {Gozzini}, R. and {Gracia Ruiz}, R. and {Graf}, K. and {Guidi}, C. and {Hallmann}, S. and {van Haren}, H. and {Heijboer}, A.~J. and {Hello}, Y. and {Hern{\'a}ndez-Rey}, J.~J. and {H{\"o}{\ss}l}, J. and {Hofest{\"a}dt}, J. and {Huang}, F. and {Illuminati}, G. and {James}, C.~W. and {Jisse-Jung}, B. and {de Jong}, M. and {de Jong}, P. and {Kadler}, M. and {Kalekin}, O. and {Katz}, U. and {Kouchner}, A. and {Kreykenbohm}, I. and {Kulikovskiy}, V. and {Lahmann}, R. and {Lamoureux}, M. and {Le Breton}, R. and {Lef{\`e}vre}, D. and {Leonora}, E. and {Levi}, G. and {Le Stum}, S. and {Lopez-Coto}, D. and {Loucatos}, S. and {Maderer}, L. and {Manczak}, J. and {Marcelin}, M. and {Margiotta}, A. and {Marinelli}, A. and {Mart{\'\i}nez-Mora}, J.~A. and {Melis}, K. and {Migliozzi}, P. and {Moussa}, A. and {Muller}, R. and {Nauta}, L. and {Navas}, S. and {Nezri}, E. and {Fearraigh}, B. {\'O}. and {P{\u{a}}un}, A. and {P{\u{a}}v{\u{a}}la{\c{s}}}, G.~E. and {Pellegrino}, C. and {Perrin-Terrin}, M. and {Pestel}, V. and {Piattelli}, P. and {Pieterse}, C. and {Poir{\`e}}, C. and {Popa}, V. and {Pradier}, T. and {Randazzo}, N. and {Real}, D. and {Reck}, S. and {Riccobene}, G. and {Romanov}, A. and {S{\'a}nchez-Losa}, A. and {Samtleben}, D.~F.~E. and {Sanguineti}, M. and {Sapienza}, P. and {Schnabel}, J. and {Schumann}, J. and {Sch{\"u}ssler}, F. and {Seneca}, J. and {Spurio}, M. and {Stolarczyk}, Th. and {Taiuti}, M. and {Tayalati}, Y. and {Tingay}, S.~J. and {Vallage}, B. and {Van Elewyck}, V. and {Versari}, F. and {Viola}, S. and {Vivolo}, D. and {Wilms}, J. and {Zavatarelli}, S. and {Zegarelli}, A. and {Zornoza}, J.~D. and {Z{\'u}{\~n}iga}, J. and {Albert}, A. and {Alvarez}, C. and {Arteaga-Vel{\'a}zquez}, J.~C. and {Babu}, R. and {Belmont-Moreno}, E. and {Caballero-Mora}, K.~S. and {Capistr{\'a}n}, T. and {Carrami{\~n}ana}, A. and {Casanova}, S. and {Cotti}, U. and {Chaparro-Amaro}, O. and {Cotzomi}, J. and {Couti{\~n}o de Le{\'o}n}, S. and {De la Fuente}, E. and {de Le{\'o}n}, C. and {Diaz Hernandez}, R. and {DuVernois}, M.~A. and {Durocher}, M. and {D{\'\i}az-V{\'e}lez}, J.~C. and {Engel}, K. and {Espinoza}, C. and {Fan}, K.~L. and {Fern{\'a}ndez Alonso}, M. and {Fraija}, N. and {Garc{\'\i}a-Gonz{\'a}lez}, J.~A. and {Garfias}, F. and {Gonz{\'a}lez}, M.~M. and {Goodman}, J.~A. and {Harding}, J.~P. and {Hernandez}, S. and {Huang}, D. and {Hueyotl-Zahuantitla}, F. and {H{\"u}ntemeyer}, P. and {Iriarte}, A. and {Joshi}, V. and {Kaufmann}, S. and {Lara}, A. and {Le{\'o}n Vargas}, H. and {Linnemann}, J.~T. and {Longinotti}, A.~L. and {Luis-Raya}, G. and {Malone}, K. and {Martinez}, O. and {Martinez-Castellanos}, I. and {Mart{\'\i}nez-Castro}, J. and {Matthews}, J.~A. and {Miranda-Romagnoli}, P.},
        title = "{Search for Gamma-Ray and Neutrino Coincidences Using HAWC and ANTARES Data}",
      journal = {\apj},
         year = 2023,
        month = feb,
       volume = {944},
       number = {2},
          eid = {166},
        pages = {166},
          doi = {10.3847/1538-4357/acafdd},
archivePrefix = {arXiv},
       eprint = {2209.13462},
 primaryClass = {astro-ph.HE},
       adsurl = {https://ui.adsabs.harvard.edu/abs/2023ApJ...944..166A}
}

@ARTICLE{bannister12,
   author = {{Bannister}, K.~W. and {Murphy}, T. and {Gaensler}, B.~M. and 
	{Reynolds}, J.~E.},
    title = "{Limits on Prompt, Dispersed Radio Pulses from Gamma-Ray Bursts}",
  journal = {\apj},
archivePrefix = "arXiv",
   eprint = {1207.6399},
 primaryClass = "astro-ph.HE",
     year = 2012,
    month = sep,
   volume = 757,
      eid = {38},
    pages = {38},
      doi = {10.1088/0004-637X/757/1/38},
   adsurl = {http://adsabs.harvard.edu/abs/2012ApJ...757...38B}
}

@ARTICLE{barthelmy95,
       author = {{Barthelmy}, S.~D. and {Butterworth}, P. and {Cline}, T.~L. and {Gehrels}, N. and {Fishman}, G.~J. and {Kouveliotou}, C. and {Meegan}, C.~A.},
        title = "{BACODINE, the Real-Time BATSE Gamma-Ray Burst Coordinates Distribution Network}",
      journal = {\apss},
         year = 1995,
        month = sep,
       volume = {231},
       number = {1-2},
        pages = {235-238},
          doi = {10.1007/BF00658623},
       adsurl = {https://ui.adsabs.harvard.edu/abs/1995Ap&SS.231..235B}
}

@ARTICLE{barthelmy05,
       author = {{Barthelmy}, Scott D. and {Barbier}, Louis M. and {Cummings}, Jay R. and {Fenimore}, Ed E. and {Gehrels}, Neil and {Hullinger}, Derek and {Krimm}, Hans A. and {Markwardt}, Craig B. and {Palmer}, David M. and {Parsons}, Ann and {Sato}, Goro and {Suzuki}, Masaya and {Takahashi}, Tadayuki and {Tashiro}, Makota and {Tueller}, Jack},
        title = "{The Burst Alert Telescope (BAT) on the SWIFT Midex Mission}",
      journal = {\ssr},
         year = 2005,
        month = oct,
       volume = {120},
       number = {3-4},
        pages = {143-164},
          doi = {10.1007/s11214-005-5096-3},
archivePrefix = {arXiv},
       eprint = {astro-ph/0507410},
 primaryClass = {astro-ph},
       adsurl = {https://ui.adsabs.harvard.edu/abs/2005SSRv..120..143B}
}

@ARTICLE{bellm19,
       author = {{Bellm}, Eric C. and {Kulkarni}, Shrinivas R. and {Graham}, Matthew J. and {Dekany}, Richard and {Smith}, Roger M. and {Riddle}, Reed and {Masci}, Frank J. and {Helou}, George and {Prince}, Thomas A. and {Adams}, Scott M. and {Barbarino}, C. and {Barlow}, Tom and {Bauer}, James and {Beck}, Ron and {Belicki}, Justin and {Biswas}, Rahul and {Blagorodnova}, Nadejda and {Bodewits}, Dennis and {Bolin}, Bryce and {Brinnel}, Valery and {Brooke}, Tim and {Bue}, Brian and {Bulla}, Mattia and {Burruss}, Rick and {Cenko}, S. Bradley and {Chang}, Chan-Kao and {Connolly}, Andrew and {Coughlin}, Michael and {Cromer}, John and {Cunningham}, Virginia and {De}, Kishalay and {Delacroix}, Alex and {Desai}, Vandana and {Duev}, Dmitry A. and {Eadie}, Gwendolyn and {Farnham}, Tony L. and {Feeney}, Michael and {Feindt}, Ulrich and {Flynn}, David and {Franckowiak}, Anna and {Frederick}, S. and {Fremling}, C. and {Gal-Yam}, Avishay and {Gezari}, Suvi and {Giomi}, Matteo and {Goldstein}, Daniel A. and {Golkhou}, V. Zach and {Goobar}, Ariel and {Groom}, Steven and {Hacopians}, Eugean and {Hale}, David and {Henning}, John and {Ho}, Anna Y.~Q. and {Hover}, David and {Howell}, Justin and {Hung}, Tiara and {Huppenkothen}, Daniela and {Imel}, David and {Ip}, Wing-Huen and {Ivezi{\'c}}, {\v{Z}}eljko and {Jackson}, Edward and {Jones}, Lynne and {Juric}, Mario and {Kasliwal}, Mansi M. and {Kaspi}, S. and {Kaye}, Stephen and {Kelley}, Michael S.~P. and {Kowalski}, Marek and {Kramer}, Emily and {Kupfer}, Thomas and {Landry}, Walter and {Laher}, Russ R. and {Lee}, Chien-De and {Lin}, Hsing Wen and {Lin}, Zhong-Yi and {Lunnan}, Ragnhild and {Giomi}, Matteo and {Mahabal}, Ashish and {Mao}, Peter and {Miller}, Adam A. and {Monkewitz}, Serge and {Murphy}, Patrick and {Ngeow}, Chow-Choong and {Nordin}, Jakob and {Nugent}, Peter and {Ofek}, Eran and {Patterson}, Maria T. and {Penprase}, Bryan and {Porter}, Michael and {Rauch}, Ludwig and {Rebbapragada}, Umaa and {Reiley}, Dan and {Rigault}, Mickael and {Rodriguez}, Hector and {van Roestel}, Jan and {Rusholme}, Ben and {van Santen}, Jakob and {Schulze}, S. and {Shupe}, David L. and {Singer}, Leo P. and {Soumagnac}, Maayane T. and {Stein}, Robert and {Surace}, Jason and {Sollerman}, Jesper and {Szkody}, Paula and {Taddia}, F. and {Terek}, Scott and {Van Sistine}, Angela and {van Velzen}, Sjoert and {Vestrand}, W. Thomas and {Walters}, Richard and {Ward}, Charlotte and {Ye}, Quan-Zhi and {Yu}, Po-Chieh and {Yan}, Lin and {Zolkower}, Jeffry},
        title = "{The Zwicky Transient Facility: System Overview, Performance, and First Results}",
      journal = {\pasp},
         year = 2019,
        month = jan,
       volume = {131},
       number = {995},
        pages = {018002},
          doi = {10.1088/1538-3873/aaecbe},
archivePrefix = {arXiv},
       eprint = {1902.01932},
 primaryClass = {astro-ph.IM},
       adsurl = {https://ui.adsabs.harvard.edu/abs/2019PASP..131a8002B}
}

@ARTICLE{benz10,
       author = {{Benz}, Arnold O. and {G{\"u}del}, Manuel},
        title = "{Physical Processes in Magnetically Driven Flares on the Sun, Stars, and Young Stellar Objects}",
      journal = {\araa},
         year = 2010,
        month = sep,
       volume = {48},
        pages = {241-287},
          doi = {10.1146/annurev-astro-082708-101757},
       adsurl = {https://ui.adsabs.harvard.edu/abs/2010ARA&A..48..241B}
}

@ARTICLE{bezuidenhout22,
       author = {{Bezuidenhout}, M.~C. and {Barr}, E. and {Caleb}, M. and {Driessen}, L.~N. and {Jankowski}, F. and {Kramer}, M. and {Malenta}, M. and {Morello}, V. and {Rajwade}, K. and {Sanidas}, S. and {Stappers}, B.~W. and {Surnis}, M.},
        title = "{MeerTRAP: 12 Galactic fast transients detected in a real-time, commensal MeerKAT survey}",
      journal = {\mnras},
         year = 2022,
        month = may,
       volume = {512},
       number = {1},
        pages = {1483-1498},
          doi = {10.1093/mnras/stac579},
archivePrefix = {arXiv},
       eprint = {2203.00557},
 primaryClass = {astro-ph.HE},
       adsurl = {https://ui.adsabs.harvard.edu/abs/2022MNRAS.512.1483B}
}

@ARTICLE{bhardwaj24,
       author = {{Bhardwaj}, Mohit and {Palmese}, Antonella and {Maga{\~n}a Hernandez}, Ignacio and {D'Emilio}, Virginia and {Morisaki}, Soichiro},
        title = "{Challenges for Fast Radio Bursts as Multimessenger Sources from Binary Neutron Star Mergers}",
      journal = {\apj},
         year = 2024,
        month = dec,
       volume = {977},
       number = {1},
          eid = {122},
        pages = {122},
          doi = {10.3847/1538-4357/ad9023},
archivePrefix = {arXiv},
       eprint = {2306.00948},
 primaryClass = {astro-ph.HE},
       adsurl = {https://ui.adsabs.harvard.edu/abs/2024ApJ...977..122B}
}

@ARTICLE{bright20,
       author = {{Bright}, J.~S. and {Fender}, R.~P. and {Motta}, S.~E. and {Williams}, D.~R.~A. and {Moldon}, J. and {Plotkin}, R.~M. and {Miller-Jones}, J.~C.~A. and {Heywood}, I. and {Tremou}, E. and {Beswick}, R. and {Sivakoff}, G.~R. and {Corbel}, S. and {Buckley}, D.~A.~H. and {Homan}, J. and {Gallo}, E. and {Tetarenko}, A.~J. and {Russell}, T.~D. and {Green}, D.~A. and {Titterington}, D. and {Woudt}, P.~A. and {Armstrong}, R.~P. and {Groot}, P.~J. and {Horesh}, A. and {van der Horst}, A.~J. and {K{\"o}rding}, E.~G. and {McBride}, V.~A. and {Rowlinson}, A. and {Wijers}, R.~A.~M.~J.},
        title = "{An extremely powerful long-lived superluminal ejection from the black hole MAXI J1820+070}",
      journal = {Nature Astronomy},
         year = 2020,
        month = mar,
       volume = {4},
        pages = {697-703},
          doi = {10.1038/s41550-020-1023-5},
archivePrefix = {arXiv},
       eprint = {2003.01083},
 primaryClass = {astro-ph.HE},
       adsurl = {https://ui.adsabs.harvard.edu/abs/2020NatAs...4..697B}
}

@ARTICLE{bright23,
       author = {{Bright}, Joe S. and {Rhodes}, Lauren and {Farah}, Wael and {Fender}, Rob and {van der Horst}, Alexander J. and {Leung}, James K. and {Williams}, David R.~A. and {Anderson}, Gemma E. and {Atri}, Pikky and {DeBoer}, David R. and {Giarratana}, Stefano and {Green}, David A. and {Heywood}, Ian and {Lenc}, Emil and {Murphy}, Tara and {Pollak}, Alexander W. and {Premnath}, Pranav H. and {Scott}, Paul F. and {Sheikh}, Sofia Z. and {Siemion}, Andrew and {Titterington}, David J.},
        title = "{Precise measurements of self-absorbed rising reverse shock emission from gamma-ray burst 221009A}",
      journal = {Nature Astronomy},
         year = 2023,
        month = aug,
       volume = {7},
        pages = {986-995},
          doi = {10.1038/s41550-023-01997-9},
archivePrefix = {arXiv},
       eprint = {2303.13583},
 primaryClass = {astro-ph.HE},
       adsurl = {https://ui.adsabs.harvard.edu/abs/2023NatAs...7..986B}
}

@ARTICLE{bochenek20,
       author = {{Bochenek}, C.~D. and {Ravi}, V. and {Belov}, K.~V. and {Hallinan}, G. and {Kocz}, J. and {Kulkarni}, S.~R. and {McKenna}, D.~L.},
        title = "{A fast radio burst associated with a Galactic magnetar}",
      journal = {\nat},
         year = 2020,
        month = nov,
       volume = {587},
       number = {7832},
        pages = {59-62},
          doi = {10.1038/s41586-020-2872-x},
archivePrefix = {arXiv},
       eprint = {2005.10828},
 primaryClass = {astro-ph.HE},
       adsurl = {https://ui.adsabs.harvard.edu/abs/2020Natur.587...59B}
}

@ARTICLE{burns23,
       author = {{Burns}, Eric and {Svinkin}, Dmitry and {Fenimore}, Edward and {Kann}, D. Alexander and {Ag{\"u}{\'\i} Fern{\'a}ndez}, Jos{\'e} Feliciano and {Frederiks}, Dmitry and {Hamburg}, Rachel and {Lesage}, Stephen and {Temiraev}, Yuri and {Tsvetkova}, Anastasia and {Bissaldi}, Elisabetta and {Briggs}, Michael S. and {Dalessi}, Sarah and {Dunwoody}, Rachel and {Fletcher}, Cori and {Goldstein}, Adam and {Hui}, C. Michelle and {Hristov}, Boyan A. and {Kocevski}, Daniel and {Lysenko}, Alexandra L. and {Mailyan}, Bagrat and {Mangan}, Joseph and {McBreen}, Sheila and {Racusin}, Judith and {Ridnaia}, Anna and {Roberts}, Oliver J. and {Ulanov}, Mikhail and {Veres}, Peter and {Wilson-Hodge}, Colleen A. and {Wood}, Joshua},
        title = "{GRB 221009A: The Boat}",
      journal = {\apjl},
         year = 2023,
        month = mar,
       volume = {946},
       number = {1},
          eid = {L31},
        pages = {L31},
          doi = {10.3847/2041-8213/acc39c},
archivePrefix = {arXiv},
       eprint = {2302.14037},
 primaryClass = {astro-ph.HE},
       adsurl = {https://ui.adsabs.harvard.edu/abs/2023ApJ...946L..31B}
}

@ARTICLE{burrows05,
       author = {{Burrows}, David N. and {Hill}, J.~E. and {Nousek}, J.~A. and {Kennea}, J.~A. and {Wells}, A. and {Osborne}, J.~P. and {Abbey}, A.~F. and {Beardmore}, A. and {Mukerjee}, K. and {Short}, A.~D.~T. and {Chincarini}, G. and {Campana}, S. and {Citterio}, O. and {Moretti}, A. and {Pagani}, C. and {Tagliaferri}, G. and {Giommi}, P. and {Capalbi}, M. and {Tamburelli}, F. and {Angelini}, L. and {Cusumano}, G. and {Br{\"a}uninger}, H.~W. and {Burkert}, W. and {Hartner}, G.~D.},
        title = "{The Swift X-Ray Telescope}",
      journal = {\ssr},
         year = 2005,
        month = oct,
       volume = {120},
       number = {3-4},
        pages = {165-195},
          doi = {10.1007/s11214-005-5097-2},
archivePrefix = {arXiv},
       eprint = {astro-ph/0508071},
 primaryClass = {astro-ph},
       adsurl = {https://ui.adsabs.harvard.edu/abs/2005SSRv..120..165B}
}

@ARTICLE{caleb20,
       author = {{Caleb}, M. and {Stappers}, B.~W. and {Abbott}, T.~D. and {Barr}, E.~D. and {Bezuidenhout}, M.~C. and {Buchner}, S.~J. and {Burgay}, M. and {Chen}, W. and {Cognard}, I. and {Driessen}, L.~N. and {Fender}, R. and {Hilmarsson}, G.~H. and {Hoang}, J. and {Horn}, D.~M. and {Jankowski}, F. and {Kramer}, M. and {Lorimer}, D.~R. and {Malenta}, M. and {Morello}, V. and {Pilia}, M. and {Platts}, E. and {Possenti}, A. and {Rajwade}, K.~M. and {Ridolfi}, A. and {Rhodes}, L. and {Sanidas}, S. and {Serylak}, M. and {Spitler}, L.~G. and {Townsend}, L.~J. and {Weltman}, A. and {Woudt}, P.~A. and {Wu}, J.},
        title = "{Simultaneous multi-telescope observations of FRB 121102}",
      journal = {\mnras},
         year = 2020,
        month = aug,
       volume = {496},
       number = {4},
        pages = {4565-4573},
          doi = {10.1093/mnras/staa1791},
archivePrefix = {arXiv},
       eprint = {2006.08662},
 primaryClass = {astro-ph.HE},
       adsurl = {https://ui.adsabs.harvard.edu/abs/2020MNRAS.496.4565C}
}

@ARTICLE{caleb22,
       author = {{Caleb}, Manisha and {Heywood}, Ian and {Rajwade}, Kaustubh and {Malenta}, Mateusz and {Stappers}, Benjamin Willem and {Barr}, Ewan and {Chen}, Weiwei and {Morello}, Vincent and {Sanidas}, Sotiris and {van den Eijnden}, Jakob and {Kramer}, Michael and {Buckley}, David and {Brink}, Jaco and {Motta}, Sara Elisa and {Woudt}, Patrick and {Weltevrede}, Patrick and {Jankowski}, Fabian and {Surnis}, Mayuresh and {Buchner}, Sarah and {Bezuidenhout}, Mechiel Christiaan and {Driessen}, Laura Nicole and {Fender}, Rob},
        title = "{Discovery of a radio-emitting neutron star with an ultra-long spin period of 76 s}",
      journal = {Nature Astronomy},
         year = 2022,
        month = may,
       volume = {6},
        pages = {828-836},
          doi = {10.1038/s41550-022-01688-x},
archivePrefix = {arXiv},
       eprint = {2206.01346},
 primaryClass = {astro-ph.HE},
       adsurl = {https://ui.adsabs.harvard.edu/abs/2022NatAs...6..828C}
}

@ARTICLE{caleb23,
       author = {{Caleb}, M. and {Driessen}, L.~N. and {Gordon}, A.~C. and {Tejos}, N. and {Bernales}, L. and {Qiu}, H. and {Chibueze}, J.~O. and {Stappers}, B.~W. and {Rajwade}, K.~M. and {Cavallaro}, F. and {Wang}, Y. and {Kumar}, P. and {Majid}, W.~A. and {Wharton}, R.~S. and {Naudet}, C.~J. and {Bezuidenhout}, M.~C. and {Jankowski}, F. and {Malenta}, M. and {Morello}, V. and {Sanidas}, S. and {Surnis}, M.~P. and {Barr}, E.~D. and {Chen}, W. and {Kramer}, M. and {Fong}, W. and {Kilpatrick}, C.~D. and {Prochaska}, J. Xavier and {Simha}, S. and {Venter}, C. and {Heywood}, I. and {Kundu}, A. and {Schussler}, F.},
        title = "{A subarcsec localized fast radio burst with a significant host galaxy dispersion measure contribution}",
      journal = {\mnras},
         year = 2023,
        month = sep,
       volume = {524},
       number = {2},
        pages = {2064-2077},
          doi = {10.1093/mnras/stad1839},
archivePrefix = {arXiv},
       eprint = {2302.09754},
 primaryClass = {astro-ph.HE},
       adsurl = {https://ui.adsabs.harvard.edu/abs/2023MNRAS.524.2064C}
}

@ARTICLE{callingham21,
       author = {{Callingham}, J.~R. and {Vedantham}, H.~K. and {Shimwell}, T.~W. and {Pope}, B.~J.~S. and {Davis}, I.~E. and {Best}, P.~N. and {Hardcastle}, M.~J. and {R{\"o}ttgering}, H.~J.~A. and {Sabater}, J. and {Tasse}, C. and {van Weeren}, R.~J. and {Williams}, W.~L. and {Zarka}, P. and {de Gasperin}, F. and {Drabent}, A.},
        title = "{The population of M dwarfs observed at low radio frequencies}",
      journal = {Nature Astronomy},
         year = 2021,
        month = dec,
       volume = {5},
        pages = {1233-1239},
          doi = {10.1038/s41550-021-01483-0},
archivePrefix = {arXiv},
       eprint = {2110.03713},
 primaryClass = {astro-ph.SR},
       adsurl = {https://ui.adsabs.harvard.edu/abs/2021NatAs...5.1233C}
}

@ARTICLE{callister19,
       author = {{Callister}, Thomas A. and {Anderson}, Marin M. and {Hallinan}, Gregg and {D'addario}, Larry R. and {Dowell}, Jayce and {Kassim}, Namir E. and {Lazio}, T. Joseph W. and {Price}, Danny C. and {Schinzel}, Frank K.},
        title = "{A First Search for Prompt Radio Emission from a Gravitational-wave Event}",
      journal = {\apjl},
         year = 2019,
        month = jun,
       volume = {877},
       number = {2},
          eid = {L39},
        pages = {L39},
          doi = {10.3847/2041-8213/ab2248},
archivePrefix = {arXiv},
       eprint = {1903.06786},
 primaryClass = {astro-ph.HE},
       adsurl = {https://ui.adsabs.harvard.edu/abs/2019ApJ...877L..39C}
}

@ARTICLE{camilo06,
       author = {{Camilo}, Fernando and {Ransom}, Scott M. and {Halpern}, Jules P. and {Reynolds}, John and {Helfand}, David J. and {Zimmerman}, Neil and {Sarkissian}, John},
        title = "{Transient pulsed radio emission from a magnetar}",
      journal = {\nat},
         year = 2006,
        month = aug,
       volume = {442},
       number = {7105},
        pages = {892-895},
          doi = {10.1038/nature04986},
archivePrefix = {arXiv},
       eprint = {astro-ph/0605429},
 primaryClass = {astro-ph},
       adsurl = {https://ui.adsabs.harvard.edu/abs/2006Natur.442..892C}
}

@ARTICLE{camilo07,
       author = {{Camilo}, F. and {Ransom}, S.~M. and {Halpern}, J.~P. and {Reynolds}, J.},
        title = "{1E 1547.0-5408: A Radio-emitting Magnetar with a Rotation Period of 2 Seconds}",
      journal = {\apjl},
         year = 2007,
        month = sep,
       volume = {666},
       number = {2},
        pages = {L93-L96},
          doi = {10.1086/521826},
archivePrefix = {arXiv},
       eprint = {0708.0002},
 primaryClass = {astro-ph},
       adsurl = {https://ui.adsabs.harvard.edu/abs/2007ApJ...666L..93C}
}

@ARTICLE{cao16,
       author = {{Cao}, Yi and {Nugent}, Peter E. and {Kasliwal}, Mansi M.},
        title = "{Intermediate Palomar Transient Factory: Realtime Image Subtraction Pipeline}",
      journal = {\pasp},
         year = 2016,
        month = nov,
       volume = {128},
       number = {969},
        pages = {114502},
          doi = {10.1088/1538-3873/128/969/114502},
archivePrefix = {arXiv},
       eprint = {1608.01006},
 primaryClass = {astro-ph.IM},
       adsurl = {https://ui.adsabs.harvard.edu/abs/2016PASP..128k4502C}
}

@ARTICLE{chastain24,
       author = {{Chastain}, S.~I. and {van der Horst}, A.~J. and {Anderson}, G.~E. and {Rhodes}, L. and {d'Antonio}, D. and {Bell}, M.~E. and {Fender}, R.~P. and {Hancock}, P.~J. and {Horesh}, A. and {Kouveliotou}, C. and {Mooley}, K.~P. and {Rowlinson}, A. and {Vergani}, S.~D. and {Wijers}, R.~A.~M.~J. and {Woudt}, P.~A.},
        title = "{Constraints on short gamma-ray burst physics and their host galaxies from systematic radio follow-up campaigns}",
      journal = {\mnras},
         year = 2024,
        month = aug,
       volume = {532},
       number = {2},
        pages = {2820-2831},
          doi = {10.1093/mnras/stae1568},
archivePrefix = {arXiv},
       eprint = {2407.11883},
 primaryClass = {astro-ph.HE},
       adsurl = {https://ui.adsabs.harvard.edu/abs/2024MNRAS.532.2820C}
}

@ARTICLE{chastain26pp,
       author = {{Chastain}, S.~I. and {Anderson}, G.~E. and {van der Horst}, A.~J. and {Rhodes}, L. and {Morley}, C. and {Gulati}, A. and {Leung}, J.~K. and {Russel}, T.~D. and {Ryder}, S.~D.},
        title = "{GRB 240205B: A Reverse Shock Detected in Rapid Response Radio Observations}",
      journal = {arXiv e-prints},
         year = 2026,
        month = mar,
          eid = {arXiv:2603.19047},
        pages = {arXiv:2603.19047},
          doi = {10.48550/arXiv.2603.19047},
archivePrefix = {arXiv},
       eprint = {2603.19047},
 primaryClass = {astro-ph.HE},
       adsurl = {https://ui.adsabs.harvard.edu/abs/2026arXiv260319047C}
}

@ARTICLE{chomiuk2021,
       author = {{Chomiuk}, Laura and {Metzger}, Brian D. and {Shen}, Ken J.},
        title = "{New Insights into Classical Novae}",
      journal = {\araa},
         year = 2021,
        month = sep,
       volume = {59},
        pages = {391-444},
          doi = {10.1146/annurev-astro-112420-114502},
archivePrefix = {arXiv},
       eprint = {2011.08751},
 primaryClass = {astro-ph.HE},
       adsurl = {https://ui.adsabs.harvard.edu/abs/2021ARA&A..59..391C}
}

@ARTICLE{Chomiuk2021b,
       author = {{Chomiuk}, Laura and {Linford}, Justin D. and {Aydi}, Elias and {Bannister}, Keith W. and {Krauss}, Miriam I. and {Mioduszewski}, Amy J. and {Mukai}, Koji and {Nelson}, Thomas J. and {Rupen}, Michael P. and {Ryder}, Stuart D. and et al.},
        title = "{Classical Novae at Radio Wavelengths}",
      journal = {\apjs},
         year = 2021,
        month = dec,
       volume = {257},
       number = {2},
          eid = {49},
        pages = {49},
          doi = {10.3847/1538-4365/ac24ab},
archivePrefix = {arXiv},
       eprint = {2107.06251},
 primaryClass = {astro-ph.HE},
       adsurl = {https://ui.adsabs.harvard.edu/abs/2021ApJS..257...49C}
}

@ARTICLE{chu16,
       author = {{Chu}, Q. and {Howell}, E.~J. and {Rowlinson}, A. and {Gao}, H. and {Zhang}, B. and {Tingay}, S.~J. and {Bo{\"e}r}, M. and {Wen}, L.},
        title = "{Capturing the electromagnetic counterparts of binary neutron star mergers through low-latency gravitational wave triggers}",
      journal = {\mnras},
         year = 2016,
        month = jun,
       volume = {459},
       number = {1},
        pages = {121-139},
          doi = {10.1093/mnras/stw576},
archivePrefix = {arXiv},
       eprint = {1509.06876},
 primaryClass = {astro-ph.HE},
       adsurl = {https://ui.adsabs.harvard.edu/abs/2016MNRAS.459..121C}
}

@ARTICLE{chauhan21,
       author = {{Chauhan}, Jaiverdhan and {Miller-Jones}, J.~C.~A. and {Anderson}, G.~E. and {Paduano}, A. and {Sokolowski}, M. and {Flynn}, C. and {Hancock}, P.~J. and {Hurley-Walker}, N. and {Kaplan}, D.~L. and {Russell}, T.~D. and {Bahramian}, A. and {Duchesne}, S.~W. and {Altamirano}, D. and {Croft}, S. and {Krimm}, H.~A. and {Sivakoff}, G.~R. and {Soria}, R. and {Trott}, C.~M. and {Wayth}, R.~B. and {Gupta}, V. and {Johnston-Hollitt}, M. and {Tingay}, S.~J.},
        title = "{A broadband radio view of transient jet ejecta in the black hole candidate X-ray binary MAXI J1535-571}",
      journal = {\pasa},
         year = 2021,
        month = sep,
       volume = {38},
          eid = {e045},
        pages = {e045},
          doi = {10.1017/pasa.2021.38},
archivePrefix = {arXiv},
       eprint = {2107.13019},
 primaryClass = {astro-ph.HE},
       adsurl = {https://ui.adsabs.harvard.edu/abs/2021PASA...38...45C}
}

@ARTICLE{clarke25,
       author = {{Clarke}, Teagan A. and {Sarin}, Nikhil and {Howell}, Eric J. and {Lasky}, Paul D. and {Thrane}, Eric},
        title = "{Quantifying the coincidence between gravitational waves and fast radio bursts from neutron star-black hole mergers}",
      journal = {\prd},
         year = 2025,
        month = apr,
       volume = {111},
       number = {8},
          eid = {083023},
        pages = {083023},
          doi = {10.1103/PhysRevD.111.083023},
archivePrefix = {arXiv},
       eprint = {2408.02534},
 primaryClass = {astro-ph.HE},
       adsurl = {https://ui.adsabs.harvard.edu/abs/2025PhRvD.111h3023C}
}

@ARTICLE{cooper23,
       author = {{Cooper}, A.~J. and {Gupta}, O. and {Wadiasingh}, Z. and {Wijers}, R.~A.~M.~J. and {Boersma}, O.~M. and {Andreoni}, I. and {Rowlinson}, A. and {Gourdji}, K.},
        title = "{Pulsar revival in neutron star mergers: multimessenger prospects for the discovery of pre-merger coherent radio emission}",
      journal = {\mnras},
         year = 2023,
        month = mar,
       volume = {519},
       number = {3},
        pages = {3923-3946},
          doi = {10.1093/mnras/stac3580},
archivePrefix = {arXiv},
       eprint = {2210.17205},
 primaryClass = {astro-ph.HE},
       adsurl = {https://ui.adsabs.harvard.edu/abs/2023MNRAS.519.3923C}
}

@ARTICLE{cooper24,
       author = {{Cooper}, A.~J. and {Wadiasingh}, Z.},
        title = "{Beyond the Rotational Deathline: Radio Emission from Ultra-long Period Magnetars}",
      journal = {\mnras},
         year = 2024,
        month = sep,
       volume = {533},
       number = {2},
        pages = {2133-2155},
          doi = {10.1093/mnras/stae1813},
archivePrefix = {arXiv},
       eprint = {2406.04135},
 primaryClass = {astro-ph.HE},
       adsurl = {https://ui.adsabs.harvard.edu/abs/2024MNRAS.533.2133C}
}

@ARTICLE{coward17,
       author = {{Coward}, D.~M. and {Gendre}, B. and {Tanga}, P. and {Turpin}, D. and {Zadko}, J. and {Dodson}, R. and {Devog{\'e}le}, M. and {Howell}, E.~J. and {Kennewell}, J.~A. and {Bo{\"e}r}, M. and {Klotz}, A. and {Dornic}, D. and {Moore}, J.~A. and {Heary}, A.},
        title = "{The Zadko Telescope: Exploring the Transient Universe}",
      journal = {\pasa},
         year = 2017,
        month = jan,
       volume = {34},
          eid = {e005},
        pages = {e005},
          doi = {10.1017/pasa.2016.61},
archivePrefix = {arXiv},
       eprint = {1609.06445},
 primaryClass = {astro-ph.HE},
       adsurl = {https://ui.adsabs.harvard.edu/abs/2017PASA...34....5C}
}

@ARTICLE{cunningham19,
       author = {{Cunningham}, Virginia and {Cenko}, S. Bradley and {Burns}, Eric and {Goldstein}, Adam and {Lien}, Amy and {Kocevski}, Daniel and {Briggs}, Michael and {Connaughton}, Valerie and {Miller}, M. Coleman and {Racusin}, Judith and {Stanbro}, Matthew},
        title = "{A Search for High-energy Counterparts to Fast Radio Bursts}",
      journal = {\apj},
         year = 2019,
        month = jul,
       volume = {879},
       number = {1},
          eid = {40},
        pages = {40},
          doi = {10.3847/1538-4357/ab2235},
archivePrefix = {arXiv},
       eprint = {1905.06818},
 primaryClass = {astro-ph.HE},
       adsurl = {https://ui.adsabs.harvard.edu/abs/2019ApJ...879...40C}
}

@ARTICLE{curtin23,
       author = {{Curtin}, Alice P. and {Tendulkar}, Shriharsh P. and {Josephy}, Alexander and {Chawla}, Pragya and {Andersen}, Bridget and {Kaspi}, Victoria M. and {Bhardwaj}, Mohit and {Cassanelli}, Tomas and {Cook}, Amanda and {Dong}, Fengqiu Adam and {Fonseca}, Emmanuel and {Gaensler}, B.~M. and {Kaczmarek}, Jane F. and {Lanmnan}, Adam E. and {Leung}, Calvin and {Pearlman}, Aaron B. and {Petroff}, Emily and {Pleunis}, Ziggy and {Rafiei-Ravandi}, Masoud and {Ransom}, Scott M. and {Shin}, Kaitlyn and {Scholz}, Paul and {Smith}, Kendrick and {Stairs}, Ingrid},
        title = "{Limits on Fast Radio Burst-like Counterparts to Gamma-Ray Bursts Using CHIME/FRB}",
      journal = {\apj},
         year = 2023,
        month = sep,
       volume = {954},
       number = {2},
          eid = {154},
        pages = {154},
          doi = {10.3847/1538-4357/ace52f},
archivePrefix = {arXiv},
       eprint = {2208.00803},
 primaryClass = {astro-ph.HE},
       adsurl = {https://ui.adsabs.harvard.edu/abs/2023ApJ...954..154C}
}

@ARTICLE{curtin24,
       author = {{Curtin}, Alice P. and {Sirota}, Sloane and {Kaspi}, Victoria M. and {Tendulkar}, Shriharsh P. and {Bhardwaj}, Mohit and {Cook}, Amanda M. and {Fong}, Wen-Fai and {Gaensler}, B.~M. and {Main}, Robert A. and {Masui}, Kiyoshi W. and {Michilli}, Daniele and {Pandhi}, Ayush and {Pearlman}, Aaron B. and {Scholz}, Paul and {Shin}, Kaitlyn},
        title = "{Constraining Near-simultaneous Radio Emission from Short Gamma-Ray Bursts Using CHIME/FRB}",
      journal = {\apj},
         year = 2024,
        month = sep,
       volume = {972},
       number = {1},
          eid = {125},
        pages = {125},
          doi = {10.3847/1538-4357/ad5c65},
archivePrefix = {arXiv},
       eprint = {2404.09242},
 primaryClass = {astro-ph.HE},
       adsurl = {https://ui.adsabs.harvard.edu/abs/2024ApJ...972..125C}
}

@ARTICLE{dai19,
       author = {{Dai}, Shi and {Lower}, Marcus E. and {Bailes}, Matthew and {Camilo}, Fernando and {Halpern}, Jules P. and {Johnston}, Simon and {Kerr}, Matthew and {Reynolds}, John and {Sarkissian}, John and {Scholz}, Paul},
        title = "{Wideband Polarized Radio Emission from the Newly Revived Magnetar XTE J1810-197}",
      journal = {\apjl},
         year = 2019,
        month = apr,
       volume = {874},
       number = {2},
          eid = {L14},
        pages = {L14},
          doi = {10.3847/2041-8213/ab0e7a},
archivePrefix = {arXiv},
       eprint = {1902.04689},
 primaryClass = {astro-ph.HE},
       adsurl = {https://ui.adsabs.harvard.edu/abs/2019ApJ...874L..14D}
}

@article{deRuiter2023,
 adsurl = {https://ui.adsabs.harvard.edu/abs/2023MNRAS.523..132D},
 archiveprefix = {arXiv},
 author = {{de Ruiter}, Iris and {Nyamai}, Miriam M. and {Rowlinson}, Antonia and {Wijers}, Ralph A.~M.~J. and {O'Brien}, Tim J. and {Williams}, David R.~A. and {Woudt}, Patrick},
 doi = {10.1093/mnras/stad1418},
 eprint = {2301.10552},
 journal = {\mnras},
 month = {July},
 number = {1},
 pages = {132-148},
 primaryclass = {astro-ph.HE},
 title = {{Low-frequency radio observations of recurrent nova RS Ophiuchi with MeerKAT and LOFAR}},
 volume = {523},
 year = {2023}
}

@ARTICLE{dessenne96,
       author = {{Dessenne}, C.~A. -C. and {Green}, D.~A. and {Warner}, P.~J. and {Titterington}, D.~J. and {Waldram}, E.~M. and {Barthelmy}, S.~D. and {Butterworth}, P.~S. and {Cline}, T.~L. and {Gehrels}, N. and {Palmer}, D.~M. and {Fishman}, G.~J. and {Kouveliotou}, C. and {Meegan}, C.~A.},
        title = "{Searches for prompt radio emission at 151 MHz from the gamma-ray bursts GRB 950430 and GRB 950706.}",
      journal = {\mnras},
         year = 1996,
        month = aug,
       volume = {281},
       number = {3},
        pages = {977-984},
          doi = {10.1093/mnras/281.3.977},
       adsurl = {https://ui.adsabs.harvard.edu/abs/1996MNRAS.281..977D}
}

@ARTICLE{driessen24,
       author = {{Driessen}, Laura Nicole and {Pritchard}, Joshua and {Murphy}, Tara and {Heald}, George and {Robrade}, Jan and {Das}, Barnali and {Duchesne}, Stefan William and {Kaplan}, David L. and {Lenc}, Emil and {Lynch}, Christene R. and {Mitchell-Bolton}, Jackson and {Pope}, Benjamin J.~S. and {Rose}, Kovi and {Stelzer}, Beate and {Wang}, Yuanming and {Zic}, Andrew},
        title = "{The Sydney Radio Star Catalogue: Properties of radio stars at megahertz to gigahertz frequencies}",
      journal = {\pasa},
         year = 2024,
        month = nov,
       volume = {41},
          eid = {e084},
        pages = {e084},
          doi = {10.1017/pasa.2024.72},
archivePrefix = {arXiv},
       eprint = {2404.07418},
 primaryClass = {astro-ph.SR},
       adsurl = {https://ui.adsabs.harvard.edu/abs/2024PASA...41...84D}
}

@ARTICLE{dobie19,
       author = {{Dobie}, D. and {Murphy}, T. and {Kaplan}, D.~L. and {Ghosh}, S. and {Bannister}, K.~W. and {Hunstead}, R.~W.},
        title = "{An optimised gravitational wave follow-up strategy with the Australian Square Kilometre Array Pathfinder}",
      journal = {\pasa},
         year = 2019,
        month = may,
       volume = {36},
          eid = {e019},
        pages = {e019},
          doi = {10.1017/pasa.2019.9},
archivePrefix = {arXiv},
       eprint = {1903.01481},
 primaryClass = {astro-ph.IM},
       adsurl = {https://ui.adsabs.harvard.edu/abs/2019PASA...36...19D}
}

@ARTICLE{dobie22,
       author = {{Dobie}, D. and {Stewart}, A. and {Hotokezaka}, K. and {Murphy}, Tara and {Kaplan}, D.~L. and {Buckley}, D.~A.~H. and {Cooke}, J. and {Ho}, A.~Y.~Q. and {Lenc}, E. and {Leung}, J.~K. and {Gromadzki}, M. and {O'Brien}, A. and {Pintaldi}, S. and {Pritchard}, J. and {Wang}, Y. and {Wang}, Z.},
        title = "{A comprehensive search for the radio counterpart of GW190814 with the Australian Square Kilometre Array Pathfinder}",
      journal = {\mnras},
         year = 2022,
        month = mar,
       volume = {510},
       number = {3},
        pages = {3794-3805},
          doi = {10.1093/mnras/stab3628},
archivePrefix = {arXiv},
       eprint = {2109.08452},
 primaryClass = {astro-ph.HE},
       adsurl = {https://ui.adsabs.harvard.edu/abs/2022MNRAS.510.3794D}
}

@ARTICLE{dulk85,
       author = {{Dulk}, G.~A.},
        title = "{Radio emission from the sun and stars.}",
      journal = {\araa},
         year = 1985,
        month = jan,
       volume = {23},
        pages = {169-224},
          doi = {10.1146/annurev.aa.23.090185.001125},
       adsurl = {https://ui.adsabs.harvard.edu/abs/1985ARA&A..23..169D}
}

@ARTICLE{falcke14,
       author = {{Falcke}, Heino and {Rezzolla}, Luciano},
        title = "{Fast radio bursts: the last sign of supramassive neutron stars}",
      journal = {\aap},
         year = 2014,
        month = feb,
       volume = {562},
          eid = {A137},
        pages = {A137},
          doi = {10.1051/0004-6361/201321996},
archivePrefix = {arXiv},
       eprint = {1307.1409},
 primaryClass = {astro-ph.HE},
       adsurl = {https://ui.adsabs.harvard.edu/abs/2014A&A...562A.137F}
}

@ARTICLE{fender04,
       author = {{Fender}, R.~P. and {Belloni}, T.~M. and {Gallo}, E.},
        title = "{Towards a unified model for black hole X-ray binary jets}",
      journal = {\mnras},
         year = 2004,
        month = dec,
       volume = {355},
       number = {4},
        pages = {1105-1118},
          doi = {10.1111/j.1365-2966.2004.08384.x},
archivePrefix = {arXiv},
       eprint = {astro-ph/0409360},
 primaryClass = {astro-ph},
       adsurl = {https://ui.adsabs.harvard.edu/abs/2004MNRAS.355.1105F}
}

@ARTICLE{Fender2006,
       author = {{Fender}, R.~P. and {Stirling}, A.~M. and {Spencer}, R.~E. and {Brown}, I. and {Pooley}, G.~G. and {Muxlow}, T.~W.~B. and {Miller-Jones}, J.~C.~A.},
        title = "{A transient relativistic radio jet from Cygnus X-1}",
      journal = {\mnras},
         year = 2006,
        month = jun,
       volume = {369},
       number = {2},
        pages = {603-607},
          doi = {10.1111/j.1365-2966.2006.10193.x},
archivePrefix = {arXiv},
       eprint = {astro-ph/0602307},
 primaryClass = {astro-ph},
       adsurl = {https://ui.adsabs.harvard.edu/abs/2006MNRAS.369..603F}
}

@ARTICLE{fender15,
       author = {{Fender}, R.~P. and {Anderson}, G.~E. and {Osten}, R. and {Staley}, T. and {Rumsey}, C. and {Grainge}, K. and {Saunders}, R.~D.~E.},
        title = "{A prompt radio transient associated with a gamma-ray superflare from the young M dwarf binary DG CVn.}",
      journal = {\mnras},
         year = 2015,
        month = jan,
       volume = {446},
        pages = {L66-L70},
          doi = {10.1093/mnrasl/slu165},
archivePrefix = {arXiv},
       eprint = {1410.1545},
 primaryClass = {astro-ph.HE},
       adsurl = {https://ui.adsabs.harvard.edu/abs/2015MNRAS.446L..66F}
}

@ARTICLE{fender23,
       author = {{Fender}, R.~P. and {Mooley}, K.~P. and {Motta}, S.~E. and {Bright}, J.~S. and {Williams}, D.~R.~A. and {Rushton}, A.~P. and {Beswick}, R.~J. and {Miller-Jones}, J.~C.~A. and {Kimura}, M. and {Isogai}, K. and {Kato}, T.},
        title = "{Comprehensive coverage of particle acceleration and kinetic feedback from the stellar mass black hole V404 Cygni}",
      journal = {\mnras},
         year = 2023,
        month = jan,
       volume = {518},
       number = {1},
        pages = {1243-1259},
          doi = {10.1093/mnras/stac1836},
archivePrefix = {arXiv},
       eprint = {2206.09831},
 primaryClass = {astro-ph.HE},
       adsurl = {https://ui.adsabs.harvard.edu/abs/2023MNRAS.518.1243F}
}

@ARTICLE{Fender2025,
       author = {{Fender}, R.~P. and {Motta}, S.~E.},
        title = "{The connection between the fastest astrophysical jets and the spin axis of their black hole}",
      journal = {Nature Astronomy},
         year = 2025,
        month = sep,
          doi = {10.1038/s41550-025-02665-w},
       adsurl = {https://ui.adsabs.harvard.edu/abs/2025NatAs.tmp..198F}
}

@ARTICLE{fishman94,
       author = {{Fishman}, Gerald J. and {Meegan}, Charles A. and {Wilson}, Robert B. and {Brock}, Martin N. and {Horack}, John M. and {Kouveliotou}, Chryssa and {Howard}, Sethanne and {Paciesas}, William S. and {Briggs}, Michael S. and {Pendleton}, Geoffrey N. and {Koshut}, Thomas M. and {Mallozzi}, Robert S. and {Stollberg}, Mark and {Lestrade}, John Patrick},
        title = "{The First BATSE Gamma-Ray Burst Catalog}",
      journal = {\apjs},
         year = 1994,
        month = may,
       volume = {92},
        pages = {229},
          doi = {10.1086/191968},
       adsurl = {https://ui.adsabs.harvard.edu/abs/1994ApJS...92..229F}
}

@article{Giroletti2020,
 adsurl = {https://ui.adsabs.harvard.edu/abs/2020A&A...638A.130G},
 archiveprefix = {arXiv},
 author = {{Giroletti}, M. and {Munari}, U. and {K{\"o}rding}, E. and {Mioduszewski}, A. and {Sokoloski}, J. and {Cheung}, C.~C. and {Corbel}, S. and {Schinzel}, F. and {Sokolovsky}, K. and {O'Brien}, T.~J.},
 doi = {10.1051/0004-6361/202038142},
 eid = {A130},
 eprint = {2005.06473},
 journal = {\aap},
 month = {June},
 pages = {A130},
 primaryclass = {astro-ph.SR},
 title = {{Very long baseline interferometry imaging of the advancing ejecta in the first gamma-ray nova V407 Cygni}},
 volume = {638},
 year = {2020}
}

@misc{tns,
  author       = {{Transient Name Server}},
  howpublished = {\url{https://www.wis-tns.org}},
  note         = {Accessed: 2025-10-19}
}

\end{document}